\documentclass[10pt-default]{article}
\usepackage{amsmath}
\usepackage{amssymb}
\usepackage[dvips]{graphicx}
\usepackage{amsfonts}

\input{tcilatex}

\begin{document}

{\huge \ }

{\Huge Heisenberg honeycombs}

{\Huge solve }{\huge Veneziano puzzle}

$\ \ \ \ \ \ \ \ \ \ \ \ \ \ \ \ \ \ \ \ \ \ \ \ \ \ \ \ \ \ \ \ \ $

\ \ \ \ \ \ \ \ \ \ \ \ \ \ \ \ \ \ \ \ \ \ \ \ \ \ \ A.L. Kholodenko

\textit{375 H.L.Hunter Laboratories, Clemson University, Clemson, }

\textit{SC} 29634-0973, USA

\bigskip \textbf{Abstract}

In this (expository) paper and its (more technical)\ companion we
reformulate some results of the Nobel Prize winning paper by Werner
Heisenberg into modern mathematical language of honeycombs. This language
was recently developed in connection with complete solution of the Horn
problem (to be defined and explained in the text). Such a reformulation is
done with the purpose of posing and solving the following problem: is by
analyzing the (spectroscopic) experimental data it possible to recreate the
underlying microscopic model generating these data? Although in the case of
Hydrogen atom positive answer is known, to obtain an affirmative answer for
spectra of other quantum \ mechanical systems is much harder task.
Development of Heisenberg's ideas happens to be the most useful for this
purpose. It supplies\ needed \ tools to solve the Veneziano and other
puzzles. The meaning of the word "puzzle" is two-fold. On one hand, it means
(in the case of Veneziano amplitudes) to find a physical model reproducing
these amplitudes. On another, from the point of view of combinatorics of
honeycombs, it means to find explicitly fusion rules for such amplitudes.
Solution of these tasks\ is facilitated by our earlier developed
string-theoretic formalism. In this paper only qualitative arguments are
presented (with few exceptions). These arguments provide enough evidence
that the underlying model compatible with Veneziano amplitudes is the
standard (i.e. non supersymmetric!) QCD. In addition, usefulness of \ the
proposed formalism is illustrated on numerous examples such as physically
motivated solution of the saturation conjecture (to be defined in the text),
derivation of the Yang-Baxter and Knizhnik-Zamolodchikov equations,Verlinde
and Hecke algebras, computation of the Gromov-Witten invariants for small
quantum cohomology ring, etc. Finally, we discuss possible uses of these
ideas in condensed matter physics

\bigskip



\textit{Subj Class}.: String theory; noncommutative geometry and analysis


\bigskip

\pagebreak

\bigskip

\section{Introduction}

\subsection{Some general facts about Veneziano condition and its CFT analog}

To avoid repetitions, we refer our readers to earlier \ papers, Refs.[1-3],
which shall be called Parts I-III respectively\footnote{%
In this work, when we occasionally refer to equations in other parts, \ we
shall use the following convention, e.g. Eq.(I 3.28), refers to Eq.(3.28) of
Part I, etc.}. In particular, in Part I we noticed that for the 4-particle
scattering amplitude the Veneziano condition is given by 
\begin{equation}
\alpha (s)+\alpha (t)+\alpha (u)=-1,  \tag{1.1}
\end{equation}%
where $\alpha (s)$, $\alpha (t),\alpha (u)$ $\in \mathbf{Z}$ . Eq.(1.1) is a
simple statement about the energy -momentum conservation. Although the
numerical entries in this equation can be changed to make them more suitable
for theoretical treatments, actual physical values can be subsequently re
obtained by appropriate coordinate shift. Such a procedure is not applicable
to amplitudes in conformal field theories (CFT) where the periodic
(antiperiodic, etc.) boundary conditions cause energy and momenta to become
a \textit{quasi -energy} and a \textit{quasi- momenta }in terminology taken
from the solid state physics. Noticed differences between the \
combinatorial properties of CFT and high energy physics amplitudes the major
theme of this work.

To explain things better, we would like \ to rewrite Eq.(1.1) in a more
convenient form. Following Ref.[4], without loss of generality, the
homogenous equation 
\begin{equation}
\alpha (s)m+\alpha (t)n+\alpha (u)l+k\cdot 1=0  \tag{1.2}
\end{equation}%
where $m,n,l,k$ are some integers can be added to Eq.(1.1) thus producing \
the Veneziano-type equation: 
\begin{equation}
\alpha (s)\tilde{m}+\alpha (t)\tilde{n}+\alpha (u)\tilde{l}=\tilde{k}.\text{
\ \ \ \ \ }(\text{Veneziano})  \tag{1.3a}
\end{equation}%
This equation is equivalent to 
\begin{equation}
n_{1}+n_{2}+n_{3}=\hat{N},  \tag{1.3b}
\end{equation}%
where \textit{by design} all entries are nonnegative integers. For the case
of multiparticle scattering we anticipate that this equation is to be
replaced by%
\begin{equation}
n_{0}+\cdot \cdot \cdot +n_{k}=N  \tag{1.4}
\end{equation}%
as discussed in Part II. Combinatorially, the task lies in finding all
nonnegative integer combinations of $n_{0},...,n_{k}$ satisfying Eq.(1.4).
It should be noted that such a task makes sense as long as $N$ is assigned.
But the actual value of $N$ is \textit{not} \textit{fixed} and, hence, can
be chosen quite arbitrarily. As explained in Part I, the value of $N$ should
coincide with the exponent of the Fermat (hyper)surface if (in contrast with
traditional string-theoretic treatments) we interpret Veneziano amplitudes
as periods of the Fermat (hyper) surfaces living in the complex projective
space. This requirement is not too rigid, however, in view of Eq.(I.3.29).
Indeed, the mathematical statement of the type given by Eq.(1.4) \ should be
considered \textit{before} the bracket operation \TEXTsymbol{<}...%
\TEXTsymbol{>} defined in Part I is applied. This means that we shall be
working mainly with\ the \textit{precursors} of the period integrals and,
therefore, we can choose any non negative integer value for $N$.Physically
correct value of $N$ can then be reobtained by the appropriate coordinate
shift.

In CFT such shift looses its meaning due to periodicity. In this case since
the Veneziano condition, Eq.(1.3a), should be replaced by the
Kac-Moody-Bloch-Bragg\ (K-M-B-B) condition (e.g. see Eq.(I.3.22)):%
\begin{equation}
\alpha (s)\tilde{m}+\alpha (t)\tilde{n}+\alpha (u)\tilde{l}=mN+nN+lN\text{,\
\ \ \ (}K-M-B-B\text{)}  \tag{1.5}
\end{equation}%
where $N$ is the same as before but $m,n,l$ are arbitrary integers\footnote{%
Clearly we can combine them together but we do not do this to keep an
analogy with the solid state physics.}. This circumstance \ causes the
energy (in free space) to become a quasi-energy (in solids). Eq.(1.5) is the
most fundamental equation in X-ray crystallography [5\textbf{]} where it is
known as the Bragg condition. In solid state physics essentially the same
equation is known as the Bloch equation. As it follows from monographs by
Kac [7] and, much earlier, by Bourbaki [8], the affine Lie groups and the
associated with them Lie algebras are generalizations of the Weyl-Coxeter
reflection groups made by analogy with crystallographic groups in solid
state physics. In the language of solid state physics, the Weyl-Coxeter
reflection groups are "point" groups while the affine groups are made of
semidirect products of translation groups and point groups are "spatial"
groups. Unlike the solid state physics, in the present case translations can
be performed in the Euclidean, hyperbolic or spherical spaces. These spaces
need not be 3 dimensional. To make a comparison with the existing
literature, we shall be concerned only with the Euclidean-type translations.
For a quick concise review of all these concepts we refer our readers to the
Appendix A of Part II.

The arbitrariness of choosing $N$ in Eq.(1.4) represents a kind of gauge
freedom. As in gauge theories, we can fix the gauge by using some physical
considerations. These include, for example, an observation made in Part I
that the 4-particle amplitude is zero if any two entries into Eq.(1.1) (or,
which is the same, into Eq.(1.3b)) are the same. This \ fact prompts us to
arrange the entries in Eq.(1.3b) in accordance with their magnitudes, e.g. $%
n_{1}\geq n_{2}\geq n_{3}.$ More generally (in view of Eq.(1.4)), we can
write: $n_{0}\geq n_{1}\geq \cdot \cdot \cdot \geq n_{k}\geq 1\footnote{%
The last inequality: $n_{k}\geq 1,$ is chosen only for the sake of
comparison with the existing literature conventions, e.g. see Ref.[6\textbf{]%
}.}$. Provided that Eq.(1.4) holds, we shall call such a sequence a \textit{%
partition }and denote it as\textit{\ }$\mathit{\lambda }\mathit{\equiv }%
(n_{0},...,n_{k})$. If $\lambda $ is a partition of $N$,then we shall write $%
\lambda \vdash N$. It is well known from combinatorics [4,9] that there is
one-to-one correspondence between the Young diagrams and partitions. We used
this observation in Part II for designing new partition function capable of
reproducing the Veneziano (and Veneziano-like) amplitudes. Now we would like
to look at the same problem from a different angle.

\subsection{The additive Horn problem}

Consider some $k\times k$ Hermitian matrix $\mathcal{H}$ whose spectrum $%
\lambda =\{\lambda _{1}\geq ...\geq \lambda _{k}\}$ can be written as a
partition made of weakly decreasing sequence of $k$ \textsl{real\footnote{%
In accord with the existing mathematical literature we refer to this (the
most general case) \ as "classical". Accordingly, the "quantum" case
corresponds to a situation when \ all numbers are integers. We shall say
more on this topic in Section 5.}} numbers. Conversely, for every spectrum $%
\lambda $ there is a set $\mathcal{O}_{\lambda }$ of Hermitian matrices
whose spectrum $\ $is $\lambda .$ Suppose that we are given 3 such spectra: $%
\lambda ,\mu $ and $\nu ,$ then there should be matrices $\mathcal{H}%
_{\lambda }\in \mathcal{O}_{\lambda },\mathcal{H}_{\mu }\in \mathcal{O}_{\mu
}$ and $\mathcal{H}_{\nu }\in \mathcal{O}_{\nu }$ such that $\mathcal{H}%
_{\lambda }+\mathcal{H}_{\mu }=\mathcal{H}_{\nu }.$ More precisely,
according to Hermann Weyl [10], the following problem can be formulated.

\textbf{Problem 1.1. }Assuming the eigenvalues of two $k\times k$ Hermitian
matrices $\mathcal{H}_{\lambda }$ and $\mathcal{H}_{\mu }$ \ to be known,
how does one determine all possible sets of eigenvalues for the sum $%
\mathcal{H}_{\lambda }+\mathcal{H}_{\mu }?\footnote{%
Such type of questions occur, for instance, in the theory of quantum
computation, \ [11,12]. Other applications are also possible and will be
considered elsewhere.}$

For $k=1$ the answer is obvious but for $k>1$ the answer is much less
obvious. Since $tr[\mathcal{H}_{\lambda }+\mathcal{H}_{\mu }]=tr[\mathcal{H}%
_{\nu }]$ we \textit{always} have the \textit{trace condition}:%
\begin{equation}
\nu _{1}+...+\nu _{k}=\lambda _{1}+...+\lambda _{k}+\mu _{1}+...+\mu _{k}. 
\tag{1.6}
\end{equation}%
Clearly, in addition, we can expect that $\nu _{1}\leq \lambda _{1}+\mu _{1}$
since the largest eigenvalue of $\mathcal{H}_{\lambda }+\mathcal{H}_{\mu }$
\ is at most the sum of $\mathcal{H}_{\lambda }$ and $\mathcal{H}_{\mu }$
individual eigenvalues.

Let now \textbf{I}, \textbf{J} and \textbf{K} be subsets of $\{1,...,k\}$
then, what can be said about the validity of the inequality 
\begin{equation}
\dsum\limits_{i\in \mathbf{I}}\nu _{i}\leq \dsum\limits_{j\in \mathbf{J}%
}\lambda _{j}+\dsum\limits_{k\in \mathbf{K}}\mu _{k}\text{ ?}  \tag{1.7}
\end{equation}%
Such type of inequalities \ were analyzed by Horn [13] who formulated the
following.

\textbf{Conjecture 1.2}. (Horn) A triple $\lambda ,\mu $ and $\nu $ can
represent the eigenvalues of the $\tilde{k}\times \tilde{k}$ Hermitian
matrices $\mathcal{H}_{\lambda },\mathcal{H}_{\mu }$ and $\mathcal{H}_{\nu
}, $ where $\mathcal{H}_{\lambda }+\mathcal{H}_{\mu }=\mathcal{H}_{\nu },$
if and only if the trace condition, Eq.(1.6), holds and, in addition, the
inequality given by Eq.(1.7) holds \ for all sets $\{\mathbf{I},\mathbf{J}$ $%
and$ $\mathbf{K\}}\in T_{r}^{\tilde{k}}$ , $r<\tilde{k}.$ $T_{r}^{\tilde{k}}$
is defined recursively as follows. For each positive integer $k$ and $r\leq 
\tilde{k}$, let 
\begin{equation}
U_{r}^{\tilde{k}}=\{(\mathbf{I},\mathbf{J},\mathbf{K})\mid \sum\limits_{i\in 
\mathbf{I}}i+\sum\limits_{j\in \mathbf{J}}j=\sum\limits_{k\in \mathbf{K}%
}k+r(r+1)/2\}.  \tag{1.8}
\end{equation}%
In this case for $r=1$, let $T_{1}^{\tilde{k}}=U_{1}^{\tilde{k}}$ while for $%
r>1,$ let%
\begin{eqnarray*}
T_{r}^{\tilde{k}} &=&\left\{ (\mathbf{I},\mathbf{J},\mathbf{K)}\in U_{r}^{%
\tilde{k}}\mid \text{ \ for all }p<r\text{ and for all }(\mathbf{F},\mathbf{G%
},\mathbf{H})\in T_{p}^{\tilde{k}},\right. \\
&&\left. \dsum\limits_{f\in \mathbf{F}}i_{f}+\dsum\limits_{g\in \mathbf{G}%
}j_{g}\leq \dsum\limits_{h\in \mathbf{H}}k_{h}+p(p+1)/2\right\} .
\end{eqnarray*}%
The above conjecture in principle can be checked directly using the above
described recurrence. In reality, for not too large $k^{\prime }s$ the above
recurrence becomes impractical. Hence, the conjecture remained unproven for
about 36 years since its formulation. A complete solution was found
independently by Klyachko [14] and Knutson, Tao and Woodward [15, 16] (KTW).
Moreover, the infinite dimensional generalization of Klyachko results was
recently obtained also by Friedland [17]. In this work we shall discuss some
results of KTW since, in our opinion, they have some physical appeal. In
particular, they can be used for solution of the following problem.

\textbf{Problem 1.3.} Is it possible to design a diagrammatic method for
description of \textsl{fusion algebra} for the Veneziano and Veneziano-like
amplitudes ? \ \ \ \ \ \ \ \ \ \ \ \ \ \ \ \ \ \ Stated differently, suppose
we have the Veneziano\ (or Veneziano-like) amplitides $F(\lambda )$, $F(\mu
) $ and $F(\nu ),$ can we represent the product $F(\lambda )F(\mu )$ as%
\begin{equation}
F(\lambda )\cdot F(\mu )=\dsum\limits_{\nu }C_{\lambda \mu }^{\nu }F(\nu )%
\text{ }  \tag{1.9}
\end{equation}%
with coefficients $C_{\lambda \mu }^{\nu }$ \ whose calculation can be
completely described ? Using results by KTW we shall provide an affirmative
answer to this question.

\subsection{Organization of the rest of the paper}

\bigskip

In addition to the topics already mentioned we would like to provide a
summary of the content of this paper. In Section 2 we compare KTW results
with those obtained much earlier by Heisenberg [18]. Although not present in
textbooks on quantum mechanics, Heisenberg's original formulation of quantum
mechanics is very much in accord with that developed by KTW (who had
different purposes in mind). For this reason we decided to name the KTW
honeycombs ( to be introduced in Section 2) as Heisenberg honeycombs. The
whole discussion of this section is motivated by the observation that the
honeycomb condition(s) and Veneziano condition(s) are mathematically the
same. In Section 3 we review some results from Parts II and III in order to
formulate additional problems to be discussed in this (expository) paper and
later, in its more technical companion, Ref.[19]. In particular, in this
section we formulate problem about the fusion rules for the Veneziano
amplitudes. Our discussion is not limited to these amplitudes however since
combinatorially all scattering processes of high energy physics have many
things in common. In the same section we argue (in accord with Parts II\ and
III) that combinatorial considerations alone are sufficient for recovering
the microscopic model leading to Veneziano and Veneziano-like amplitudes. We
provide a qualitative evidence that such a model should coincide with the
standard QCD leaving all details to [19\textbf{]. }In Section 4 we discuss
actual calculations of the Littlewood-Richardson (fusion) coefficients using
KTW formalism of honeycombs and puzzles. Although highly nontrivial in their
design and mathematical justification, these computational tools make
calculation of fusion coefficients as simple as possible and can be compared
in their simplicity with Feynman's diagrams. The advantage of utilization of
such methods of calculation becomes especially apparent after we introduce
and discuss the saturation conjecture and its solution in Section 5. In
short, the solution of the saturation conjecture allows us to significantly
enlarge number of fusion coefficients using as an input just few. Moreover,
the mathematical proof of this conjecture imposes some unexpected extra
constraints on fusion coefficients which can be checked experimentally.
Section 6 provides necessary background for the multiplicative Horn problem
to be discussed in Section 7. Section 6 is also of independent interest
since it provides the most economical and physically convincing way to
arrive at the classical and quantum Yang-Baxter equations,
Knizhnik-Zamolodchikov equations, Dunkl operators and Hecke algebra. The
multiplicative Horn problem discussed in Section 7 emerges in various
physical contexts. For instance, it emerges in connection with study of
solutions of Knizhnik-Zamolodchikov equations, study of spectra of quantum
spin chains, study of band structure of solids, etc. Although these problems
are of no immediate use for develpment of our string-theoretic formalism,
they are logically compatible with this formalism and are of independent
interest as we explain in Section 7. The treatment of these physically
interesting problems depends upon computation of the 3-point Gromov-Witten
(G-W) invariants which are structure constants of small "quantum" cohomology
ring. We enclose the word "quantum" in quotation marks since, actually, it
should be called "double quantum" or, better, the "deformed quantum" as we
shall explain in the text. Computation of these invariants in Section 7 is
non traditional in the sence that it is made for people with standard
physical education (that is for people who are experts in fields other than
string theory). It is hoped, that the provided background should be
sufficient for proper understanding of current physical and mathematical
literature on these subjects thus leading to their further uses in condensed
matter physics. To keep the main text focused, some computations are
presented in Appendices A through C.

\section{Heisenberg Honeycomb and Veneziano condition}

Following Knutson and Tao [15,16,20] (KT), we would like to describe \ the
construction of a honeycomb graph-a precursor of the Veneziano puzzle-which
is used for calculation of $C_{\lambda \mu }^{\nu }$. In addition, \ the
main mathematical purpose of such a honeycomb is to provide constructive
solution of the Horn conjecture. In this paper we mainly keep focus on the
physical task of calculating $C_{\lambda \mu }^{\nu }$. By doing so we
provide a sketch of how the Horn conjecture was solved with help of
honeycombs. Details can be found in the literature already cited.

To begin, we need to work out the one dimensional example (a 1-honeycomb) in
some detail in order to proceed inductively. It is helpful to replace the
equation $\mathcal{H}_{\lambda }+\mathcal{H}_{\mu }=\mathcal{H}_{\nu }$ by
the analogous equation $\mathcal{H}_{\lambda }+\mathcal{H}_{\mu }+\mathcal{H}%
_{\nu }=0,$ provided that $\lambda +\mu +\nu =0.$ Under such circumstances
the inequality $\nu _{1}\leq \lambda _{1}+\mu _{1}$ is replaced by $\ 0\leq
\lambda _{1}+\mu _{1}+\nu _{1}.$ \ Although these results are sufficient for
construction of 1-honeycomb, before doing so we would like to put them in
some historical perspective.

In his Nobel Prize winning paper [18] Werner Heisenberg made the following
observation. Influenced by Bohr's ideas, he looked at the famous equation
for the energy levels difference: 
\begin{equation}
\omega (n,n-\alpha )=\frac{1}{h}(E(n)-E(n-\alpha )),  \tag{2.1}
\end{equation}%
where both $n$ and $n-\alpha $ are some integers. He noticed that this
definition allows him to write the following \ fundamental composition law%
\begin{equation}
\omega (n-\beta ,n-\alpha -\beta )+\omega (n,n-\beta )=\omega (n,n-\alpha
-\beta ),  \tag{2.2}
\end{equation}%
or, since $\omega (k,n)=-\omega (n,k),$ the above equation can be rewritten
in a more symmetric form 
\begin{equation}
\omega (n,m)+\omega (m,k)+\omega (k,n)=0  \tag{2.3}
\end{equation}%
which explains its relevance to the KT results. Most likely, being aware of
Heisenberg's unpublished letter to Kr\"{o}nig$\footnote{%
Dated by June 5th 1925 (Ref.[18], page 331).},$ Dirac in his paper,
Ref.[21], of October 7th 1925 noticed that using the above combinatorial law
for the frequencies in the Fourier expansions of observables leads to the
multiplication rule for the Fourier amplitudes:%
\begin{equation}
a(nm)b(mk)=ab(nk).  \tag{2.4}
\end{equation}%
By noticing that, in general,%
\begin{equation}
ab(nk)\neq ba(nk),  \tag{2.5}
\end{equation}%
he concluded (in accord with Heisenberg's \ earlier cited letter) that the
above multiplication rule is characteristic for matrices. Hence, the Fourier
amplitudes are actually matrices! After this observation, to arrive formally
at the famous quantization condition 
\begin{equation}
\lbrack \hat{x},\hat{p}]=i\hbar  \tag{2.6}
\end{equation}%
using Eq.s (2.4) and (2.5) is quite straightforward. For instance, by using
the famous Dirac prescription, e.g. see Eq.(11) of his paper or his book,
Ref.[22, page 86], 
\begin{equation}
i\hbar \{x,p\}_{P.B.}=[\hat{x},\hat{p}],  \tag{2.7}
\end{equation}%
where \{ , \}$_{P.B.}$ is the classical Poisson bracket. Such quantization
prescription is \textsl{very formal} as Dirac freely admits in his paper,
Ref.[21], Section 4. Its correctness is based on the historical curiosity
which we discuss in detail in Section 5 in connection with physical
(Heisenberg-style) solution of the saturation conjecture. In the meantime,
we would like to return to Eq.(2.3) in order to analyze it from the point of
view of modern mathematics.

Having in mind the KT results [15,16,20], we would like to discuss briefly
some earlier efforts by Lidskii, Ref.[23], aimed at solution of the Horn
conjecture. Lidskii considered $3n-1$ dimensional \textsl{real} vector space
whose coordinates ($x_{1},...,x_{n};y_{1},...,y_{n};z_{1},...,z_{n})$ are
constrained by the equation 
\begin{equation}
\tsum\limits_{j=1}^{n}(x_{j}+y_{j}-z_{j})=0.  \tag{2.8}
\end{equation}%
Let $W_{n}$ be a subspace of \ \textbf{R}$^{3n}$ determined by the following
inequalities: $x_{1}$ $\geq x_{2}\cdot \cdot \cdot \geq x_{n};y_{1}$ $\geq
y_{2}\cdot \cdot \cdot \geq y_{n};z_{1}$ $\geq z_{2}\cdot \cdot \cdot \geq
z_{n}.$ Furthermore, let $S_{n}$ be a subset of $W_{n}$ determined by the
additional conditions. If $x=(a_{1},...,a_{n};b_{1},...,b_{n}$ ; $%
c_{1},...,c_{n})$ and $x\in S_{n}$, then this is possible if and only if
there are linear Hermitian operators \ $A,B$ and $C$ acting in $n-$%
dimensional complex space whose eigenvalues are $%
(a_{1},...,a_{n}),(b_{1},...,b_{n})$ and ($c_{1},...,c_{n})$ respectively,
provided that, in addition, $A+B=C$. The problem lies in describing $S_{n}.$
Lidskii proves that $S_{n}$ is described by the inequalities given by
Eq.(1.7) supplemented by the ordering requirements on $x_{i},y_{i}$ and $%
z_{i}$ and the Horn conditions, Eq.(1.8). According to Zelevinsky [24],
Lidskii actually have not supplied a complete proof. Nevertheless, his
results apparently have a stimulating effect on the KT work.

Knutson and Tao solved this problem differently in ingenious geometric way
by replacing $3n-1$ space with specially chosen 2 dimensional plane in $%
\mathbf{R}^{3}$ defined by 
\begin{equation}
\mathbf{R}_{\sum =0}^{3}:=\{\{x,y,z\}\in \mathbf{R}^{3}:x+y+z=0\}.  \tag{2.9}
\end{equation}%
The rationale for this plane can be understood using method of induction. In
one dimensional case the condition $\lambda +\mu +\nu =0$ corresponds to
some particular point in $\mathbf{R}_{\sum =0}^{3}$ (which we shall \
temporarily denote as $(\lambda ,\mu ,\nu )$ ). In their original work,
Ref.[15], KT introduce both the \textit{honeycomb} \textit{tinkertoys} and 
\textit{honeycombs}. The tinkertoy is some abstract \textit{directed} graph
placed in $\mathbf{R}_{\sum =0}^{3}$ plane whose configuration is determined
by (encoded by) the honeycomb. In order to illustrate these concepts we need
to establish several additional rules defining honeycombs. For instance,
with each tinkertoy in $\mathbf{R}_{\sum =0}^{3}$ plane we associate another
2 dimensional (honeycomb) plane (2-plane for short) where lines can be drawn
only in 3 possible directions: northeast-southwest (Ne-Sw), north-south
(N-S) and northwest--southeast (Nw-Se) directions. On such 2-plane we can
place a Y-shaped tripod made of nonoriented labeled semiinfinite edges as
depicted in Fig.1a).The topological information encoded in such 1-honeycomb
is used for construction of 1-honeycomb tikertoy, Fig.1b). The location of
the vertex on such tikertoy $\mathbf{R}_{\sum =0}^{3}$ plane is determined
by the condition $\bar{\lambda}+\bar{\mu}+\bar{\nu}=0$ imposed on the
vectors $\bar{\lambda},\bar{\mu}$ and $\bar{\nu}$ in accord with Eq.(2.9).


\begin{figure}[tbp]
\begin{center}
\includegraphics[width=3.00 in]{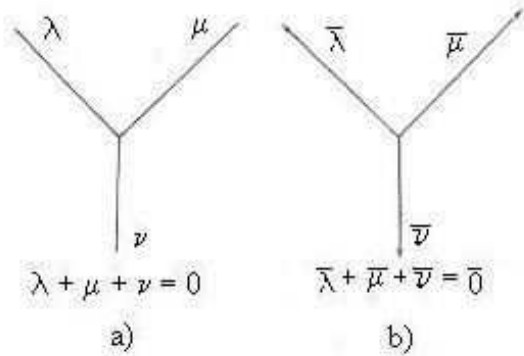}
\end{center}
\caption{The 1-honeycomb a) and its 1-honeycomb tinkertoy b)}
\end{figure}

Looking at this figure more general rules for $n-$honeycombs can be
developed. These are:

\bigskip a) there is one-to-one correspondence between the points in $%
\mathbf{R}_{\tsum =0}^{3}$ \ plane

\ \ \ \ and the vertices in 2-plane;

b) there should be only finitely many vertices inside the honeycomb diagram;

c) the semiinfinite lines at the boundary of the honeycomb's boundary are

\ \ \ allowed to go only in the Ne, Nw and S directions in the 2-plane. \ \
\ \ \ \ \ \ \ \ \ \ \ \ \ \ \ \ \ \ \ \ \ \ \ \ \ \ \ \ \ \ \ \ \ \ \ \ \ \
\ \ \ \ 

\bigskip

\textbf{Remark 2.1}. To make sense out of the defining rules for honeycombs,
our readers are encouraged to look at the following web site:

http://www.math.ucla.edu/\symbol{126}tao/java/Honeycomb.html \newline
from which they can get an idea of how rules just described are implemented.

\textbf{Remark 2.2} By comparing the Veneziano condition, Eq.(1.3b), with
the definition of $\mathbf{R}_{\tsum =0}^{3}$ plane, Eq.(2.9), it is clear
that the condition, Eq.(1.3b), can be brought to the form given in Eq.(2.9).
Since the Heisenberg frequency condition, Eq.(2.3), is exactly the same as
condition in Eq.(2.9) and since it historically was formulated much earlier,
we shall call just described \ honeycombs as \textsl{Heisenberg honeycombs}.

Assuming that our readers looked at the suggested web site, we still would
like to describe some details about how these honeycombs are constructed in
order to provide some sketch of KT way of solving the Horn conjecture.

Although the case of 1-honeycomb is seemingly trivial, already description
of 2-honeycombs provides much more information useful for inductive
analysis.\ The suggested web link allows our readers to recreate a
2-honeycomb and the associated with it 2-honeycomb tinkertoy. Fig.2 provides
additional details.


\begin{figure}[tbp]
\begin{center}
\includegraphics[width=3.00 in]{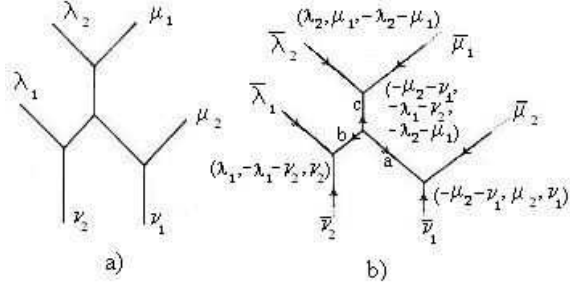}
\end{center}
\caption{The 2-honeycomb a) and its 2-honeycomb tinkertoy b). The euclidean
coordinates (x,y,z) of the vertices of such tinkertoy are shown explicitly}
\end{figure}

By discussing these details the rationale for keeping both the \ honeycomb
tinkertoys and honeycombs should become apparent. In particular, both the
tinkertoy and honeycomb are uniquely determined by their boundary values,
i.e. by the prescribed set ($\lambda ,\mu ,\nu )\equiv (\lambda _{1},\lambda
_{2};\mu _{1},\mu _{2};\nu _{1},\nu _{2})$ for which the equality like that
given by Eq.(1.6) is satisfied$.$That is we must have%
\begin{equation}
\lambda _{1}+\lambda _{2}+\mu _{1}+\mu _{2}+\nu _{1}+\nu _{2}=0.  \tag{2.10}
\end{equation}%
In addition, however, for the tinkertoy, the length of the inner edges
(determined as the \textit{Euclidean distance in }$\mathbf{R}_{\tsum =0}^{3}$
plane between the adjacent vertices) matters too. For instance, the
Euclidean length of the vector \textbf{b }(may be, up to\textbf{\ }%
irrelevant factor of $\sqrt{2})$\textbf{\ }is given by $\lambda _{1}+\mu
_{2}+\nu _{1}\geq 0$ thus providing us with the 1st new nontrivial
inequality. Clearly, the Euclidean lengths of vectors $\mathbf{a}$ and $%
\mathbf{c}$ \ are obtained in the same way and are given respectively by $%
\lambda _{2}+\mu _{1}+\nu _{1}\geq 0$ and $\lambda _{1}+\mu _{1}+\nu
_{2}\geq 0.$ In view of Eq.(2.10) these inequalities can be restated as
triangle inequality for the triangle made of sides whose lengths are $%
\lambda _{2}-\lambda _{2},\mu _{1}-\mu _{2}$ and $\nu _{1}-\nu _{2}$
respectively. Thus, the Horn conjecture in this case is solved completely by
simple geometrical means. For $n>2$ the $n-$honeycomb is determined by its $%
3n$ boundary values ($\lambda ,\mu ,\nu )\equiv (\lambda _{1},...,\lambda
_{n};\mu _{1},...,\mu _{n};\nu _{1},...,\nu _{n})$ subject to the constraint
analogous to Eq.(2.10). In this case, however, the boundary values \textbf{do%
} \textbf{not} determine the $n-$honeycomb uniquely. Even though they do not
determine the $n-$honeycomb uniquely, there is only finite number of
different honeycombs still. Evidently, each of these more complex honeycombs
will look like that depicted in Fig.3


\begin{figure}[tbp]
\begin{center}
\includegraphics[width=3.00 in]{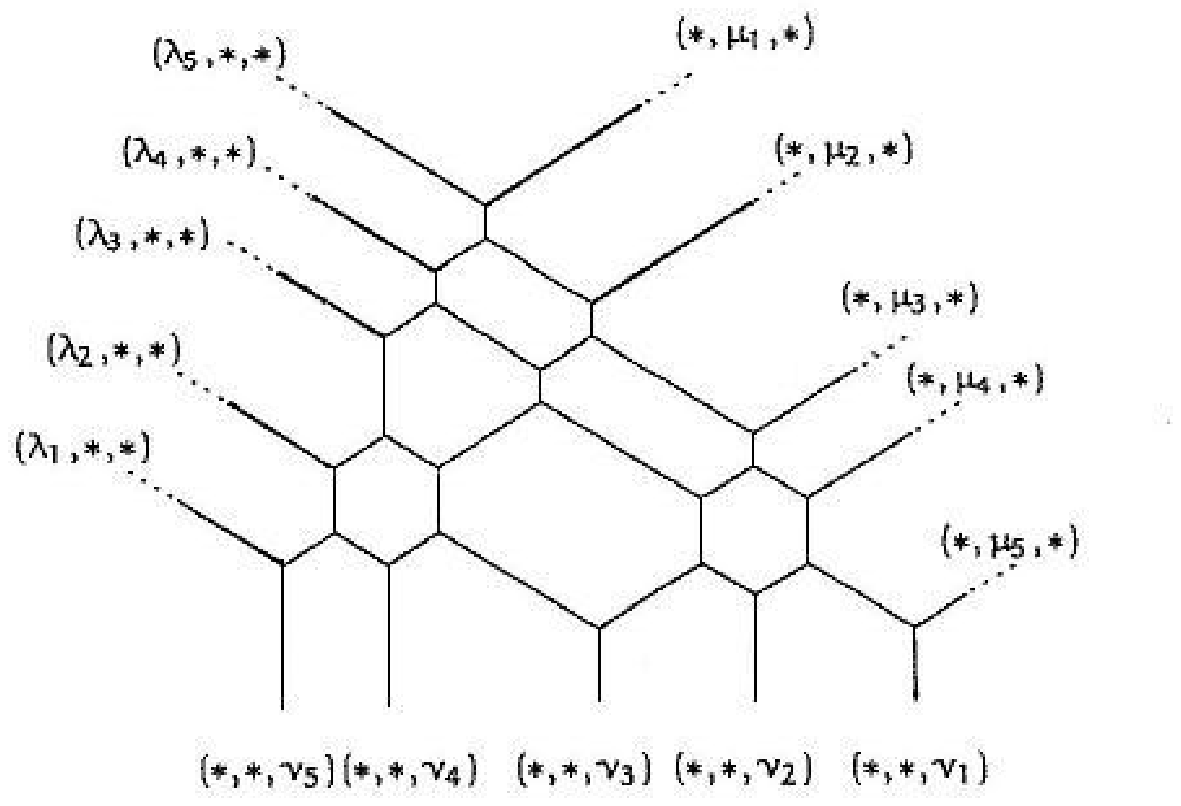}
\end{center}
\caption{A typical 5-honeycomb}
\end{figure}

Already in Part I we noticed that the multiparticle Veneziano amplitude can
be factorized into product of 4-particle amplitudes, e.g. see Eq.(I.3.28).
Since with each such factorized amplitude we can associate a 1-honeycomb, it
is only natural to expect that the higher order honeycombs can be built from
an assembly of 1-honeycombs. This indeed happens to be the case and provides
a compelling reason for our use of honeycombs for description of the
combinatorics of Veneziano (and Veneziano-like) amplitudes.

To explain how this happens we need to pay attention to the labeling pattern
, e.g. that for the 5-honeycomb depicted in Fig.3. One notices that
numeration of external indices in the Ne-Sw direction goes from the
bottom-up while in the Nw-Se direction-from the top-down. In the S direction
the numeration goes from the right to left. This observation is essential
for designing honeycombs. We would like to illustrate it by designing, say,
a 4-honeycomb. For this purpose we have to assemble 4 1-honeycombs in the
way depicted in Fig.4.


\begin{figure}[tbp]
\begin{center}
\includegraphics[width=2.20 in]{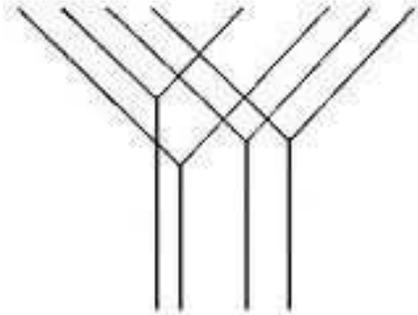}
\end{center}
\caption{A precursor of a typical 4-honeycomb is made of an assembly of 4
1-honeycombs}
\end{figure}

What is depicted in this figure is still a precursor of the honeycomb. To
convert this precursor into the honeycomb we need to resolve each 4-vertex
into two 3-vertices as depicted in Fig.5.


\begin{figure}[tbp]
\begin{center}
\includegraphics[width=3.00 in]{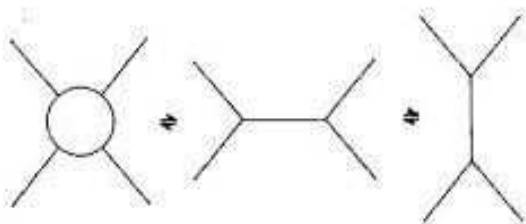}
\end{center}
\caption{Resolution of a typical 4-vertex into 3-vertices}
\end{figure}

Everybody familiar with string theory at this point will be able to
recognize the famous duality property depicted diagrammatically, e.g. read
Ref.[25], page 265. Unlike known particle physics case, such a resolution
must be performed in accord with the rules \ for honeycombs defined earlier.
It is also evident that the existence of such a resolution is primary cause
\ for the higher order honeycombs not to be determined only by the boundary
values.

To illustrate these ideas we apply the rules just discussed to Fig.4 (but
adopted to 2-honeycombs). In this case to make the duality of Fig.5
compatible with the rules defining honeycombs we have to give preference to
only one type of resolution of 4-vertex. This fact explains why 2-honecomb
is determined uniquely by its boundary data. This also explains why the
higher order honeycombs are not determined by their boundary data uniquely.
The honeycomb rules, although quite natural, still \ are not too physically
illuminating. To bring some physics into this discussion requires several
steps. These are described below.

\section{Complex Grassmannians \ and their relation to the Veneziano-like
amplitudes}

\subsection{ Designing partition function reproducing Veneziano (and
Veneziano-like) amplitudes}

In Parts II and III we constructed the partition function reproducing
Veneziano amplitudes. In this paper and its companion we would like to
provide additional details needed for establishing firm links \ between this
partition function and the microscopic model, e.g. QCD. \ In \ Part II we
used the one-to-one correspondence between partitions and the Young tableaux
in order to arrive at correct result for the partition function. These
simple arguments were reinforced by much deeper results taken from the
theory of group invariants for pseudo-reflection groups. \ In this paper we
follow the same philosophy by providing simple qualitative arguments prior
to more sophisticated ones.

We would like to recall some facts from Part II since they will be used
later, in Section 7, when we shall discuss details of computation of the
Gromov-Witten invariants. For this purpose we choose a square lattice and
place on it the Young diagram (tableaux) related to some partition $\lambda $
such that $\lambda \vdash N$. To do so, we choose some $\tilde{n}\times 
\tilde{m}$ rectangle\footnote{%
The parameters $\tilde{n}$ and $\tilde{m}$ will be specified below.} so that
the Young diagram occupies the left part of this rectangle. We choose the
upper left vertex of the rectangle as the origin of the $xy$ coordinate
system whose $y$ axis (south direction) is directed downwards and $x$ axis
is directed eastwards. Then, the south-east boundary of the Young diagram
can be interpreted as directed (that is without self-intersections) random
walk which begins at $(0,-\tilde{m})$ and ends at $(\tilde{n},0).$ Clearly,
such a walk completely determines the diagram. The walk can be described by
a sequence of $0^{\prime }$s and $1^{\prime }$s,\ say, $0$ for the $x-$step
move and $1$ for the $y-$step move. We shall use a notation $\omega (\lambda
)$ for such a walk so that $\omega (\lambda ):=(\omega _{1}$ $,...,\omega
_{N})$\ where the random "occupation" numbers $\omega _{i}=0,1$ are
analogous to those used in the Fermi statistics. Evidently $\omega _{1}$ +$%
\omega _{2}+\cdot \cdot \cdot +\omega _{N}=\tilde{m}$. The totality $%
\mathcal{N}$ of Young diagrams which can be placed into rectangle is in
one-to-one correspondence with the number of arrangements of 0's and 1's
whose number $N$ is $\tilde{m}+\tilde{n}$. The logarithm of the number $%
\mathcal{N}$ of possible combinations of 0's and 1's \ is just the entropy
associated with the Fermi statistic (or, equivalently, the entropy of mixing
for the binary mixture) used in physics literature. The number $\mathcal{N}$
is given by $\mathcal{N}=(\tilde{m}+\tilde{n})!/\tilde{m}!\tilde{n}!$. It
can be represented in two equivalent ways%
\begin{eqnarray}
(\tilde{m}+\tilde{n})!/\tilde{m}!\tilde{n}! &=&\frac{(\tilde{n}+1)(\tilde{n}%
+2)\cdot \cdot \cdot (\tilde{n}+\tilde{m})}{\tilde{m}!}\equiv \left( 
\begin{array}{c}
\tilde{n}+\tilde{m} \\ 
\tilde{m}%
\end{array}%
\right)  \notag \\
&=&\frac{(\tilde{m}+1)(\tilde{m}+2)\cdot \cdot \cdot (\tilde{n}+\tilde{m})}{%
\tilde{n}!}\equiv \left( 
\begin{array}{c}
\tilde{m}+\tilde{n} \\ 
\tilde{n}%
\end{array}%
\right) .  \TCItag{3.1}
\end{eqnarray}%
In Part I, Eq-s (I.1.21)-(I. 1.23) \ explain how the factor $\mathcal{N}$ is
entering the Veneziano amplitude. Additional significance of this number in
connection with Veneziano amplitudes is discussed at length in both in Parts
II and III.

Let now $p(N;\tilde{k},\tilde{m})$ be the number of partitions of $N$ into $%
\leq \tilde{k}$ \ non negative parts, each not larger than $\tilde{m}$.
Consider the generating function of the following type: 
\begin{equation}
\mathcal{F}(\tilde{k},\tilde{m}\mid q)=\dsum\limits_{N=0}^{S}p(N;\tilde{k},%
\tilde{m})q^{N},  \tag{3.2}
\end{equation}%
where the upper limit $S$\ will be determined shortly below. It is shown in
Refs.[3-5,9] that $\mathcal{F}(\tilde{k},\tilde{m}\mid q)=\left[ 
\begin{array}{c}
\tilde{k}+\tilde{m} \\ 
\tilde{m}%
\end{array}%
\right] _{q}$

$\equiv \left[ 
\begin{array}{c}
\tilde{k}+\tilde{m} \\ 
\tilde{k}%
\end{array}%
\right] _{q}$ where, for instance,$\left[ 
\begin{array}{c}
\tilde{k}+\tilde{m} \\ 
\tilde{m}%
\end{array}%
\right] _{q=1}=\left( 
\begin{array}{c}
\tilde{k}+\tilde{m} \\ 
\tilde{m}%
\end{array}%
\right) \footnote{%
On page 15 of the book by Stanley, Ref.[9], one can find that the number of
solutions $N(n,k)$ in \textit{positive} integers to $y_{1}+...+y_{k}=n+k$ is
given by $\left( 
\begin{array}{c}
n+k-1 \\ 
k-1%
\end{array}%
\right) $ while the number of solutions in \textit{nonnegative} integers to $%
x_{1}+...+x_{k}=n$ is $\left( 
\begin{array}{c}
n+k \\ 
k%
\end{array}%
\right) .$Careful reading of Page 15 indicates however that the last number
refers to solution in nonnegative integers of the equation $%
x_{0}+...+x_{k}=n $. We have used this fact in Part I, e.g. see Eq.(I.1.21).}%
.$ The expression $\left[ 
\begin{array}{c}
\tilde{k}+\tilde{m} \\ 
\tilde{m}%
\end{array}%
\right] _{q}$ is a $q-$analog of the binomial coefficient $\left( 
\begin{array}{c}
\tilde{k}+\tilde{m} \\ 
\tilde{m}%
\end{array}%
\right) .$ In the literature [3-5,9] this $q-$ analog is known as the 
\textit{Gaussian} coefficient. Explicitly,%
\begin{equation}
\left[ 
\begin{array}{c}
k \\ 
m%
\end{array}%
\right] _{q}=\frac{(q^{k}-1)(q^{k-1}-1)\cdot \cdot \cdot (q^{k-m+1}-1)}{%
(q^{m}-1)(q^{m-1}-1)\cdot \cdot \cdot (q-1)}.  \tag{3.3}
\end{equation}%
From this definition it should be intuitively clear that the sum defining
generating function $\mathcal{F}(\tilde{k},\tilde{m}\mid q)$ in Eq.(3.2)
should have only \textit{finite} number of terms. Eq.(3.3) allows easy
determination of the upper limit $S$ in the sum, Eq.(3.2). It is given by $%
\tilde{k}\tilde{m}$. This is just the area of $\tilde{k}\times \tilde{m}$
rectangle. Evidently, in view of the definition of $p(N;\tilde{k},\tilde{m})$%
, the number $\tilde{m}=N-\tilde{k}$. Using this fact, Eq.(3.2) can be
rewritten as: $\mathcal{F}(N,N-\tilde{k}\mid q)=\left[ 
\begin{array}{c}
N \\ 
\tilde{k}%
\end{array}%
\right] _{q}.$ This expression happens to be the Poincare$^{\prime }$
polynomial for the complex Grassmannian $Gr(\tilde{m},\tilde{k})$. This can
be found on page 292 of the famous book by Bott and Tu, Ref.[26]\footnote{%
To make a comparison it is sufficient to replace parameters $t^{2}$ and $n$
in \ Bott and Tu book by $q$ and $N.$}. From this point of view the
numerical coefficients, i.e. $p(N;\tilde{k},\tilde{m}),$ in the $q$
expansion of Eq.(3.2) should be interpreted as Betti numbers of this
Grassmannian. They can be determined recursively using the following
property of the Gaussian coefficients [9, page 26]%
\begin{equation}
\left[ 
\begin{array}{c}
n+1 \\ 
k+1%
\end{array}%
\right] _{q}=\left[ 
\begin{array}{c}
n \\ 
k+1%
\end{array}%
\right] _{q}+q^{n-k}\left[ 
\begin{array}{c}
k \\ 
m%
\end{array}%
\right] _{q}  \tag{3.4}
\end{equation}%
and taking into account that $\left[ 
\begin{array}{c}
n \\ 
0%
\end{array}%
\right] _{q}=1.$ To connect this result with partition function reproducing
Veneziano (and Veneziano-like) amplitudes we notice that, in view of
relation $\tilde{m}=N-\tilde{k},$ it is more advantageous \ for us to use
parameters $\tilde{m}$ and $\tilde{k}$ than $N$ and $\tilde{k}$. With this
in mind we obtain\footnote{%
From now on we shall drop the tildas for $k$ and $m$.}, 
\begin{eqnarray}
&&\left[ 
\begin{array}{c}
k+m \\ 
k%
\end{array}%
\right] _{q}  \notag \\
&=&\frac{(q^{k+m}-1)(q^{k+m-1-1}-1)\cdot \cdot \cdot (q^{m+1}-1)}{%
(q^{k}-1)(q^{k-1}-1)\cdot \cdot \cdot (q-1)}  \notag \\
&=&\dprod\limits_{i=1}^{k}\frac{1-q^{m+i}}{1-q^{i}}\equiv \mathcal{F}%
(k,m\mid q).  \TCItag{3.5}
\end{eqnarray}%
This result is of \textit{central importance} since it represents the
partition function capable of reproducing the Veneziano and Veneziano-like
amplitudes as we have explained at length in Parts II and III of our work.
In the limit : $q\rightarrow 1,$ Eq.(3.2) reduces to the number $\mathcal{N}$
as required. To use this information in the context of honeycombs we need to
remind our readers some basic facts about the Schubert calculus

\subsection{Representation theory, Grassmannians, Schubert calculus and
physics of orthogonal polynomials}

The irreducible polynomial representations of general linear group $GL(k,%
\mathbf{C})$ are parametrized by the integer partitions $\lambda $ with at
most $k$ parts. Given any two such polynomial representations $V^{\lambda }$
and $V^{\mu }$ one can construct the tensor product $V^{\lambda }\otimes
V^{\mu }$ which is expected to be decomposable according to the rule 
\begin{equation}
V^{\lambda }\otimes V^{\mu }=\tsum\limits_{\nu }C_{\lambda \mu }^{\nu
}V^{\nu }  \tag{3.6}
\end{equation}%
into irreducible representations $V^{\nu }$ of $GL(k,\mathbf{C}).$
Evidently, since in Eq.(1.9) the combinatorics is the same as in Eq.(3.6),
the coefficients $C_{\lambda \mu }^{\nu }$ (known in literature as the 
\textit{Littlewood-Richardson} \ (L-R) coefficients) should \ also be the
same.

Let $\left\vert \lambda \right\vert =\lambda _{1}+...+\lambda _{k}$ then, it
is known [27] that the sum in the r.h.s. of Eq.(3.6) is over partitions $\nu 
$ for which $\left\vert \lambda \right\vert +\left\vert \mu \right\vert
=\left\vert \nu \right\vert .$ Consider the standard flag $\mathcal{F}$ of
complex subspaces $\mathcal{F}$: $\mathbf{C}^{1}\subset \mathbf{C}%
^{2}\subset \cdot \cdot \cdot \subset \mathbf{C}^{N}$, where $N$ is related
to $k$ via $k=N-m$ as before. The Schubert cell $\ \Omega _{\lambda }$ is
made out of $\ k-$dimensional subspaces \ $V\subset \mathbf{C}^{N}$ with
prescribed dimensions of intersections with elements of $\mathcal{F}$ . More
accurately, they can be described as subspaces for which $\dim (V\cap 
\mathbf{C}^{i})=\omega _{N}+\omega _{N-1}+\cdot \cdot \cdot +\omega _{N-i+1}$
for $i=1,2,...,m.$ Since the notion of the \textit{closure} $\bar{\Omega}%
_{\lambda }$ for the Schubert cell is a bit technical (e.g. read page 122 of
Ref.[28]) we shall skip it without much damage to physics \footnote{%
Effectively, the closure means that the space $V$ defined by $\dim (V\cap 
\mathbf{C}^{i})=\omega _{N}+\omega _{N-1}+\cdot \cdot \cdot +\omega _{N-i+1}$
should be replaced by an assembly of spaces for which $\dim (V\cap \mathbf{C}%
^{i})\geq \omega _{N}+\omega _{N-1}+\cdot \cdot \cdot +\omega _{N-i+1}.$}.
Such closures are called \textit{Schubert} \textit{varieties}. It happens,
that the fundamental cohomology classes $\sigma _{\lambda }=[\bar{\Omega}%
_{\lambda }]$ of such varieties form a \textbf{Z}-basis of the cohomology
ring H$^{\ast }$($Gr(m,k))$ of the complex Grassmannian $Gr(m,k)$ [28-30%
\textbf{]. }The dimension of such a ring was determined in the previous
subsection as $\left( 
\begin{array}{c}
N \\ 
k%
\end{array}%
\right) $ and the multiplication rule for the product of two cohomology
classes given by 
\begin{equation}
\sigma _{\lambda }\cdot \sigma _{\mu }=\tsum\limits_{\nu }C_{\lambda \mu
}^{\nu }\sigma _{\nu },  \tag{3.7}
\end{equation}%
provided that $\left\vert \lambda \right\vert +\left\vert \mu \right\vert
=\left\vert \nu \right\vert .$

The obtained correspondence can be explained based on "physically intuitive"
arguments. Indeed, if the Veneziano (and Veneziano-like) amplitudes are
periods of the Fermat hypersurfaces (varieties), as explained in Part I,
they should be naturally associated with the differential forms generating
the cohomology ring for these varieties. When these varieties are embedded
into complex projective space, where the complex Grassmannian is also
embedded (via the Pl\"{u}cker embedding as explained in Part II), the
cohomology ring for both of these varieties become interrelated.
Mathematical details supporting such nonrigorous "physical" arguments can be
found in the paper by Tamvakis, Ref.[31].

The results presented above are pretty standard. We would like now to inject
some physics into them.To do so, we notice that combinatorially the fusion
rule, Eq.(3.7), is described with help of the Schur polynomials $s_{\lambda
}(\mathbf{x})$ [32], that is (omitting the $x-$dependence) by 
\begin{equation}
s_{\lambda }\cdot s_{\mu }=\tsum\limits_{\nu }C_{\lambda \mu }^{\nu }s_{\nu
}.  \tag{3.8}
\end{equation}%
These functions are orthogonal polynomials. That is for partitions $\mu $
and $\lambda $\ and for the appropriately defined scalar product $<,>$ we
obtain: 
\begin{equation}
\left\langle s_{\lambda },s_{\mu }\right\rangle =\delta _{\lambda ,\mu }. 
\tag{3.9}
\end{equation}%
By combining Eq.s(3.8),(3.9) we obtain as well 
\begin{equation}
C_{\lambda \mu }^{\nu }=\left\langle s_{\lambda }\cdot s_{\mu },s_{\nu
}\right\rangle ,  \tag{3.10a}
\end{equation}%
which, in view of Eq.(3.7), is equivalent to%
\begin{equation}
C_{\lambda \mu }^{\nu }=\left\langle \sigma _{\lambda }\cdot \sigma _{\mu
},\sigma _{\nu }\right\rangle .  \tag{3.10b}
\end{equation}%
Although this result is very important from the point of view of algebraic
geometry (since it describes intersection of Schubert cycles), to use it
physically requires more work as we would like to explain now.

In Sections 7 and 8 of Part II we discussed why the Schur polynomials
associated with KdV hierarchy (and, hence, with the Virasoro algebra
(through method of coadjoint orbits) are \textbf{not} relevant for Veneziano
amplitudes. At this point we are ready to provide additional explanation why
this is so.

First,\ we would like to recall the definition of \ one of the basic
symmetric function $m_{\lambda }$ [32]. For this we have to define the
monomials, e.g. \textbf{x}$^{\lambda }=x_{1}^{\lambda _{1}}x_{2}^{\lambda
_{2}}\cdot \cdot \cdot ,$ associated with partition $\lambda .$ With thus
defined monomials, $m_{\lambda }$ is just the sum of these monomials made of
all possible permutations of $x^{\prime }s$. When this definition is
combined with the results of Part I desribing Veneziano amplitudes as period
of Fermat varieties, it should become clear that the fully symmetrized
Veneziano amplitude can be obtained by using $m_{\lambda }$ in the numerator
of the period integral.

\textbf{Remark.3.1}. It is known that $m_{\lambda }$ are eigenfunctions of
the Calogero-Sutherland\ (C-S) model [33]. It is also known that 2
dimensional QCD is reducible to the C-S model [$34,35$]. Thus, it should be
not too surprising that Veneziano amplitudes had been successful in
describing scattering of mesons.

Second, since we are interested not only in Veneziano amplitudes but in
general combinatorial properties of the scattering amplitudes of high energy
physics, we would like to develop things a bit further. For instance, in the
next subsection we shall demonstrate that combinatorial data contained in
Veneziano amplitudes are quite sufficient for calculation of $C_{\lambda \mu
}^{\nu }$. From this fact it is easy to make a mistake and to use the Schur
polynomials $s_{\lambda }(\mathbf{x})$ in the subsequent developments. This
is known and well developed pathway to the traditional string theory
formalism. One should keep in mind however that, like in ordinary quantum
mechanics, \textit{any} symmetric function can be Fourier decomposed into
Fourier series whose basis is made of orthogonal polynomials. In view of
Eq.(3.9), the Schur polynomials provide a basis for such an expansion but
this basis is not unique. There are \textit{other} orthogonal polynomials
which can be used for such a purpose [36,37]. Very much like in quantum
mechanics, where all exactly solvable problems possess a complete orthogonal
set of eigenfunctions, different for different problems, one can think of
the corresponding exactly solvable (many-body) problems associated with
orthogonal polynomials, \ also different for different problems. More
interesting, however, is to be able to solve the inverse problem.

\textbf{Problem 3.2}. For a given \ set of orthogonal polynomials find the
corresponding many-body operator for which such a set of orthogonal
polynomials forms a complete set of eigenfunctions.

In view of the Remark 3.1., this task can be solved completely in the case
of Veneziano amplitudes. In this paper and its companion we make an attempt
at providing \ more general outlook at solution of the Problem 3.2. We hope,
that by rasing these issues more works will follow enabling to solve this
problem completely.

\section{ Designing and solving Veneziano puzzles}

\subsection{General remarks}

With the background just provided we are ready to connect the combinatorics
of honeycombs with computation of the L-R coefficients $C_{\lambda \mu
}^{\nu }$. As a by product we shall introduce another honeycomb-related
construction for calculating these coefficients which KTW call a
"puzzle"[16]. Combinatorics of honeycombs is connected with the L-R
coefficients in view of the following theorem, Ref.[15], page 1053.

\ \ 

\textbf{Theorem 4.1. }\textit{Let }$\lambda ,\mu $\textit{\ and }$\nu $%
\textit{\ be three pre assigned (boundary) partitions (e.g. like those
depicted in Fig.3 for the 5-honeycomb) for the }$k-$\textit{honeycomb. Then
the number of different honeycombs with such pre assigned boundary
conditions is given by the L-R coefficient }$C_{\lambda \mu }^{\nu }$

\textit{\ \ }

Although the above theorem hints at physical relevance of honeycombs, e.g.
for practical calculation of the L-R coefficients, actual use of honeycombs
for such a purpose based on the information provided is somewhat
problematic. We would like to correct this deficiency now. Firstly,
following KT [15], we would like to illustrate general principles by using a
simple example. In particular, in the case of 3-honeycomb the decomposition
of the tensor product%
\begin{equation*}
V_{(2,1,0)}\otimes V_{(2,1,0)}=V_{(4,2,0)}\oplus V_{(3,2,1)}^{\otimes
2}\oplus V_{(4,1.1)}\oplus V_{(3,3,0)}\oplus V_{(2,2,2)}
\end{equation*}%
is graphically depicted in Fig.6.


\begin{figure}[tbp]
\begin{center}
\includegraphics[width=4.70 in]{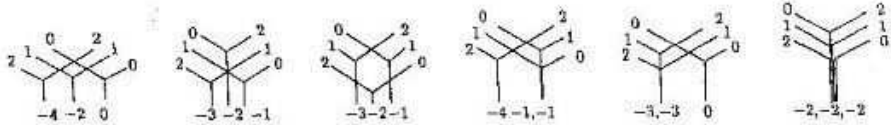}
\end{center}
\caption{A typical example of a graphical calculation of the L-R
coefficients with help of 3-honeycombs}
\end{figure}
Evidently, the L-R coefficients can be read off from such a decomposition
straightforwardly. Moreover, as Fig.6 indicates, in actual calculations of
these coefficients it is sufficient to use the precursor, e.g. see Fig.4,
rather than the full blown $n-$honeycomb. This observation is especially
helpful for the Veneziano (and Veneziano-like) amplitudes in view of already
noticed factorization property provided by Eq.(I.3.28). It is more
questionable if we are interested in the most general form of multiparticle
scattering amplitude compatible with the energy-momentum conservation laws.

For the sake of such generality, we would like to discuss yet another method
of computation of the L-R coefficients\footnote{%
There are many methods of calculating these coefficients. In fact, the
number of these methods is so large that even the most reputable monographs,
e.g. Ref.[37], are unable to provide the complete list. We only mention K-T 
\textit{hives} to be discussed in Section 5.3. which are just a slight
modifications of the Berenstein-Zelevinskii (BZ) \textit{patterns} nicely
described in Ref.[37\textbf{],} page\textbf{\ }437. We discuss KT variant of
constructing the L-R coefficients mainly because of the factorization
property, Eq.(I.3.28), of the Veneziano amplitudes.}. This method requires
designing and solving puzzles associated with honeycombs. The simplest
puzzle associated with 3-honeycomb is depicted in Fig.7.


\begin{figure}[tbp]
\begin{center}
\includegraphics[width=3.30 in]{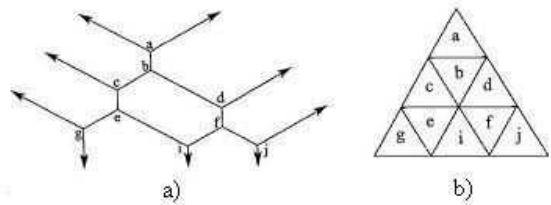}
\end{center}
\caption{The labeling order of vertices of the 3-honeycomb tinkertoy a) is
in one-to one correspondence with labeling of triangles (precursors of) in a
puzzle b) associated with such a tinkertoy}
\end{figure}

This picture only provides a hint that honeycombs and puzzles are
interconnected but not much insight into rationale for switching from
honeycombs to puzzles. We would like to discuss this rationale now. For this
purpose, we need to use the correspondence between the partitions and
directed random walks already discussed. For each partition triple $\lambda
,\mu $ and $\nu $ there is its realization in terms of such walks$:\omega
(\lambda ),\omega (\mu )$ and $\omega (\nu ).$ Consider a particular
directed random walk. It is described by the Fermi-type variable $\omega
_{i}(\lambda )$ such that the constraint $\tsum\nolimits_{i=1}^{N}\omega
_{i}(\lambda )=m=N-k$ holds. Consider now an equilateral \ triangle whose
sides are divided into $N$ segments of equal length. Furthermore, let \ us
put these segments \ in correspondence with $\omega _{i}(\lambda ),\omega
_{i}(\mu )$ and $\omega _{i}(\nu )$, respectively for each side of the
triangle. Finally, consider the set of puzzle pieces depicted in Fig.8.


\begin{figure}[tbp]
\begin{center}
\includegraphics[width=3.50 in]{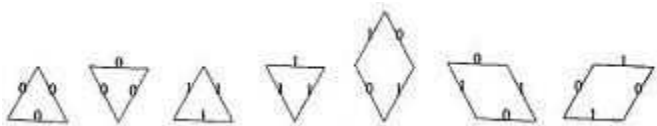}
\end{center}
\caption{Basic building blocks of a puzzle}
\end{figure}

They are made of equilateral triangles and rhombi whose sides all have the
same lengths equal to that of the segment on the side of the larger
triangle. Since these puzzle pieces are labeled, the task is to fill in the
large equilateral triangle with the puzzle pieces provided that these pieces
can be rotated but not reflected when they are used to solve the puzzle. The
final result looks like that depicted in Fig.9.


\begin{figure}[tbp]
\begin{center}
\includegraphics[width=2.35 in]{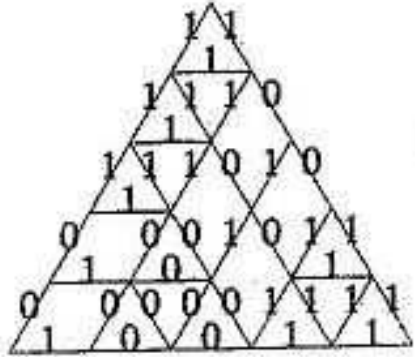}
\end{center}
\caption{A typical assembled puzzle}
\end{figure}

By analogy with Theorem 4.1. it is possible to prove the following theorem.

\ 

\textbf{Theorem 4.2}. (Knutson-Tao-Woodward [16]) \textit{Let partitions }$%
\lambda ,\mu $\textit{\ and }$\nu $\textit{\ be encoded by random walks }$%
\omega (\lambda ),\omega (\mu )$\textit{\ and }$\omega (\nu )$\textit{\
whose particular realization is described in terms of 0's and 1's on the
sides of \ the equilateral triangle encoded from left to right on each of
its sides in a clockwise order. Then, the number of puzzles constructed with
such boundary data equals to the L-R coefficient }$C_{\lambda \mu }^{\nu }$.

\ 

The easiest way to prove this theorem is through graphical bijection between
the honeycombs and puzzles which we would like now to describe. At this
point, in view of Fig.7, we know already that such a bijection does exist.
Using results of KTW [16], it can be made more accurate now. For this
purpose, the following steps should be made:

a) one should place a solved puzzle on the $\mathbf{R}_{\tsum =0}^{3}$ plane
in such a way that the bottom right corner is at the origin. Next one should
rotate this puzzle around origin by $30^{0}$ counterclockwise;

b) at each boundary segment, which is labeled by 1(1-region), attach a
rhombus (outside the puzzle), then another (parallel to fist) and so on ad
infimum. Fill in the rest of the plane with 0-trianges (see Fig.10);

c) deflate thus obtained extended puzzle, while keeping the right corner at
the origin. The result will be the honeycomb\ whose vertices originate from
such deflated 1-regions \ and whose edges are labeled \ by the thickness of
the original rhombus region.

Fig.10 illustrates such described reduction procedure.


\begin{figure}[tbp]
\begin{center}
\includegraphics[width=4.99 in]{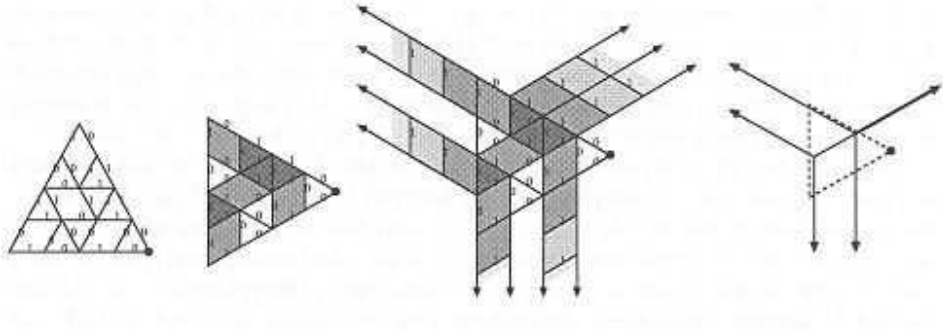}
\end{center}
\caption{Successive stages of conversion of a puzzle into honeycomb}
\end{figure}

This completes our description of KTW puzzles and the associated with them
Heisenberg honeycombs.

\subsection{Some comments about solution of the Horn conjecture}

\bigskip\ 

The results described \ previously are incomplete without further discussion
of the Horn and saturation conjectures. In this subsection we would like to
discuss some additional details related to the Horn conjecture. These may be
of some importance in organizing experimental data for the mass spectrum of
hadrons.

Since the case of 2-honeycombs can be solved completely, the following
problem emerges.

\textbf{Problem 4.3}. To what extent can one use methods developed for
2-honeycombs to obtain similar type inequalities for more complex honeycombs?

We can develop our intuition by looking at Fig.4 and asking a question: is
it permissible to make, say, 3-honeycomb out of 2 and 1-honeycombs as
depicted in Fig.11.


\begin{figure}[tbp]
\begin{center}
\includegraphics[width=3.50 in]{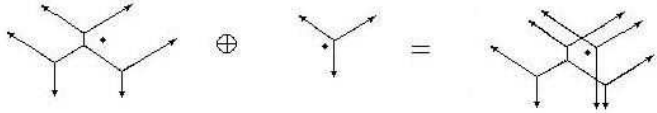}
\end{center}
\caption{Designing larger honeycombs (honeycomb tinkertoys) from the smaller
ones. The origin (0,0,0) is marked for each case by the black dot}
\end{figure}

Stated more formally, suppose we have $h$ and $h^{\prime }$ honeycombs what
is the meaning of their direct sum $h\oplus h^{\prime }?$ If we are talking
about $n-$honeycomb $h$ and $m-$honeycomb $h^{\prime },$ then the direct sum 
$h\oplus h^{\prime }$ must evidently correspond to the direct sum of $%
n\times n$ and $m\times m$ matrices combined together to form an $%
(m+n)\times (m+n)$ block-diagonal matrix. With such clarifications, it is
possible, following KT, to give a purely geometric proof (involving $n$%
-honeycomb) of the inequality $\lambda _{i}+\mu _{j}+\nu _{k}\geq 0$ (known
already to Weyl, Ref.[10]) provided that $i+j+k=n+2.$ For $n=1$ this
inequality is obviously correct. For $n>1,$ the requirement $\ i+j+k=n+2$
should hold while in order to prove that $\lambda _{i}+\mu _{j}+\nu _{k}\geq
0,$ the "physical" arguments can be used as follows. Consider some vertex $P$
inside the $n-$honeycomb as depicted in Fig.12.


\begin{figure}[tbp]
\begin{center}
\includegraphics[width=2.65 in]{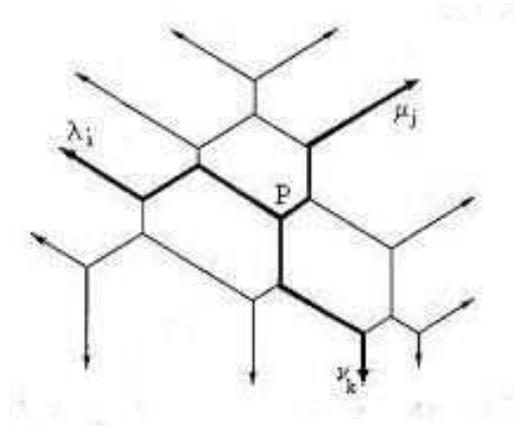}
\end{center}
\caption{Using honeycomb for proving Weyl's inequality}
\end{figure}

Next, we connect this vertex with the boundary lines marked respectively as $%
\lambda _{i},\mu _{j}$ and $\nu _{k}.$ Assume that our honeycomb is made of
wire and assume that constant currents with intensity $\lambda _{i},\mu _{j}$
and $\nu _{k}$ are applied at the boundary lines while the rest of the
boundary lines serve as sinks (i.e. they are grounded). Remembering the
Kirkhoff's rule for each vertex\ (that the total algebraic sum of all
currents entering a given vertex must be zero) we apply this rule to the
selected vertex $P$. If now the edges emanating from $P$ are labeled
respectively by $\lambda _{i}^{\prime },\mu _{j}^{\prime }$ and $\nu
_{k}^{\prime }$ then, the Kirkhoff rule requires: $\lambda _{i}^{\prime
}+\mu _{j}^{\prime }+\nu _{k}^{\prime }=0.$ At the same time, since the
current goes to other vertices too, it should be clear that: $\lambda
_{i}\geq \lambda _{i}^{\prime },$ $\mu _{j}\geq \mu _{j}^{\prime }$ and $\nu
_{k}\geq \nu _{k}^{\prime }.$ \ From here \ we obtain: $\lambda _{i}+\mu
_{j}+\nu _{k}\geq 0,$ \ in accord with Weyl. Since we would like to stay
focused on our immediate (physical) tasks, we refer our readers to the
already cited literature for additional details on solution of the Horn
problem.

\section{Solution of the saturation conjecture: from Heisenberg to Knutson
and Tao and beyond}

\subsection{Statement of the problem}

Before providing exact definitions, following KT we would like to discuss
some general issues for the sake of presenting them subsequently in a proper
physical context. In particular, if there exist Hermitian matrices $A,B$ and 
$C$ with eigenvalue sets $\lambda ,\mu ,$ and $\nu $ respectively, one can
associate to the matrix equation $A+B+C=0$ the symbolic relation of the type 
\begin{equation}
\lambda \boxplus \mu \boxplus \nu \sim _{c}0\Longleftrightarrow \lambda +\mu
+\nu =0.  \tag{5.1a}
\end{equation}%
Here the subscript $c$ means "classical". Although the hermiticity of
matrices $A$,$B$ and $C$ makes them suitable for quantum mechanical
interpretation, KT introduce another relation%
\begin{equation}
\lambda \boxplus \mu \boxplus \nu \sim _{q}0\Longleftrightarrow \lambda +\mu
+\nu =0  \tag{5.1b}
\end{equation}%
where the subscript $q$ means "quantum". KT consider "classical" problems as
those involving Hermitian matrices with \textit{real} spectrum while they
consider problems as "quantum" if the spectrum involves only \textit{integral%
} $\lambda ,\mu ,$ and $\nu ^{\prime }s.$ Based on these definitions, KT
formulate and solve the following theorem.

\ 

\textbf{Theorem 5.1.} (Knutson-Tao [20]) \textit{Let }$\lambda ,\mu ,$%
\textit{\ and }$\nu $\textit{\ be weakly decreasing sequences of }$n$\textit{%
\ integers. Then, if Eq.(5.1b) holds, Eq.(5.1.a) holds as well. On another
hand, if Eq.(5.1a) holds, then, there exists an integer }$\ N>0$\textit{\
such that }$N\lambda \boxplus N\mu \boxplus N\nu \sim
_{q}0\Longleftrightarrow \lambda \boxplus \mu \boxplus \nu \sim _{c}0$%
\textit{. (That is asymptotically, i.e. for some large }$N^{\prime }s$%
\textit{, quantum and classical results coincide)}

\ 

Based on this result KT formulate and prove the following

\ 

\textbf{Conjecture 5.2.} (Saturation conjecture) The above classical-quantum
equivalence persist even for $N=1$.

\bigskip

\textbf{Remark 5.3. }The above conjecture can be restated in terms of the
L-R coefficients $C_{\lambda \mu }^{\nu }.$ Following Fulton, Ref.[38], page
238, we expect that\ if $\lambda ,\mu ,$ and $\nu $ are a triple of
partitions and $C_{N\lambda N\mu }^{N\nu }\neq 0,$ then $C_{\lambda \mu
}^{\nu }\neq 0$ as well. In particular, if $C_{\lambda \mu }^{\nu }=1$ we
should expect $C_{N\lambda N\mu }^{N\nu }=1$.

It should be noted, however, that in the case if $C_{\lambda \mu }^{\nu
}\neq 1$ there is no reason to expect that $C_{N\lambda N\mu }^{N\nu }=$\ $%
C_{\lambda \mu }^{\nu }.$ Since such an observation can be checked
experimentally, it makes sense to discuss it in some detail in this section.
Mathematically, the proof of saturation conjecture was made not only by KT
but by several other authors as well . They are listed on page 238 of
Ref.[38]. In this section we would like to discuss the physics of this
proven conjecture based on arguments used by Heisenberg in his key paper on
quantum mechanics, Ref.[18]. Some auxiliary results are presented in
Appendix A.

\subsection{Heisenberg's proof of the saturation conjecture}

We begin by noticing that use of Hermitian operators in quantum mechanics is
motivated by the requirements that the observables which these operators
represent are real numbers. In particular, for isolated stable physical
system the spectrum of its eigenvalues should be real. This fact is in
apparent contradiction with the KT definition of \ what is "classical" and
what is "quantum". For instance, the famous Hydrogen atom energy spectrum is
known to behave as $E_{n}\sim 1/n^{2},n=1,2,...$ \ The contradiction,
nevertheless is only apparent. Before going into detailed explanations, we
recall that famous semiclassical approximation of quantum mechanics \
relates quantum results to classical in the limit of large quantum numbers.
This fact can be taken as \textit{physical} proof of Theorem 5.1. This makes
sense only if we accept that the Hermitian operators producing \textit{real}
spectra have something in common with the "classical" world. \
Alternatively, we can try to prove that such a spectra cannot belong to any
quantum mechanical system. Clearly, such a proof will still be insufficient
to place such type of spectra into "classical" world since in classical
world there are no operators and everything commutes. Hence, we would like
to approach the saturation conjecture somewhat pragmatically using physical
arguments.

For this purpose, we would like to bring \ to attention of our readers some
important quotations from the classical book by Dirac, Ref.[22]. On page 177
of this book one reads: " In fact it was the idea of replacing classical
Fourier components by matrix elements\footnote{%
E.g. see Eq.s (2.4), (2.5).}which lead Heisenberg to the discovery of
quantum mechanics in 1925. Heisenberg assumed that the formulas describing
the interaction with radiation of a system in the quantum theory can be
obtained from the classical formulas by substituting for the Fourier
components of the total electric displacement\footnote{%
The dipole moment.} of the system \ the corresponding\footnote{%
I.e.quantum.} matrix elements". Further in the text Dirac elaborates on this
quantum-classical correspondence. Specifically, on pages 245-246 we read: "
Thus, the elementary theory\footnote{%
That is completely classical (e.g. see Appendix A)}...in which the radiation
is treated as an external perturbation,\footnote{%
That is classically!} gives the correct value for the absorption coefficient%
\footnote{%
Which is calculated quantum mechanically}. This agreement between the
elementary theory and the present theory could be inferred from general
arguments. The two theories differ only in that the field quantities all
commute with one another in the elementary theory and satisfy definite
commutation relations in the present theory\footnote{%
That is quantum.}, and this difference \textit{becomes unimportant} for
strong fields. Thus the two theories must give the same absorption and
emission when strong fields are concerned. Since both theories give the rate
of absorption proportional to the intensity of the incident beam, the
agreement must hold also for the weak fields in the case of adsorption. In
the same way the stimulated part of emission in the present theory must
agree with the emission in the elementary theory".

We brought such extensive quotations from Dirac only to emphasize that,
actually, at least in some cases, the quantum-classical correspondence can
be pushed way down into the \ seemingly quantum domain thus providing a
"proof" of the saturation conjecture. Since the existing literature on
quantum mechanics (including Dirac's book) for some reason does not discuss
these issues in sufficient detail, we would like to provide such details in
this section.

Following the logic of Heisenberg's original paper, Ref.[18], Eq.(2.1), when
it is treated classically, can be rewritten as%
\begin{equation}
\omega (n,\alpha )=\alpha \omega (n)\simeq \alpha \frac{1}{\hbar }\frac{%
\partial E}{\partial n}.  \tag{2.1a}
\end{equation}%
Next, Heisenberg notices that, actually, the famous Bohr-Sommerfeld (B-S)
quantization rule%
\begin{equation}
\oint pdq=nh,\text{ }n=0,1,2,...,  \tag{5.2}
\end{equation}%
is not exact ! It is determined only with accuracy up to some constant
(unknown at the time of \ his writing). He argues, that if such a constant
would be known, this rule \textit{would become exact}, that is valid for 
\textit{any} $n$'s. From the point of view of our present understanding of
quantum mechanics his intuition was correct: the old fashioned
Bohr-Sommerfeld rule is valid rigorously in the limit of large $n$'s while
the calculation of the constant can be done, for instance, with help of
either the WKB or of \ much more sophisticated \ theory of Maslov index
[39]. These arguments although plausible are superficial nevertheless as can
be found from the book by Arnold, Ref.[40], page 246. From it we find that 
\textit{already at the classical level} the adiabatic invariant $\oint pdq$
is determined only with accuracy up to some constant. This observation makes
Heisenbeg's arguments less convincing.

In particular, following Heisenberg we claim that if the B-S quanization
rule, when corrected, makes sense fully quantum mechanically, one can get,
in principle, the additional information out of it. For this purpose
Heisenberg introduces the Fourier decomposition of the generalized
coordinate $q$, i.e.%
\begin{equation}
q(n,t)=\sum\limits_{\alpha =-\infty }^{\infty }a_{\alpha }(n)\exp (i\omega
(n,\alpha )t)  \tag{5.3}
\end{equation}%
where we used Eq.(2.1a)\footnote{%
In the original Heisenberg keeps $\alpha \omega (n)$ instead of $\omega
(n,\alpha ).$ Our notations happen to be more convenient as we shall
demostrate shortly.} thus causing us to keep our calculations with respect
to the some pre assigned energy level $n$ (e.g. see Appendix A). The
velocity can be readily obtained now as%
\begin{equation}
\dot{q}(n,t)=\sum\limits_{\alpha =-\infty }^{\infty }ia_{\alpha }(n)\omega
(n,\alpha )\exp (i\omega (n,\alpha )t)  \tag{5.4a}
\end{equation}%
so that calculation of the velocity square averaged over the total period is
given by 
\begin{equation}
\oint \left[ \dot{q}(n,t)\right] ^{2}dt=2\pi \sum\limits_{\alpha =-\infty
}^{\infty }\left\vert a_{\alpha }(n)\right\vert ^{2}\omega (n,\alpha )^{2}. 
\tag{5.4b}
\end{equation}%
At this point it should be noted that the original of Heisenberg's paper,
Ref.[18], contains (perhaps) a typographical error: instead of having $%
\omega (n,\alpha )^{2}$ Heisenberg writes $\omega (n,\alpha ).$ This fact
was noticed by the editors of his collected papers, Ref.[18]. Now we use
this result in the B-S quantization rule, i.e. we have%
\begin{equation}
\oint pdq=\oint m\dot{q}dq=\oint m\dot{q}^{2}dt=2\pi m\sum\limits_{\alpha
=-\infty }^{\infty }\left\vert a_{\alpha }(n)\right\vert ^{2}\omega
(n,\alpha )^{2}=nh+const.  \tag{5.5}
\end{equation}%
Next, Heisenberg argues as follows. Since the \textit{const} is unknown, it
is of interest to obtain results which are constant-independent. At the same
time, since the result, Eq.(5.5), is assumed to be exact, we have to use
instead of scalars $\left\vert a_{\alpha }(n)\right\vert ^{2}$ the matrices
in accord with Eq.(2.4). This would lead us to matrices of the type $%
\left\vert a(n,n+\alpha )\right\vert ^{2}$ and $\left\vert a(n,n-\alpha
)\right\vert ^{2}$ depending on the actual sign of $\alpha .$ In addition,
he silently \ had assumed that the $n-$dependence of the amplitudes is much
weaker than that for the frequencies $\ \omega (n,n+\alpha )$ and $\omega
(n,n-\alpha )$ so that it can neglected completely. Under such conditions he
treats $n$ as continuous variable and differentiates both sides of Eq.(5.5)
with respect to $n$ thus obtaining (recall that $\omega (mn)=-\omega (nm)):$ 
\begin{equation}
h=4\pi m\sum\limits_{\alpha =0}^{\infty }\{\left\vert a(n,n+\alpha
)\right\vert ^{2}\omega (n,n+\alpha )-\left\vert a(n,n-\alpha )\right\vert
^{2}\omega (n,n-\alpha )\}.  \tag{5.6}
\end{equation}%
The validity of this result depends upon the additional assumption about the
ground state energy. If $n_{0}$ represents such a state, then one must
require that 
\begin{equation}
a(n_{0},n_{0}-\alpha )=0\text{ \ }\forall \alpha >0.  \tag{5.7}
\end{equation}

In Ref.[18] Heisenberg acknowledges that this result was inspired by the
earlier result of Kramers who calculated the induced dipole moment of
electrons in atom assuming rules of quantum mechanics (in fact \textit{before%
} it was officially inaugurated !), e. g. see Appendix A. Results of the
Appendix A then lead us directly to the famous commutation rule 
\begin{equation}
\lbrack \hat{x},\hat{p}]=i\hbar .  \tag{5.8}
\end{equation}%
Heisenberg argues that his reasonings are correct since the frequency of the
incoming (scattered) light is much higher than that for characteristic
"rotational" frequencies in the atom so that the electrons can be treated as
"free" and independent.

The discussion we just presented is aimed to underscore the differences
between physical reality and mathematical correctness. It can be considered
as the Heisenberg-style proof \ of the saturation conjecture and provides us
with rationale for discussion of an alternative formulation of quantum
mechanics (to be presented in the next section). Before doing so we would
like to discuss the saturation conjecture and its proof in connection with
results of our earlier published Parts II and III. This will enable us to
make some physical sense out of recently obtained mathematically interesting
results.

\subsection{Combinatorics of L-R coefficients and the Ehrhart polynomial}

In this subsection we would like to address the following problem: suppose
we are given a L-R coefficient $C_{\lambda \mu }^{\nu }$, can this
information be used for calculation of $C_{N\lambda N\mu }^{N\nu }?$
Although calculations of L-R coefficients have a rather long history [41],
the full answer to this question was obtained only quite recently in
connection with positive solution of the saturation conjecture. It should be
noted that\ although the attempts in this direction were made a bit earlier
by Berenstein and Zelevinsky [42\textbf{]}, the actual numerical results
were obtained much later by King et al [43]. These authors were inspired by
the results of KT, Refs.[15,16], where, in addition to the honeycomb model,
the hive model was introduced which we have not discussed thus far. An 
\textit{n-hive} is a triangular array of numbers $a_{ij}$ with $0\leq
i,j\leq n$. Say, for $n=4$ a typical arrangement looks as follows

\ \ \ \ \ \ \ \ \ \ $%
\begin{array}{ccccccccc}
&  &  &  & a_{00} &  &  &  &  \\ 
&  &  & a_{10} &  & a_{01} &  &  &  \\ 
&  & a_{20} &  & a_{11} &  & a_{02} &  &  \\ 
& a_{30} &  & a_{21} &  & a_{12} &  & a_{03} &  \\ 
a_{40} &  & a_{31} &  & a_{22} &  & a_{13} &  & a_{04}%
\end{array}%
$ \ \ \ \ \ \ \ \ \ \ \ \ \ \ \ \ \ \ \ \ \ \ \ \ \ \ \ \ \ \ \ \ \ \ \ \ \
\ \ \ \ \ \ \ \ \ \ \ \ \ 

\bigskip\ 

Such an $n-$hive is an \textit{integer hive} if all of its entries are
non-negative integers. The numbers $a_{ij}$ in the hive are subject to the
hive conditions given symbolically \ by R1: $%
\begin{array}{ccc}
a & b &  \\ 
& c & d%
\end{array}%
;$ R2: $%
\begin{array}{ccc}
& a &  \\ 
b &  & c \\ 
& d & 
\end{array}%
$ and R3: $%
\begin{array}{ccc}
& b & d \\ 
a & c & 
\end{array}%
$ implying $b+c\geq a+d$ for $a,b,c$ and $d$ being the neighboring entries
in the $n-$hive. We shall call such type of inequalities the \textit{hive
condition} (HC). Based \ on these results, the following definition can be
made

\ 

\textbf{Definition 5.4}. A L-R hive is an integer hive satisfying the HC for
all constituent rhombi R1-R3. For such a hive the border entries are
determined by partitions $\lambda ,\mu $ and $\nu $ in such a way that $%
a_{00}=0,a_{0j}=\lambda _{1}+\lambda _{2}+...+\lambda _{j\text{ }%
},j=1,2,...,n,a_{i0}=\nu _{1}+\nu _{2}+...+\nu
_{i},i=1,2,...,n,a_{k,n-k}=a_{0n}+\mu _{1}+\mu _{2}+...+\mu _{k}$ , $%
k=1,2,...,n$, provided that $\left\vert \lambda \right\vert +\left\vert \mu
\right\vert =\left\vert \nu \right\vert $ and the number of parts $\QTR{sl}{l%
}$($\lambda ),\QTR{sl}{l}(\mu )$ and \textit{l}($\nu )$in partitions $%
\lambda ,\mu $ and $\nu $ is bounded by $n$.

\ 

Based on this definition, and motivated by Theorems 4.1. and 4.2. Fulton\
[44] proved the following theorem

\ 

\textbf{Theorem 5.5. }\textit{The L-R coefficient} $C_{\lambda \mu }^{\nu }$ 
\textit{is the number of LR hives with border labels determined by }$\lambda
,\mu $\textit{\ and }$\nu .$

\ \ \ 

Alternative (simpler) proof of this theorem as well as many other useful
results can be found in the recent paper by Pak and Vallejo [45]. In both
cases the proof is based on careful solution of the rhombus constraints (or
HC) for the integer hives. In the case of $n$-hive there are $m=(n-1)(n-2)/2$
interior vertex labels for $a_{ij}$.The corresponding set of linear
constraints with integer coefficients defines a convex rational (not
integral! as in Part III) (hive) polytope $\mathcal{P}$ living in \textbf{R}$%
^{m}.$ As in Part III we can define the Ehrhart polynomial $\mathfrak{P}$%
\textit{(}$N$\textit{, m)} for such polytope and the associated with it
generating\ (partition) function $\mathcal{F}(\mathcal{P},x)$ via%
\begin{equation}
\mathcal{F}(\mathcal{P},x)=\tsum\limits_{N=0}^{\infty }\mathfrak{P}\mathit{(}%
N\mathit{,\ m)x}^{N}.  \tag{5.9}
\end{equation}%
In the present case, $\mathfrak{P}\mathit{(}N\mathit{,\ m)=}C_{N\lambda N\mu
}^{N\nu }$ \footnote{%
A typical example is shown on page 14 of Ref.[43].}. Unlike the integral
polytope $\mathcal{P}$ for which the generating function $\mathcal{F}(%
\mathcal{P},x)$ can be written in closed form given by Eq.(III.1.14), in the
present case, since the polytope is only rational (that is not all of its
vertices lie at the nodes of \textbf{Z}$^{m})$, such closed form universal
result for $\mathcal{F}(\mathcal{P},x)$ cannot be used. Using method of
quivers, Derksen and Weyman [46] were able to prove that, nevertheless, the
universal form given by Eq.(III.1.14) still holds (provided that the
dimensionality $m$ of the lattice \textbf{Z}$^{m}$ is replaced by \~{m} and
the indeterminate $x$ is replaced by $x^{\alpha }$ where both $\tilde{m}$
and $\alpha $ are some known (in principle) nonnegative integers.

In view of such an interpretation of stretched L-R coefficients, the whole
chain of arguments of Part III can be used practically unchanged. This makes
Eq.(III.4.30) especially relevant demonstrating that combinatorics of the
L-R coefficients can be obtained field-theoretically in the spirit of
earlier treated Witten-Kontsevich model [47]. These results will be extended
further in Section 7 in which we discuss (among other things) the
Gromov-Witten invariants.

\textbf{Remark 5.6.} Obtained connections between the L-R coefficients $%
C_{\lambda \mu }^{\nu }$ and their inflated counterparts $C_{N\lambda N\mu
}^{N\nu }$ is very important from the experimental point of view. Indeed, by
looking at Eq.(1.4) and recalling its relevance to Veneziano amplitudes
described in Parts II and III, one would naively expect that for \textit{any}
$N$ one should have $C_{\lambda \mu }^{\nu }=C_{N\lambda N\mu }^{N\nu }$ .
Mathematics tells us that this may happen only if $C_{\lambda \mu }^{\nu
}=1. $ It remains to check experimentally if the constraint $C_{\lambda \mu
}^{\nu }=1$ is a valid physical constraint. These arguments clearly not
restricted to the Veneziano-type amplitudes and should be taken into account
for \textit{any} amplitude of high energy physics.

\textbf{Problem 5.7. }In the case if the hypothetical scattering processes
requires use of supersymmetric QCD, how supersymmetry can be detected
through study of \ combinatorics of experimental data, say, of the LR fusion
coefficients ?\footnote{%
Even though methods developed in Part III include use of supersymmetry, it
was demonstrated \ in Parts II and III that its use is not essential. Hence,
one should not confuse these results with the supersymmetric extension of
the underlying microscopic QCD model and with calculations of observables
for such type of models.}

\section{From Heisenberg back to Maxwell,Tait and Kelvin}

\subsection{General remarks}

In view of the results of previous section and those of the Appendix A we
would like to formulate the following problem

\textbf{Problem 6.1}. Given the experimental origin (discussed in Appendix
A) and the associated with it inevitable computational approximations
leading to the basic commutation rule, Eq.(5.8), is it possible to develop
quantum mechanics without this rule being put at its foundation?
Alternatively said, can we recover this basic rule without use of light
scattering experiments and the B-S quantization rule?

\textbf{Remark 6.2}. Since \textit{all} experimental data for\textit{\ any }%
quantum mechanical system (atom, molecule, solid, etc.) are spectroscopic in
nature we know about the system as much as the combinatorics of experimental
data provides. From such point of view there is not much difference between,
say, biological problems, computer science problems, astrophysical problems,
etc. and those in the high energy physics.

A superficial answer to just posed problem can be made like this: should the
above rule be wrong we would not be able to recover the spectrum of Hydrogen
atom with such an amazing accuracy using established methods of quantum
mechanics. This remarkable agreement between theory and experiment is
possible only if such commutation rule is correct. Clearly, it is correct!
Nevertheless, we can\ still argue against its up-front use as follows.
Heisenberg uses the \ B-S quanization rule (perhaps adjusted) to obtain his
results. This rule makes sense only if classically there is a complete
separation of variables done with help of the Hamilton-Jacobi formalism.
When this happens, the system is considered to be completely integrable.
Hence, any completely integrable system is just the set of independent
harmonic oscillators\footnote{%
This explains why Heisenberg was able to do his formal differentiation (over 
$n$) of Eq.(5.5) and arrived at correct result. This also explains why KT
call system "quantum" if it has an integral spectrum according to the B-S
quantization prescription.}. The Hydrogen atom is surely a good candidate
for such procedures but what about the Helium ? The B-S rule cannot be
applied strictly speaking already to the Helium\footnote{%
A very interesting detailed discussion of this fact is given in the
monograph by Max Born, Ref.[48], pages 286-299, published in 1924, i.e.prior
to the official birth of modern quantum mechanics.} so that Heisenberg's
chain of reasoning leading to the commutator, Eq.(5.8), formally breaks down%
\footnote{%
In the paper by Pauli and Born written in 1922 [49] it is noted that Bohr
conceded that only the Hydrogen atom is quantizable but the rest of atoms
are not. Therefore the spectral lines of elements other than Hydrogen must
be noticeably wider. This expectation is in disagreement with what is
observed specrosopically. The spectroscopic data for most of elements of
periodic table were available already in 1905 [50], that is long before the
quantum mechanics was formulated.}. Besides, since the B-S quantization
cannot be used for spin quantization (since formally there is no classical
analog of spin, i.e. B-S rule does not account for the half integers), the
spin has no place in the Schr\"{o}dinger's formalism. Since Schr\"{o}dinger
have demonstrated the equivalence of his formalism to that developed by
Heisenberg as described in Dirac's book, Ref.[22], apparently, there is no
room for the spin in the Heisenberg formalism as well. Surely, this happens
to be \textit{only apparently} true as we would like to explain now. In
doing so we \textit{do not} need to use the relativistic formalism developed
by Dirac.

\subsection{Heisenberg's paper revisited}

We begin with discussion of some consequences of Heisenberg's results
following \ the1925 paper by Dirac, Ref.[21]. We selected this paper in view
of its remarkable completeness: all quantum mechanical formalism used today
can be traced back to this paper\footnote{%
It should be noted, nevertheless, that Dirac's paper [21] was received by
the Editorial office on 7th of November of 1925 while on November 16th of
the same year the paper by Born, Heisenberg and Jordan, Ref.[51], was
registered by the Editors. It contained practically the same results as
Dirac's paper and many other results in addition.}. Dirac acknowledges,
though, that his paper was written as consequence of Heisenberg's results.
In particular, the famous Dirac quantization rule, Eq.(2.7), is just
restatement of the results by Heisenberg. As good as it is, its use is
questionable in general. Indeed, in comments to his Eq.(11) Dirac states
that the difference of Heisenberg's products of two quantum observables $x$
and $y$ is equal to the (classical!) Poisson bracket of their classical
counterparts multiplied by $\frac{ih}{2\pi }$ or, symbolically, 
\begin{equation}
\hat{x}\hat{y}-\hat{y}\hat{x}=\frac{ih}{2\pi }\{x,y\}_{p.b.}  \tag{6.1}
\end{equation}%
This expression makes sense for $x\leftrightarrows \hat{x}$ and $%
y\leftrightarrows \hat{p}$ . But, in general, for arbitrary classical
observables $x$ and $y$ \ we are dealing with the Lie algebra which requires
this Poisson bracket to be expanded into linear combinations of \textit{%
classical} obseravbles so that the l.h.s and the r.h.s of Eq.(6.1) do not
match. Hence, we come back to the Heisenberg-Kramers result presented in the
Appendix A as the only justification. This difficulty was recently noticed
and discussed in the book by Adler who suggests to treat classical Poisson
bracket quantum mechanically in the style of Heisenberg, e.g. see Eq.(1.13b)
of Ref.[52\textbf{]}.

In this paper, we choose another way to by pass the Heisenberg-Dirac
quantization prescription. For this purpose, we would like to make few
additional comments regarding traditional formulations. Following Dirac we
introduce the evolution operator $U(t)$ bringing the initial state wave
function $\psi _{0}$ to its final state $\psi (t),$ i.e. $\psi (t)=U(t)\psi
_{0}.$ For the time-independent Hamiltonian $\hat{H}$ the formal solution of
the Schr\"{o}dinger-type equation%
\begin{equation}
i\hbar \frac{d}{dt}\hat{U}^{-1}(t)=\hat{H}\hat{U}^{-1}(t)  \tag{6.2}
\end{equation}%
is known to be given by $\hat{U}^{-1}(t)=\exp (-\frac{i}{\hbar }\hat{H}t).%
\footnote{%
We write $U^{-1}$ instead of $U$ to be in accord with mathematical
literature. This will be of immediate use shortly below.}$ Now, Heisenberg
considered the quantum Fourier transform by replacing the usual Fourier
amplitudes by matrices, i.e. he used quantities like $a(mn)exp(\frac{i}{%
\hbar }\omega (mn)t).$ In the modern language this can be rewritten as
follows. Let $\hat{O}$ be some quantum mechanical operator whose evolution
is described by $\hat{U}(t)\hat{O}\hat{U}^{-1}(t)=\hat{O}(t).$ This operator
leads to the matrix elements : $<m\mid \hat{O}\mid n>exp(\frac{i}{\hbar }%
\omega (mn)t)$ with $\omega (mn)$ defined by Eq.(2.1) with $<m\mid $ and $%
\mid n>$ being time-independent wave functions of the Hamiltonian $\hat{H}$.
Clearly, if the observable $\hat{O}(t)$ is an identity element in the
algebra of observables, we obtain : $<m(t)\mid $ $n(t)>=<m\mid n>,$ where $%
\mid n(t)>=U(t)\mid n>.$ The requirement for the observables to be real
leads to the Hermitian type operators whose eigenfunctions are mutually
orthogonal. \ Under such conditions we may or may not require these mutually
orthogonal functions to be normalized to 1. By doing so we \textit{do not
insist} on the probabilistic interpretation of quantum mechanics. Such an
interpretation emerges anyway within quantum statistical mechanics.

In view of earlier posed Problem 6.1\textbf{. }it makes sense to replace the
Dirac quantization rule, Eq.(6.1), by the requirement of orthogonality for
the wave functions. Under such a rule we need to have a supply of orthogonal
functions (if the spectrum is countably infinite) or orthogonal basis in
some complex finite dimensional vector space. In the case of orthogonal
functions, it is known that \textit{all} one variable orthogonal functions
used in quantum mechanical exactly solvable problems are obtainable from the
one variable Gauss-type hypergeometric functions [53]. These functions are
expressible in the form of period integrals. The Veneziano and
Veneziano-like scattering amplitudes considered in Part I belong to \textit{%
the same} family of hypergeometric- type period integrals \ initially
considered by Aomoto [54] and subsequently by many others [55]. The
cohomological meaning of such integrals is explained in detail in Ref.[56].
By the \ principle of complementarity all many-body\ exactly solvable
quantum mechanical problems should be related to the hypergeometric
functions of multiple arguments. More importantly for us is that these
hypergeometric functions produce sets of all known orthogonal polynomials
replacing one-variable orthogonal functions of usual quantum mechanics%
\footnote{%
In view of Eq(60) on Page 128 of the book by Dirac [22], the corresponding
path integrals can now be easily constructed. In this paper for the sake of
space we are not going to take advantage of this observation.We refer our
readers to Ref.[59] for an illustrative example.}. Hence, they are also
period integrals. A nice summary of developments in this area can be found
in Refs.[57,58\textbf{]}. \ The finite dimensional cases (including spin) \
technically present no difficulties under such circumstances.

At this stage we are ready to provide additional \ arguments in support of
our point of view on quantum mechanics. These arguments will be also of use
in the next section. Traditionally, in the Heisenberg interpretation of
quantum mechanics equations of motion for the operators $\hat{O}_{i}(t)$ can
be obtained by simple differentiation of $\hat{U}(t)\hat{O}_{i}\hat{U}%
^{-1}(t)=\hat{O}_{i}(t)$. This procedure formally leads to the Heisenberg's
equation of motion%
\begin{equation}
i\hbar \frac{\partial \hat{O}}{\partial t}=[\hat{O},\hat{H}]  \tag{6.3}
\end{equation}%
for the operator $\hat{O}$. The rationale for such writing comes from the
analogy of this equation with that known in classical Hamiltonian mechanics.
This makes sense only if the Dirac quantization prescription makes sense.
But it does not as we just discussed! Instead of repairing this situation
using known mathematical methods of geometric quantization [60,61], we
follow Heisenberg's philosophy based on careful analysis of spectroscopic
data. From his point of view we have the set of classical observables $%
\{O_{i}(t)\}$ which is supposedly complete. This means that treating the
Poisson brackets as Lie brackets we have 
\begin{equation}
\{O_{i},O_{j}\}=\sum\limits_{k}C_{ij}^{k}O_{k}.  \tag{6.4}
\end{equation}%
Accordingly, quantum mechanically, instead of Eq.(6.3) we need to consider
the result%
\begin{equation}
\lbrack \hat{O},\hat{H}]=\sum\limits_{k}\tilde{C}_{oh}^{k}\hat{O}_{k} 
\tag{6.5}
\end{equation}%
valid for any $t$! So that under such circumstances (quantum) dynamics 
\textit{formally} disappears! This observation can be strengthened due to
the following chain of arguments. In mathematics (see Part III, Section 3.2)
expression like $\hat{U}(t)\hat{O}_{i}\hat{U}^{-1}(t)=\hat{O}_{i}(t)\equiv
Ad_{\hat{U}}\hat{O}_{i}$ defines an \textit{orbit} for the operator $\hat{O}%
_{i}$ in the Lie algebra (made of operators $\{\hat{O}_{i}\})$ caused by the
action of elements $\hat{U}$ from the associated with it Lie group. At the
same time, the mathematics of Lie groups and Lie algebras produces for $[%
\hat{O},\hat{H}]$= $\mathit{ad}_{\hat{H}}\hat{O}$ where both $\hat{O}$ and $%
\hat{H}$ are in the Lie algebra $\{\hat{O}_{i}\}.$Evidently, we can obtain
the same (or even greater) information working with $Ad$ operators instead
of $ad$. In particular, we would like to consider the trace, i.e. tr\{$Ad_{U}%
\hat{O}_{i}\}=\chi (\hat{O}_{i}),$ which is just the character for $\hat{O}%
_{i}$. Clearly, it is time-independent. If this is so, then, what is the
meaning of an orbit ? This \ topic was discussed at length in Parts II and
III of our work. To avoid repetitions we refer our readers to these papers.
\ If there is no time evolution for the character, superficially, nothing
happens. This is not true, however as was recognized already by Dirac,
Ref.[22]. In Chapter 9 he writes: " The Hamiltonian is a symmetrical
function of the dynamic variables and thus commutes with every permutation.
It follows that each permutation is a constant of motion. This happens even
if the Hamiltionian is not constant\footnote{%
That is time-dependent.}." Hence, the orbit $Ad_{\hat{U}}\hat{O}_{i}$ is
caused by permutations. These can be analyzed with help of the torus action
thus leading to the Weyl-Coxeter reflection group $W=N/T$ described in
Section 3.1 of Part III and to the associated with them Lie algebras
discussed in section 3.2. of Part III. Representations of these Lie algebras
\ (including the affine Lie algebras) produce all known quantum mechanical
results as well as those of conformal field theories (CFT). This was
explained in Part III.\footnote{%
Many quantum mechanical problems do involve time evolution, e.g. decay of
the metastable state, etc. To account for such phenomena we should consider
random walks on groups. An excellent introduction to this topic can be found
in the monograph by Diaconis[62].} At this point we would like to do more.

\subsection{Symmetric group and its relatives}

\bigskip As is well known, the symmetric group $S_{n}$ has the following
presentation in terms of generators $s_{i}$ and (Coxeter) relations\footnote{%
A quick introduction can be found in Appendix A of Part II.}%
\begin{eqnarray}
s_{i}^{2} &=&1  \notag \\
s_{i}s_{j} &=&s_{j}s_{i\text{ }}\text{for }\left\vert i-j\right\vert \geq 2,
\notag \\
s_{i}s_{i+1}s_{i} &=&s_{i+1}s_{i}s_{i+1}.  \TCItag{6.6}
\end{eqnarray}%
If there is a set of $n$ elements (say, the Weyl roots arranged in a certain
order) the generator $s_{i}$ interchanges an element $i$ with $i+1$ so that $%
s_{1},...,s_{n-1}$ generate $S_{n.}$Clearly, there are $n!$ permutations in
the set of $n$ elements. If we assign the initial ordered state, then any
other state can be reached by successful application of permutational
generators to this state so that the word $w=s_{a_{1}}s_{a_{2}}\cdot \cdot
\cdot s_{a_{l}}$ (where the indices $a_{1},...,a_{l}$ represent a subset of
the set of $n-1$ elements) can be identified with such a state. Since one
can reach this state in many ways, it makes sense to introduce the \textit{%
reduced word} $w$ whose \textit{length} $\mathit{l(w)}$ is minimal. With
these definitions, we would like to complicate matters a bit. We would like
the generators of $S_{n}$ to act on monomials $\mathbf{x}^{\mathbf{a}%
}=x_{1}^{a_{1}}x_{2}^{a_{2}}\cdot \cdot \cdot x_{n}^{a_{n}}.$ Following
Lascoux and Sc\"{u}tzenberger (L-S), Ref.[63], we introduce an operator $%
\partial _{i}$ via rule%
\begin{equation}
\partial _{i}:=\frac{(1-s_{i})}{x_{i}-x_{i+1}}.  \tag{6.7}
\end{equation}%
It acts on monomials such as \textbf{x}$^{\mathbf{a}}$ in such a way that
the generator $s_{i}$ acting on the combination $%
x_{i}^{a_{i}}x_{i+1}^{a_{i+1}}$ converts it into $%
x_{i}^{a_{i+1}}x_{i+1}^{a_{i}}.$ By design an action of this operator on
monomial is zero if $a_{i}=a_{i+1}$, otherwise it diminishes the degree of
the monomial by $1$. In addition, these authors introduce the operators $%
\bar{\pi}_{i}$%
\begin{equation}
\bar{\pi}_{i}=\frac{(1-s_{i})}{x_{i}-x_{i+1}}x_{i+1}  \tag{6.8a}
\end{equation}%
and%
\begin{equation}
\pi _{i}=1+\bar{\pi}_{i}=x_{i}\frac{(1-s_{i})}{x_{i}-x_{i+1}}.  \tag{6.8b}
\end{equation}%
Evidently, in view of Eq.(6.8b), it is sufficient to use just one of these
operators. Because of this, following Ref.[63], we introduce an operator $%
D_{i}(p,q.r)=p\partial _{i}+q\bar{\pi}_{i}+rs_{i}$ with $p,q,r$ being some
numbers. L-S demonstrated that such an operator obeys the braid-type
relations (the 2nd and third of Eq.s(6.6)) while the relation $s_{i}^{2}=1$
in Eq.(6.6) is replaced by 
\begin{equation}
D_{i}^{2}=qD_{i}+r(q+r).  \tag{6.9a}
\end{equation}%
As is well known, the last relationship (with constants $q$ and $r$ properly
chosen) defines the Hecke algebra \ $H_{n}$ of the symmetric group $S_{n}.$
For the future use we shall rewrite it in the commonly used form as%
\begin{equation}
D_{i}^{2}=(1-Q)D_{i}+Q.  \tag{6.9b}
\end{equation}%
$H_{n}$ should be considered as a deformation of $S_{n}$. To be precise,
such defined Hecke algebra is of the $A_{n-1}$ type in the Coxeter -Dynkin
classification scheme\footnote{%
Because of this, one should be aware of existence of Hecke algebras for
other type of reflection groups [66].We are not going to use them in this
work.}. Since connection of Hecke algebra with knot theory \ is well known
[64], we are not discussing it in this work. Instead, we would like to
connect these results with traditional quantum mechanics thus bringing back
the spirit of old ideas of Maxwell, Kelvin and Tait [65]. From such point of
view the differences between quantum mechanics, quantum field theory and
string theory practically disappear.

Following Kirillov [58], we begin with relabeling previously defined
operator $\partial _{i}$ as $b_{ij}.$ Next, let $\partial _{i}$ be the usual
operator of differentiation, i.e. $\partial _{i}=\frac{\partial }{\partial
x_{i}},$ then we define the Dunkl operator $\mathcal{D}_{i}$ by%
\begin{equation}
\mathcal{D}_{i}=\partial _{i}+k\sum\limits_{j\neq i}b_{ij},  \tag{6.10}
\end{equation}%
where $k$ is some (known) constant. Such an operator acts on monomials
(polynomials). It possess the property $w\mathcal{D}_{i}w^{-1}=\mathcal{D}%
_{w(i)}$ $\forall w\in S_{n}.$Consider now the commutator $[\mathcal{D}_{i},%
\mathcal{D}_{j}]$. It can be rather easily demonstrated [58] that such a
commutator is zero if $b_{ij}$ satisfy the classical Yang-Baxter equations
(CYBE)%
\begin{equation}
\lbrack b_{12},b_{13}]+[b_{12},b_{23}]+[b_{13},b_{23}]=0.  \tag{6.11}
\end{equation}%
Conversely, Eq.(6.11) can be taken as a definition of $b_{ij}$. In such a
case we no longer need its explicit form given by Eq.(6.7).This is
facilitated by the designing of the so called \textit{degenerate affine
Hecke algebra}. Such an algebra is made as a semidirect product of $S_{n}$
with the familiar commutator algebra%
\begin{equation}
x_{i+1}s_{i}-s_{i}x_{i}=h,\text{ \ \ }x_{i}s_{j}=s_{j}x_{i\text{ \ }}\forall
i\neq j,j+1  \tag{6.12}
\end{equation}%
where $h$ is some constant analogous to $\hbar $\footnote{%
In fact, it is equal to $\hbar $ in most of cases known in literature. In
this work we do not impose such a requirement.}. It should be clear at this
point that Eq.s (6.12) are discrete analogs of the Heisenberg commutaton
rule, Eq.(5.8). Let us introduce yet another operator $\hat{s}%
_{i}=s_{i}+hb_{i,i+1}.$ It is designed in such a way that it obeys the braid
relations:%
\begin{equation}
\hat{s}_{1}\hat{s}_{2}\hat{s}_{1}=\hat{s}_{2}\hat{s}_{1}\hat{s}_{2}. 
\tag{6.13}
\end{equation}%
Moreover, if we define $R_{12}=s_{1}\hat{s}_{1},R_{23}=s_{2}\hat{s}%
_{2},R_{13}=s_{1}R_{23}s_{1}=s_{2}R_{12}s_{2},$ then the above Eq.(6.13)
becomes equivalent to the standard Yang-Baxter (Y-B) equation for $%
R_{ij}=1+hb_{ij}$ ( or $R_{ij}\simeq \exp (hb_{ij})$ \ for $h\rightarrow 0)$%
, i.e.%
\begin{equation}
R_{12}R_{13}R_{23}=R_{23}R_{13}R_{12}.  \tag{6.14}
\end{equation}%
Based on this logic, it follows, that the quantum Y-B equation, Eq.(6.14)
for $R_{ij}$ implies the classical Y-B Eq.(6.11).

All this discussion looks a bit formal. Indeed, why to introduce the
operator $\mathcal{D}_{i}?$ Why to be concerned about the commutator $[%
\mathcal{D}_{i},\mathcal{D}_{j}]?$ What the Yang-Baxter equations have to do
with all results of earlier sections? We would like to provide answers to
these questions now and in the next section.

First, consider an equation $\mathcal{D}_{i}f=0.$ It can be written
alternatively as%
\begin{equation}
\varkappa \frac{\partial }{\partial z_{i}}f(\mathbf{z})=\sum\limits_{j\neq i}%
\frac{\Omega _{ij}}{z_{i}-z_{j}}f(\mathbf{z})  \tag{6.15}
\end{equation}%
which is just the celebrated Knizhnik-Zamolodchikov (K-Z) equation. This
means that: a) the operator $\mathcal{D}_{i}$ is effectively a covariant
derivative (the Gaus-Manin connection [53] in the formalism of fiber
bundles) and b) that the vanishing of the commutator $[\mathcal{D}_{i},%
\mathcal{D}_{j}]$ is just the zero curvature condition [53,67\textbf{]}. The
question still remains : how $\Omega _{ij}$ in Eq.(6.15) is related to $%
b_{ij}$ ? The answer was found by Belavin and Drinfeld [68] and summarized
in Ref.[69], page 46. In the simplest "rational" case we have $b_{ij}(z)=%
\frac{\Omega _{ij}}{z}$ as expected. More complicated trigonometric and
elliptic cases were found in Ref.[68] and summarized in Ref.[69]. \ From
these references it should be clear that since solutions to the K-Z
equations are expressible in terms of hypergeometric functions of single and
multiple arguments, all examples of exactly solvable quantum mechanical
problems (including those involving the Dirac equation) found in textbooks
on quantum mechanics are covered by the formalism just described.

At this point it is legitimate to ask: all this is interesting but not new.
How these results are related to the honeycombs and puzzles discussed in
earlier sections? We provide an answer to this question below.

\section{Back to fusion}

\subsection{Motivation}

In this subsection we would like to study the following.

\textbf{Problem 7.1.} To what extent the fusion rule, Eq.(3.7), valid for
characters $s_{\lambda }$ of symmetric group $S_{n}$ should be modified if
instead of this group we consider its deformation caused by our use of \ the
Hecke algebra $H_{n}$?

We provide an answer to this problem \ having the following goal in mind.
Almost simultaneously with publications of KT \ honeycomb papers there
appeared a publication by Gleizer and Postnikov (G-P), Ref.[70], where
graphical methods alternative to those developed by KT were used,
essentially for the same purpose of calculating the L-R coefficients. These
alternative graphical methods involve braids, the Y-B and the tetrahedron
equations. Since these equations play only an auxiliary role in G-P's work,
many things where left unexplained. For instance, as soon as one introduces
these equations one leaves the domain of symmetric group $S_{n}$ and enters
the domain of Hecke algebra for this group. From G-P work it follows that
the fusion rule, Eq.(3.7), is expected to remain the same. This happens to
be the case most of the time but not always! The proof can be found in the
paper by Wenzl, Ref.[71\textbf{]}, Theorem 2.2. In the case if $Q$ in
Eq.(6.9) is the $m$-th root of unity the fusion \ rule, Eq.(3.7), should be
replaced by more elaborate fusion rule to be discussed in the subsection
7.4. The diagrammatical methods developed in G-P work provide no clues
regarding \ the possibility of such an alternative. It should be noted
though that the KT graphical methods also fail under the same circumstances.
In the following subsections we provide evidence that the "anomalous" case
corresponds to the situation when already familiar L-R coefficients should
be replaced by the Gromov-Witten (G-W) coefficients (invariants). \ In some
cases (to be specified) the fusion algebra, Eq.(3.7), is replaced by \ the
Verlinde-type algebra. Nevertheless,\ since the L-R coefficients obtained
with help of KT \ diagrammatic methods can be used as an input into more
complicated expressions for the G-W invariants, e.g. read Appendix B, this
justifies their place in this work. \ \ 

\textbf{Remark 7.2. }In view of Eq.(6.9) and the fact that $%
R_{ij}=1+hb_{ij}, $ it should be clear that representations for both the
Hecke algebra and Yang-Baxter equations are interrelated (and even
coincide!). This is indeed the case as demonstrated by Jimbo, Ref.[72].
Alternative derivations can be found in the pedagogically written paper by
Ram, Ref.[73]\footnote{%
See also [74].}. For non exceptional (generic) Q's calculation of characters
of Hecke algebra is nicely explained in the paper by King and Wybourne,
Ref.[75]. Since these are deformations of Schur functions $s_{\lambda }(%
\mathbf{x})$ that are smoothly dependent on Q, the fusion rule, Eq.(3.7),
remains unchanged.

The information just described is sufficient to bring us to our next topic.

\section{Mapping class group}

To understand better what follows, some facts about the mapping class group
are helpful at this time. We discuss them here using pedagogically written
paper by Jones [76]. Consider some Riemann surface $\mathcal{R}_{g}$ of
genus $g$. \ Every orientation-preserving homeomorphism of $\mathcal{R}$ \
is isotopic to the product of Dehn twists [77]. As is well known, e.g. see
[78], every $\mathcal{R}_{g}$ admits pants decomposition into collection of
the trice punctured (holed) spheres. This decomposition can be made along $%
c_{1},...,c_{3g-1}$ simple (non intersecting) closed curves. Every Dehn
twist can be represented as a combination of Dehn twists around just
described set of "basis" curves \ as demonstrated by Dehn. Subsequently it
was realized that it is sufficient to have just 2g+1 basic curves for this
purpose [77]. The mapping class group $\mathcal{M}_{\mathcal{R}}\mathcal{(}%
g) $ is generated by the collection of Dehn twists around these basis curves
modulo twists isotopic to identity. To understand properties of this group
it is convenient to consider $\mathcal{R}_{g}$ as branched covering (2-to-1)
of the sphere $S^{2}$ with branching done at $2g+2$ points as depicted in
Fig.13.


\begin{figure}[tbp]
\begin{center}
\includegraphics[width=2.00 in]{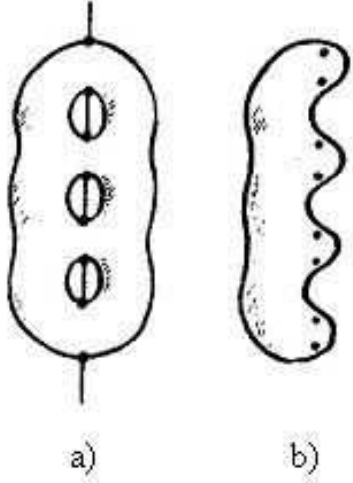}
\end{center}
\caption{Riemann surface of genus g a) is a two-fold branched covering of a
two sphere b)}
\end{figure}

Thus, $S^{2}=\mathcal{R}_{g}/i$, where $\QTR{sl}{i}$ is involution depicted
in Fig.13. Let $Q_{g}=\{q_{1},...q_{2g+2}\}$ be the branching set of points
on $S^{2}$ while $\tilde{Q}_{g}=\{\tilde{q}_{1},...\tilde{q}_{2g+2}\}$ the
corresponding set of points on $\mathcal{R}_{g}$. From Birman, Ref.[77],
page 182, one can find how the Dehn twists on $\mathcal{R}_{g}$ are related
to the set $\tilde{Q}.$

In particular, to each Dehn twist on $\mathcal{R}_{g}$ one associates points 
$\tilde{q}_{2i-1}$ and $\tilde{q}_{2i}$ on $\mathcal{R}_{g}$ \ $(1\leq i\leq
g)$ through which the $c_{i}$-th basis curve is passing. The Dehn twist on $%
\mathcal{R}_{g}$ is projected into $S^{2}$ in the form of the homeomorphism $%
\omega _{i}$ $(1\leq i\leq 2g+1)$ of $S^{2}$ resulting in exchange between
the points $q_{2i-1}$ and $q_{2i}$ which leaves the rest of points fixed. If 
$\theta _{i}$ denotes the isotopy class of the Dehn twist about $c_{i}$, it
can be demonstrated that such class obeys the braid group $\mathcal{B}_{n}$
relations given by $\theta _{i}\theta _{j}=\theta _{j}\theta _{i}$ if $%
\left\vert i-j\right\vert \geq 2$ and $\ \theta _{i}\theta _{i+1}\theta
_{i}=\theta _{i+1}\theta _{i}\theta _{i+1}$ otherwise$.$ In view of this,
the mapping class group $\ $of the $2g+2$ punctured sphere $\mathcal{M}%
_{S^{2}}(2g+2)$ \ is generated by a homomorphic image of these generators
which can be represented by $2g+2$ strings (braids) generating the braid
group. A presentation for $\mathcal{M}_{S^{2}}(2g+2)$ \ is given by the
Theorem 4.5. of Birman's book, Ref.[77]. Explicitly, it is given by%
\begin{equation}
\begin{array}{c}
\omega _{i}\omega _{j}=\omega _{j}\omega _{i}\text{ \ \ \ \ if }\left\vert
i-j\right\vert \geq 2\text{ }, \\ 
\omega _{i}\omega _{i+1}\omega _{i}=\omega _{i+1}\omega _{i}\omega _{i+1},
\\ 
(\omega _{1},...,\omega _{2g+1})^{2g+2}=1, \\ 
\omega _{1}\cdot \cdot \cdot \omega _{2g}\omega _{2g+1}^{2}\cdot \cdot \cdot
\omega _{1}=1.%
\end{array}
\tag{7.1}
\end{equation}%
The \ homomorphism just described sends $\theta _{i}$ to $\omega _{i}$. In
view of the involution depicted in Fig.13, the kernel of this homomorphism
is of order $2$. We are interested in finding out wether the Hecke algebra $%
H_{2g+1}$ can be associated with the presentation given by Eq.(7.1).

If $s_{i}$ is the generator for the symmetric group $S_{2g+1}$ (associated
with exchange of \ 2 points on $S^{2})$ we are interested in mappings of $\ $%
generators $T_{i}$ of Hecke algebra $H_{2g+1}$ into $s_{i}$ and $s_{i}$ into 
$\omega _{i}.$ This happens to be possible but nontrivial as discussed in
the Jones paper, Ref.[76]. The nontriviality comes from the fact that
representation $T_{i}^{\prime }s$ depends on $Q$ so that one has to satisfy
the constraint $(\omega _{1},...,\omega _{2g+1})^{2g+2}=1$ for arbitrary $%
Q^{\prime }$s. Let $\pi _{Y}^{\prime }(s_{i})=\omega _{i}$ be the desired
mapping (already from $T_{i}$ to $\omega _{i})$ with $Y$ indicating the
Young tableaux associated with representation of the symmetric group. Lemma
9.2. by Jones allows us to make it in such a way that (irrespective to the
actual value of $Q$) one obtains 
\begin{equation}
\pi _{Y}^{\prime }(s_{1},...,s_{2g+1})^{2g+2}=1  \tag{7.2a}
\end{equation}%
where the prime for $\pi _{Y}^{\prime }$ indicates \ the needed "adjustment"
to make presentation Q-independent. Jones argues that the relation $\omega
_{1}\cdot \cdot \cdot \omega _{2g}\omega _{2g+1}^{2}\cdot \cdot \cdot \omega
_{1}=1$ \ in Eq.(7.1) is equivalent to $(\omega _{2},...,\omega
_{2g+1})^{2g+1}=1.$ Accordingly, one obtains,%
\begin{equation}
\pi _{Y}^{\prime }(s_{2},...,s_{2g+1})^{2g+1}=1.  \tag{7.2b}
\end{equation}%
Based on this observation by Jones, one can continue this downsizing process
thus obtaining the flag of Hecke algebras $\ H_{1}(Q)\subset H_{2}(Q)\subset
\cdot \cdot \cdot H_{2g+1}(Q).$ The important theorem by Jones (to be used
below) can be stated now as follows

\ 

\textbf{Theorem 7.3.(}Jones\textbf{\ [}76], page 361\textbf{)} \textit{Let Y
be a Young diagram and let }$\pi _{Y}^{\prime }$\textit{\ be the
corresponding representation of \ }$B_{2g+1}$\textit{\ designed in such a
way that Eq.(7.2a) holds (for any Q's). Then }$\pi _{Y}^{\prime }$\textit{\
defines a representation of }$M_{S^{2}}(2g+2)$\textit{\ via }$\omega
_{i}\rightarrow \pi _{Y}^{\prime }(s_{i})$\textit{\ if and only if Y is
rectangular.}

\ 

In other words, one begins with the rectangular Young tableaux of size $%
m\times n$ as discussed in Section 3 and obtains all Young tableaux which
can fit into this rectangle by sequentially deleting boxes (one at the
time).To finish this subsection we need to discuss difference between the
mapping class group of the sphere $S^{2}$ and that of the disc $D^{2}.$ \
Since the disc can be viewed as a sphere with just one point deleted the
generators $\omega _{i}$ described before can be used in the present case as
well. \ Thus we obtain the following theorem

\ 

\textbf{Theorem 7.4.} (Birman [77\textbf{],} page 32).\textit{\ Let M be a
group of automorphisms of }$\pi _{1}(D^{2}-Q_{n})$\textit{\ which are
induced by the homeomorphisms of }$D^{2}-Q_{n}$\textit{\ which keep the
boundary of }$D^{2}$\textit{\ fixed poinwise. Then M is precisely the group }%
$B_{n}\footnote{%
That is $B_{n}$ $is$\ the braid group and the set $Q_{n}$ is analogous to
that determined immediately after Fig.13.}.$

\textit{\bigskip\ }

\textbf{Remark 7.5. }The fundamental group\textbf{\ }$\pi _{1}(D^{2}-Q_{n})$
is a free group of rank $n$. This group is made of loops anchored at some
point in $D^{2}$ and such that each loop encloses only one puncture. In the
next subsection this group will pay a role of monodromy group for the K-Z
equation.

\bigskip \textbf{Remark 7.6}. In the case of a sphere there is no boundary
which is fixed pointwise. As result, one gets relations 3 and 4 in Eq.(7.1)
which are simple consequences of rotational invariance of $S^{2}$ $($e.g.
see Fig.13 of Ref.[77], page 157).

Being armed with these facts we are ready to discuss the next topic.

\subsection{Monodromy \ group of the Knizhnik-Zamolodchikov equations and
the Riemann-Hilbert problem}

The above discussion about the mapping class group seems to be detached from
the rest of this paper. We would like to correct this deficiency now For
this purpose, we need to remind to our readers about some basic facts from
the theory of K-Z- type equations. According to Refs.[67,79] the K-Z
Eq.(6.15) is equivalent to the following system of equations%
\begin{equation}
\sum\limits_{i=1}^{n}\frac{\partial }{\partial z_{i}}f=0\text{ and }\kappa
(\sum\limits_{i=1}^{n}z_{i}\frac{\partial }{\partial z_{i}}%
)f=(\sum\limits_{i<j}\Omega _{ij})f.  \tag{7.3}
\end{equation}%
The first of these equations comes from the fact that $\Omega _{ij}=\Omega
_{ji}$. The second is obtained from Eq.(6.15) by multiplying both sides of
this equation by $z_{i}$, summing over $i^{\prime }s$ and again taking into
account that $\Omega _{ij}=\Omega _{ji}.$ The first of these equations
indicates that solutions should be translationally invariant so that only
differences of arguments must be used. The second implies that solutions
must be homogenous. To extract this homogeneity we follow Ref.[69],page 37,
and introduce new variables $\varsigma _{i}=\dfrac{z_{i+1}}{z_{i}}$, $%
i=1,...,n-1$, $\varsigma _{n}=z_{n}.$ In terms of these variables the K-Z
equations acquire the following standard form%
\begin{equation}
\varsigma _{i}\frac{\partial }{\partial \varsigma _{i}}f=A_{i}(\varsigma
_{1},...,\varsigma _{n-1})f\text{ , }i=1,...,n,  \tag{7.4}
\end{equation}%
where the $A_{i}^{\prime }s$ are holomorphic functions in the domain $D:=$\{$%
\left\vert \varsigma _{j}\right\vert <1,1\leq j\leq n\}.$ Such form of the
K-Z equations allows us to use general local theory of linear differential
equations of Fuchsian-type. In particular, the theorem \ which is formulated
and proved on page 121 of Ref.[80] provides the desired solution. It is
given by 
\begin{equation}
f=F_{0}(\varsigma _{1},...,\varsigma _{m})\varsigma _{1}^{A_{1}(0)}\cdot
\cdot \cdot \varsigma _{m}^{A_{m}(0)},  \tag{7.5}
\end{equation}%
where $F_{0}(\varsigma _{1},...,\varsigma _{m})$ is an $n\times n$ matrix-
valued function holomorphic in the domain $\hat{D}:=\{\left\vert \varsigma
_{j}\right\vert <1,1\leq j\leq m\}$ and such that $F_{0}(0)=Id.$These
results allow us to mention about the Riemann-Hilbert\ (R-H) problem and by
doing so to make a connection with previous subsection. The results which
follow will bring us directly to the discussion of the G-W invariants.

To discuss the R-H problem in the domain $D$ we would like to rewrite
Eq.s(7.4) in the following equivalent Fuchsian-type form [80]%
\begin{equation}
\frac{\partial }{\partial \varsigma _{i}}f=\tilde{A}_{i}f,  \tag{7.6}
\end{equation}%
where $\tilde{A}_{i}=A_{i}(\varsigma _{1},...,\varsigma _{n-1})/\varsigma
_{i}.$ Consider now the limiting case: $\zeta _{i}\rightarrow 0$ for a
subdomain $\bar{D}_{i}$ where presence\ of other singularities can be
neglected. Without loss of generality we can subdivide $D$ into such
subdomains so that the total (global) solution in $D$ is made of local
solutions \ in respective subdomains. The requirement that these solutions
must agree in the overlapping regions is the essence of the R-H problem
stated in simple terms. The $n=$ $2g+2$ punctured sphere discussed in the
previous subsection can be looked upon as $2g+3$ punctured disc. Hence, we
can initially develop our discussion for the disc $D$. In both cases: $D$ or 
$S^{2},$ the motion of subdomains $\bar{D}_{i}$ \ on these surfaces is
controlled by the action of the respective braid groups as explained in the
previous subsection.

Consider now a monodromy (holonomy) around given singularity. The existence
of this monodromy is assured by the fact that Eq.(7.6) is linear equation
whose solution is defined with accuracy up to a constant matrix which we
shall call $T_{i}$. This matrix can be found by noticing that the matrix $%
\tilde{A}_{i}$ has only the first order (Fuchsian) pole as singularity.
Going around this pole once \ will pic up a phase resulting in the monodromy
matrix $T_{i}$. It is given by $T_{i}=\exp (2\pi iA_{i}(0)).$ This can be
easily understood with help of Eq.(7.5) taking into account the analytical
properties of the logarithmic function. Consider now Eq.(7.5) in the overlap
of two domains $\bar{D}_{i}$ and $\bar{D}_{j}.$ Evidently, in such domains
we can use just one local coordinate so that at the overlap of these domains
one has 
\begin{equation}
F_{0}(\varsigma _{i})F_{0}^{-1}(\varsigma _{j})=\varsigma
_{i}^{A_{i}(0)}\varsigma _{j}^{-A_{j}(0)}.  \tag{7.7}
\end{equation}%
With help of Eq.(7.7) the\ R-H problem can be formulated now as a problem of
finding \ of a holomorphic vector function $\mathbf{f}(\varsigma )$ with
good behavior at $\infty $ \ and such that in the complex plane \textbf{C}$%
\varsigma $ it obeys an equation $\mathbf{f}_{-}(\varsigma )=M(\varsigma )$ $%
\mathbf{f}_{+}(\varsigma )$ for some prescribed $n\times n$ matrix $%
M(\varsigma )$ and \ for a contour (closed) $C$ in $\varsigma -$plane such
that $\mathbf{f}_{+}(\varsigma )$ and $\mathbf{f}_{-}(\varsigma )$ lie
respectively inside and outside of the domain enclosed by $C.$

This problem can be reformulated a bit differently as follows. Given a set
of $n$ points $\{a_{1},...,a_{n}\}$ in $\mathbf{C}\varsigma $ and $n\times n$
matrices $A_{1},...,A_{n}$ representing the monodromy group $\mathcal{G}$ of
these points find all equations of the type given by Eq.(7.6) which have the
monodromy group $\mathcal{G}.$ We can complicate matters further by making
the set of points to move in $\mathbf{C}\varsigma $ (or $S^{2}).$ This leads
to the \textit{isomonodromic} deformation problem. It can then be formulated
as follows.

\textbf{Problem 7.7}. (Isomonodromy problem) For a given representation of
the monodromy group $\mathcal{G}$ find dependence of matrices $\tilde{A}_{i}$
in Eq.(7.6) on location of poles given by the (moving) set $%
\{a_{1},...,a_{n}\}.$

This problem was solved by Schlesinger [81]. We would like to provide some
needed details within the context of K-Z equations. In particular, taking
into account that $\Omega _{ij}=\Omega _{ji}$ Eq.(6.15) can be rewritten as
follows 
\begin{eqnarray}
df &=&\Gamma f,\text{ where}  \notag \\
\Gamma &=&\sum\limits_{1\leq i\leq j\leq n}\frac{\kappa ^{-1}\Omega _{ij}}{%
z_{i}-z_{j}}(dz_{i}-dz_{j}).  \TCItag{7.8}
\end{eqnarray}%
The previously imposed requirement $[\mathcal{D}_{i},\mathcal{D}_{j}]=0$,
e.g. see Eq.s(6.10),(6.11) can be now rewritten as $d(df-\Gamma f)=0$
implying the Frobenius- type equation $\Gamma \wedge \Gamma =0$ which holds
only if%
\begin{equation}
\lbrack \Omega _{ij},\Omega _{kl}]=0\text{ for }i\neq j\neq k\neq l 
\tag{7.9a}
\end{equation}%
and%
\begin{equation}
\lbrack \Omega _{ij},\Omega _{ik}+\Omega _{jk}]=[\Omega _{ij}+\Omega
_{ik},\Omega _{jk}].  \tag{7.9b}
\end{equation}%
According to Kohno, Ref.[64], these are the infinitesimal pure braid
relations\footnote{%
The difference between the pure braid and braid groups is exactly the same
as the difference between the statistics of distinguishable (colored) and
indistiguishable (colorless) particles [77]. Since at the fundamental level
elementary particles, say electrons, are indistinguishable, in 2 dimensions
their motion is described by the braid group $B_{n}$. Thus, the braid group
is made of a semidirect product of pure braid group $F_{n}$ and the
permutation group $S_{n}.$}. For $n=3$ they coincide with earlier obtained
CYBE, Eq.s(6.11). Kohno, Ref.[64], demonstrated that, at least in the case
of rational solutions of the CYBE's, the results, Eq.s (7.9 a,b), can be
brought into correspondence with the CYBE's for $n>3$, i.e. for \ $1\leq
i\leq n.$ This result provides an independent support of earlier obtained
Eq.(6.11) and connects us with the mapping class group presentation,
Eq.(7.1). Evidently, the braid relations in Eq.(7.1) (adopted for $S^{2}$)
become the YBE's. Since the obtained isomorphism involves earlier discussed
monodromy matrices we come to the conclusion that the monodromy
representation for the K-Z equation is equivalent to the YBE representation.
This is known as the Kohno-Drinfeld theorem, \ Ref.[67], Thm 19.4.1. The
introduced concepts even though being useful, play only an auxiliary role in
this work. They were introduced mainly for the sake of the discussion
presented in the next subsection.

\subsection{The multiplicative Horn problem}

\subsubsection{ Emergence of \ Gromov-Witten invariants}

In view of results we just obtained\ and, taking into account Eq.s(7.1), the
monodromy matrices for the punctured sphere $S^{2}$ should be subjected to
the following constraint 
\begin{equation}
\prod\limits_{i=1}^{n}\exp (i2\pi A_{i})=\mathbf{I,}  \tag{7.10}
\end{equation}%
where $\mathbf{I}$ is the unit matrix and $n=2g+1$. \ Taking into account
Remark 7.5., this equation has a simple geometrical meaning. It represents
loops (holonomies or monodromies) around $n+1$ points, that is it represents
the fundamental group $\pi $ of $S^{2}$ with points $\{a_{1},...,a_{n+1}\}$
deleted. As it is written, this equation suffers from the fact that it is
not reflecting the differences between the topology of the disc D and that
for the sphere S$^{2}$. Because of the Remark 7.5., this \ equation cannot
be used as such for the disc since the fundamental group $\pi
_{1}(D^{2}-Q_{n})$ is \textit{free} group of rank $n$. Even though it can be
used for $S^{2}$ it does not reflect the constraint Eq.(7.2) adequately.
Fortunately, this deficiency is easily correctable if we rewrite Eq.(7.10)
in the alternative form \ as follows 
\begin{equation}
\prod\limits_{i=1}^{n}\exp (i2\pi A_{i})=\exp (i2\pi d\mathbf{I}) 
\tag{7.11a}
\end{equation}%
with $d=0,1,2,..$. or, even more generally, as 
\begin{equation}
\prod\limits_{i=1}^{n}\exp (i2\pi \frac{A_{i}}{d_{i}})=\exp (i2\pi \mathbf{I}%
).  \tag{7.11b}
\end{equation}

Matrices $\exp (i2\pi A_{i})$ are unitary by design and each of $%
A_{i}^{\prime }s$ is diagonalizable so that $\lambda (A_{i})=\{\lambda
_{1}(A_{i}),...,\lambda _{k}(A_{i})\}$ represents the eigenvalue set for the
matrix $A_{i}$. Since matrices $\exp (i2\pi A_{i})$ are unitary their
determinant is 1. This leads to the requirement $\lambda _{1}(A_{i})+\cdot
\cdot \cdot +\lambda _{k}(A_{i})=0(N\func{mod}d_{i})$ $\forall i$ provided
that $\sum\nolimits_{i}d_{i}=d.$ For the sake of comparison with earlier
sections we would like to consider (without loss of generality) the case of $%
n=3$. Then, instead of Eq.(1.6), we obtain, 
\begin{equation}
\lambda _{1}+\cdot \cdot \cdot \lambda _{k}+\mu _{1}+\cdot \cdot \cdot +\mu
_{k}=\nu _{1}+\cdot \cdot \cdot +\nu _{k}+Nd_{1}+Nd_{2}+Nd_{3}  \tag{7.12a}
\end{equation}%
which is essentially the K-M-B-B condition, Eq.(1.5). The same result can be
rewritten as 
\begin{equation}
\left\vert \lambda \right\vert +\left\vert \mu \right\vert =\left\vert \nu
\right\vert +Nd.  \tag{7.12b}
\end{equation}%
In view of this relation, the fusion rule, Eq.(3.7), should be modified
accordingly. Following Ref.[$27$\textbf{]}, we write%
\begin{equation}
\sigma _{\lambda }\ast \sigma _{\mu }=\tsum\limits_{d,\nu }q^{d}C_{\lambda
\mu }^{\nu }(d)\sigma _{\nu }.  \tag{7.13}
\end{equation}%
The star symbol represents the product of "quantum" cohomology classes. For $%
d=0$ this symbol becomes again the usual dot symbol \ used in Eq.(3.7). The
"quantum" \ L-R coefficients $C_{\lambda \mu }^{\nu }(d)$ are in fact the
genus zero 3-point Gromov-Witten invariants which are structure \ constants
in the "small" quantum cohomology ring [82].

The multiplication law, Eq.(7.13) takes place only if \ Eq.(7.12b) holds.
This equation replaces earlier equation $\left\vert \lambda \right\vert
+\left\vert \mu \right\vert =\left\vert \nu \right\vert $ used for
computations in Eq.(3.7). The r.h.s. of Eq.(7.13) is a polynomial in $q$
(where $q$ stays for "quantum "). Its physical role is clear from Eq.(7.13):
it plays a role of fugacity associated with the degree of mapping\footnote{%
From $\ $the punctured sphere $S^{2}$ to the Grassmannian $G(m,k)$ as can be
seen from Eq.s(7.2),(3.7) and (3.8)} $d$. More generally, in view of
Eq.(7.11b), we should replace $q^{d}$ in Eq.(7.13) by $q^{d}=\tprod%
\limits_{i}q_{i}^{d_{i}}.$ To avoid unnecessary complications we shall be
working just with $d$ from now on$.$ Evidently, under such circumstances the 
$q$ indeterminate becomes an analog of $Q$ in the Hecke algebra, Eq.(6.9b),
of the symmetric group. Connections with \ the Hecke algebra are highly
nontrivial. They are discussed \ below and in Appendix C. In order to
prepare our readers for this discussion we would like to proceed with actual
computation of the G-W invariants.

There are many ways to compute these invariants. We would like to discuss
only those which are logically compatible with results presented in previous
sections. \ In particular, we would like to connect the results presented in
Section 5.3 \ with what has been discussed now. Following Walton, Ref.[83],
we begin with the Weyl character formula, our (III.2.28). It is given by 
\begin{equation}
\chi (\lambda )=\sum\limits_{w\in W}n_{w}(\lambda )e(w)  \tag{7.14}
\end{equation}%
so that the fusion rule (Walton's Eq.(1.5)) reads: 
\begin{equation}
\chi (\lambda )\cdot \chi (\mu )=\sum\limits_{\nu \in \Delta _{+}}C_{\lambda
\mu }^{\nu }\chi (\nu ).  \tag{7.15}
\end{equation}%
For the sake of space we refer to our Parts II (Appendix A) and Part III,
Sections 1 and 2, for all definitions and notations. In Humphrey's book,
Ref. [84], on page 140 our readers can find (our) Eq.(7.15) with details of
its derivation. Some of these details are discussed below. Clearly, since
the constants $C_{\lambda \mu }^{\nu }$ will be different for different
Weyl-Coxeter groups, the formulas above include the fusion rule, Eq.(3.8),
as special case (for characters of symmetric group $S_{n}$ which \ is of the
type $A_{n-1\text{ }}$in Dynkin's classification of the Weyl-Coxeter
reflection groups). The Kostant multiplicity formula, Eq.(III.2.31), plays a
very important role in Walton's analysis. In particular, he uses it in order
to arrive at the Steinberg's formula for $C_{\lambda \mu }^{\nu }$ 
\begin{equation}
C_{\lambda \mu }^{\nu }=\sum\limits_{w,\mathfrak{v}\in W}\varepsilon (%
\mathfrak{v}w)P(w\lambda +\mathfrak{v}\mu -\nu ).  \tag{7.16}
\end{equation}%
In arriving at this result Eq.(III.2.31) for the Kostant multiplicity
formula was used. Detailed derivation of the Steinberg's formula can be
found on page 141 of Ref.[84]. Explicit use of this formula is inconvenient
though. It is given here to emphasize the combinatorial and symplectic
nature of the L-R coefficients in accord with Section 5.3.\footnote{%
More details \ on symplectic interpretation of Eq.(7.16) can be found either
in \ Part III or in recent paper by Guillemin and Rassart, Ref.[85].}. Since
symplectic nature of $C_{\lambda \mu }^{\nu }$ \ was emphasized in Section
5.3.,\ we brought \ Eq.(7.16) to the attention of our readers \ with
additional purposes in mind. Specifically, Eq.(7.16) can be used not only
for computations involving more traditional Weyl-Coxeter reflection groups
but also for their affine extensions (e.g. see Appendix A for Part II for
definitions). In this case Eq.(7.16) formally stays the same, except now the
Weyl group $W$ is replaced by the affine Weyl group $W^{(k)}$ (or $W^{(m)})$
where $k$ (or $m$) have the same meaning as in our Eq.(3.3) (or (3.4)). This
fact is not self-obvious and will be explained. Before doing so we note the
following. Using Eq.(III.2.16) we know that $\varepsilon (\mathfrak{v}%
w)=(-1)^{l(\mathfrak{v}w)}=(-1)^{l(\mathfrak{v})}(-1)^{l(w)}.$ This
observation allows us to rewrite the affine version of Eq.(7.16) \ in
equivalent form given by%
\begin{equation}
C_{\lambda \mu }^{\nu (k)}=\sum\limits_{w\in W^{(k)}}\varepsilon
(w)C_{\lambda \mu }^{w\cdot v}  \tag{7.17}
\end{equation}%
with $C_{\lambda \mu }^{w\cdot v}$ representing standard L-R coefficient
which can be calculated with help of honeycombs and puzzles as previously
discussed. The fact that this is the case, unfortunately, is not sufficient
for the efficient calculation of $C_{\lambda \mu }^{\nu (k)}.$ Hence, the
task lies in finding the efficient scheme for such calculations. By doing so
connections between the coefficients $C_{\lambda \mu }^{\nu (k)}$ and $%
C_{\lambda \mu }^{\nu }(d)$ will become apparent.

We begin with few definitions. In particular, in connection with the Young
diagram defined by partition $\lambda ,$ we define an $n$ rim- hook of this
partition. It is a connected subset of $n$ boxes of $\lambda $ such that it
does not contain a $2\times 2$ square. A rim-hook is \textit{legal }if by
removing it from the Young diagram $\lambda $ the remainder is still \ be a
valid Young diagram. Othervise, the $n$ rim- hook is considered to be\textit{%
\ illegal} as depicted in Fig. 14.

\begin{figure}[tbp]
\begin{center}
\includegraphics[width=3.50 in]{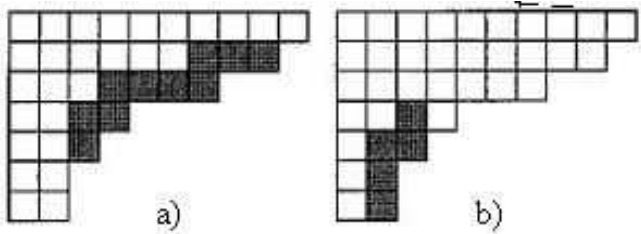}
\end{center}
\caption{Legal a) and illegal b) rim hooks}
\end{figure}

An \textit{n-core} for the partition $\lambda $ corresponds to a partition
obtained by sequential removal of legal $n$-rim hooks till one obtains the
configuration for which such removals are no longer possible. Let $%
\left\vert \lambda \right\vert $ be the \textit{weight} of partition $%
\lambda ,$ i.e. $\left\vert \lambda \right\vert =\lambda _{1}+\lambda
_{2}+\cdot \cdot \cdot .$ Let then the weight of the resulting core
partition be $\left\vert core_{n}\lambda \right\vert .$ \ The number $%
r_{n}(\lambda )$ =$\left( \left\vert \lambda \right\vert -\left\vert
core_{n}\lambda \right\vert \right) /n$ is the number of $n-$ rim hooks
removed in this process. Furthermore, define the \textit{width}$\mathit{(R}%
_{i})$ of an $i$-th $n$-rim hook $R_{i}$ as the number of columns it
occupies in the Young diagram. Let \textit{w}$_{i}$=\textit{width}$(R_{i})$
denote this number. As in Section 3.1, we place our original Young diagram
into $m\times k$ rectangle so that $m+k=N$. Because of this, we write $%
C_{\lambda \mu }^{\nu }(d)=C_{\lambda \mu }^{\nu }(d;m,k)$ so that if, say, $%
\check{\lambda}$ is the partition conjugate to $\lambda ,$ one can show \
[86] that 
\begin{equation}
C_{\lambda \mu }^{\nu }(d;m,k)=C_{\check{\lambda}\check{\mu}}^{\check{\nu}%
}(d;k,m)  \tag{7.18}
\end{equation}%
in accord with the fact that $C_{\lambda \mu }^{\nu }=C_{\check{\lambda}%
\check{\mu}}^{\check{\nu}}$ for the ordinary L-R coefficients. To make
actual computations using these definitions we need to use the following

\ 

\textbf{Lemma 7.8.} (Bertram, Ciocan-Fontanine, Fulton [86], page 733). 
\textit{If }$\mu $\textit{\ is the partition resulting from removing an
n-rim hook from }$\lambda $\textit{, then the (quantum) cohomology class }$%
\sigma _{\lambda \text{ }}$\textit{is related to the (classical) cohomology
class }$\sigma _{\mu \text{ }}$\textit{\ through the following relation}%
\begin{equation}
\sigma _{\lambda \text{ }}=(-1)^{k-w}q\sigma _{\mu \text{ }}  \tag{7.19a}
\end{equation}%
\textit{where }$w$\textit{\ is the width of the n-rim hook\ and }$k$\textit{%
\ is the width of the Young tableaux containing partition }$\lambda .$%
\textit{If }$\lambda $\textit{\ contains an illegal n-rim hook or if }$%
\lambda _{m+1}>0$\textit{\ and }$\lambda $\textit{\ contains no n-rim hooks
or if }$\lambda _{1}>k,$\textit{\ then }$\sigma _{\lambda \text{ }}=0.$

\bigskip

The above lemma allows us to use previously discussed fusion rule, Eq.(3.7),
for calculation of the Gromov-Witten coefficients. We would like to show
step-by -step how this is done. Let both partitions $\lambda $ and $\mu $
belong to the $m\times k$ rectangle and $R_{1},...,R_{r_{n}(\lambda )}$ be
the respective $n-$rim hooks removed, then, instead of Eq.(7.19a), we obtain:%
\begin{equation}
\sigma _{\lambda \text{ }}=\varepsilon (\lambda /\mu )q^{r_{n}(\lambda
)}\sigma _{\mu \text{ }},  \tag{7.19b}
\end{equation}%
where $\varepsilon (\lambda /\mu )=\prod\limits_{i=1}^{r_{n}(\lambda
)}(-1)^{k-w_{i}}.$ If the partition $\mu $ is not contained in $m\times k$
rectangle, then $\sigma _{\mu \text{ }}=0.$ Next, we combine the fusion
rule, Eq.(3.7), with just obtained result in order to obtain the
prescription for calculation of the G-W invariants. Appendix B contains an
example of calculation of G-W invariants in which the KT honeycombs are used
for calculations of $C_{\lambda \mu }^{\nu }$ as an input. In actual \
illustrative calculations done in Appendix B \ we took into account that $%
d\equiv r_{N}(\lambda )$ and $n\rightarrow N$ in accord with Eq.(7.12b).
Thus, in general, we obtain: 
\begin{equation}
\sigma _{\lambda }\ast \sigma _{\mu }=\tsum\limits_{\nu \in k\times
m}\tsum\limits_{d=0}q^{d}C_{\lambda \mu }^{\nu }(d;m,k)\sigma _{\nu }\text{%
,\ where }C_{\lambda \mu }^{\nu }(d;m,k)=\tsum\limits_{\rho }\varepsilon
(\rho /\nu )C_{\lambda \mu }^{\rho }.  \tag{7.20}
\end{equation}%
This result admits a somewhat different interpretation. For instance, let us
introduce the "quantum" cohomology class $\sigma _{\rho \text{ }}$ via 
\begin{equation}
\sigma _{\rho \text{ }}=\tsum\limits_{d=0}q^{d}\tsum\limits_{\nu \in k\times
m}\varepsilon (\rho /\nu )\sigma _{\nu },  \tag{7.21}
\end{equation}%
then we can formally rewrite Eq.(7.20) as 
\begin{equation}
\sigma _{\lambda }\ast \sigma _{\mu }=\tsum\limits_{\rho }C_{\lambda \mu
}^{\rho }\sigma _{\rho \text{ }},  \tag{7.22}
\end{equation}%
where $C_{\lambda \mu }^{\rho }$ is the classical L-R coefficient computable
with help of KT honeycombs, puzzles or hives. This result is only in formal
agreement with earlier obtained Eq.(3.7) since in Eq.(3.7) there is no
restrictions on summation over $\rho $ while in the present case $C_{\lambda
\mu }^{\rho }=$ $C_{\lambda \mu }^{\rho }(m,k)$ and is zero otherwise.
Significance of such a restriction is discussed in the Appendix C. Obtained
result is easy to understand using physical arguments. Indeed, using
Eq.(7.12b) we notice that with $N\rightarrow \infty $ the only way to
satisfy this equation is to let $d=0$. Hence, in view of the Theorem 5.1.
Eq.(7.22) makes sense. Clearly, it \ should be understood only as
qualitative result since summation over $\rho $ is actually restricted.

We \ have accumulated enough results enabling us to inject more physics into
what was obtained thus far. This is done in the next subsection.

\subsubsection{Verlinde algebra and Hecke algebra at the root of unity}

While the mathematical meaning of G-W invariants, especially for small
quantum cohomology ring, is discussed in many places, e.g. see
Refs.[82,87,88], to our knowledge, this literature does not contain any
information about physical significance of these invariants. We would like
to correct this deficiency. By doing so we also will be able to sketch some
missing links between these topics and those discussed in previous sections.
We hope that our readers will use this material along with that of Appendix
C as a point of departure for much deeper and thorough study.

We begin with observation that there are two ways to define the L-R
coefficients: one is through the composition (fusion) law, Eq.(3.8), while
the other through the Fourier series-type expansion of the skew Schur
function $s_{\lambda /\mu }$ [37]. Specifically,%
\begin{equation}
s_{\lambda /\mu }=\tsum\limits_{\nu }C_{\mu \nu }^{\lambda }s_{\nu }. 
\tag{7.23}
\end{equation}%
In view of Eq.(7.22) it is only natural to anticipate that there must be a
"quantum" analog of Eq.(7.23). Such an analog was indeed recently found by
Postnikov [27]. We would like to connect his results with those in Ref.[86]
using some results from the paper by McNamara [89].This will allow us to
inject some physics into our discussion.

Recall that if $\lambda $ and $\mu $ are two partitions such that $\mu
\subseteq \lambda $ if $\mu _{i}\leq \lambda _{i}$ $\forall i,$ then a pair $%
(\mu ,\lambda )$ is called \textsl{skew partition. }For such a partition the
Young diagram is made of diagram for $\lambda $ with $\mu $ removed.
Traditionally used notation for such obtained diagram is given by $\lambda
/\mu =(\lambda _{1},...,\lambda _{k})/(\mu _{1},...,\mu _{l}).$ Postnikov 
replaced the $m\times k$ rectangle used in calculations in Appendix B by the
torus obtained from this rectangle by identification of its boundaries in a
usual way. He considered skew partitions on such a torus and proved that
Eq.(7.23), when adapted to this toroidal topology produces $C_{\lambda \mu
}^{\nu }(d;m,k)$ as required. His derivation of this result \ makes \ the
meaning of parameter $d$ especially transparent and suitable for potential
physical applications (to be discussed below).

\textbf{Remark 7.9}. In view of Eq.s (7.11), (7.12) replacement of the
rectangle by torus is completely natural. Moreover, in view of the fact that 
$\ $in the most general case $q^{d}=\tprod\limits_{i}q_{i}^{d_{i}}$, it is
surprising that instead of using the multidimensional torus the two
dimensional torus is sufficient for reproduction of the G-W invariants. This
peculiarity is explained below and in the Appendix C.

Following McNamara [89\textbf{]}, for illustrative purposes we choose the
basic rectangle $\boldsymbol{R}$ with parameters $k=3$ and $m=4.$The $\mu $
partition is chosen to be $\mu =(2,1)$ and the \textsl{original} $\lambda $
partition is chosen to be $\lambda =(4,4,4,4,2,1,1).$ Using the French style
of writing of the Young tableaux (that is from the bottom-up) the skew
tableaux $\lambda /\mu $ is depicted in Fig.15a). \ Multiple copies of the
same skew tableaux are depicted in the universal cover of the torus made of
the basic rectangle in Fig.15b). \ To recognize them in such setting one
should pay attention to the vertical lines denoted as $V$ and the horizontal
line denoted as $H$. The lower left corner of the original rectangle $%
\boldsymbol{R}$ coincides with the intersection of $H$ and $V$ lines. After
that, the partition $\mu =(2,1)$ to be removed is easily recognizable. The
boxes labeled by $x$ help us to identify the multiple copies of $\boldsymbol{%
R}$ \ so that the partition $\lambda $ can be readily identified

\begin{figure}[tbp]
\begin{center}
\includegraphics[width=3.45 in]{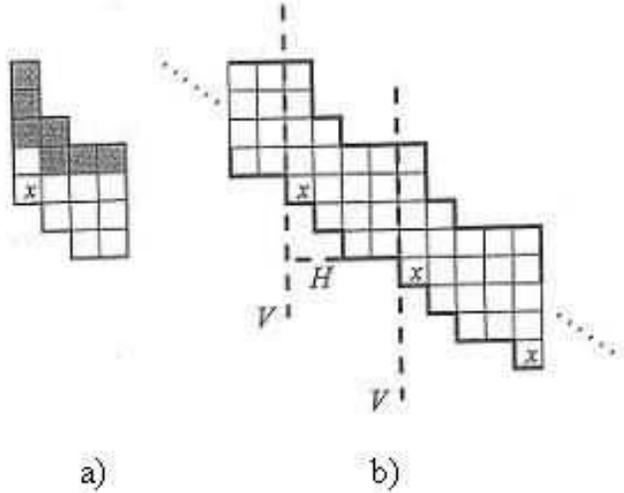}
\end{center}
\caption{The skew tableaux a) and its torus equivalent b)}
\end{figure}

Next, we remove a shaded boxes from $\lambda $ which is equivalent of
removal of one 7-rim hook (since\ in this example $k+m=7$). This leaves us
with the partition $\lambda ^{\prime }=(4,4,4,1).$ As it follows from the
Fig.15, such a partition still does not fit the $\boldsymbol{R}$ so that we
have to remove yet another 7-rim hook (just like in the Appendix B but in
reverse). The resulting (core) partition $\lambda ^{\prime \prime }=(3,3)$
does fit the rectangle $\boldsymbol{R}$. \ Such obtained skew partition can
be encoded by (3,3)/2/(2,1) or, more generally, $\lambda /d/\mu =(\lambda
_{1},...,\lambda _{k})/d/(\mu _{1},...,\mu _{l}).$ Hence, $d=2.$

With such defined \ toric skew partitions one can replace Eq.(7.23) by 
\begin{equation}
s_{\lambda /d/\mu }=\tsum\limits_{\nu \in k\times m}C_{\mu \nu }^{\lambda
}(d;m,k)s_{\nu }.  \tag{7.24}
\end{equation}%
If in the $d-$sum in Eq.(7.20) only one term is nonzero, then $C_{\mu \nu
}^{\lambda }(d;m,k)$ in Eq.(7.24) is the desired G-W invariant. In Ref.[90],
page 379, Witten argued that \ for dimensional reasons in the $d-$sum in
Eq.(7.20) only one term should be nonzero. If this is the case, then $C_{\mu
\nu }^{\lambda }(d;m,k)$ in Eq.(7.24) is the G-W invariant \ and coincides
with that given in Eq.(7.20). Moreover, if in Eq.(7.20) \ we formally choose 
$q=1$ thus obtained fusion rule coincides exactly with that for the Hecke
algebra, Eq.(6.9b), for which $Q$ is the $m$-th root of 1, i.e. $Q^{m}=1($%
recall that $N=m+k)$ as it was demonstrated by Goodman and Wenzl [91]. These
authors noticed that under such circumstances the fusion coefficients
coincide with those for the Verlinde algebra for $su(k)$ Wess-Zumino-Witten
\ (WZW) CFT model at the level $m\footnote{%
In view of symmetry, Eq.(7.18), evidently, one can talk as well about $su(m)$%
WZW model at the level $k$.}$.

\textbf{Remark.7.10.} Since connections with representations of Hecke
algebra are insured by previous stated Theorem 7.3. by Jones, while
connections between the K-Z type equations and theWZW-type models were
established by Knizhnik and Zamolodchikov [92], this explains (in view of
Goodman-Wenzl results) why in calculations of G-W invariants presented above
and in Appendix B the rectangular boxes are used.

\textbf{Remark 7.11. }\ These facts are not sufficient for explanation of
the mathematical meaning of the rim-hook removals (additions) in such
calculations. This deficiency is corrected to a some extent in Appendix C.

Obtained results give us an opportunity to discuss some additional physical
applications.

\subsubsection{Possible applications to solid state physics}

In Part I, Eq.(I,3.22), or in Eq.(1.5) of this work, we noticed that such an
equation can be interpreted within a context of solid state physics. In view
of earlier obtained results we would like to discuss this connection in some
detail now. In solid state physics the Bloch theorem [5] reflects the
difference between the wave function in the vacuum and that in the periodic
solid. Account for periodicity requires the wave function to be written in
the form $\Psi _{\mathbf{k}}(\mathbf{r})=e^{i\mathbf{k\cdot r}}u_{\mathbf{k}%
}(\mathbf{r}),$ where $u_{\mathbf{k}}(\mathbf{r})$ is some periodic
function. The vector $\mathbf{k}$ is determined by the condition 
\begin{equation}
e^{i\mathbf{k\cdot l}}=1,  \tag{7.25}
\end{equation}%
where $\mathbf{l}$ is the minimal translation vector of the \textit{direct}
lattice. Since this type of equation we have encountered already in Part II,
e.g. see Eq.(II.9.18), we know that solution is given by 
\begin{equation}
\mathbf{k\cdot l=0\func{mod}}\text{ }2\pi d  \tag{7.26}
\end{equation}%
which implies that the vector $\mathbf{k}$ \ should belong to the \textit{%
reciprocal (or dual)} lattice and $d$ should be some integer. \ In Part I we
called this equation as the Kac-Moody-Bloch-Bragg condition (K-M-B-B
condition). Clearly, Eq.(7.11) is of the same type. This fact underscores
its solid state physics relevance. In particular, it causes all physically
interesting quantities to change accordingly so that, for instance, the
momentum vectors for electrons are defined only with accuracy \ up to some
vectors of the reciprocal lattice. This causes energy of the electron not to
be well defined (as explained below). Eq.(7.26) suggests strong links with
number theory. The book by Terras, Ref.[93], especially Chr.10, provides an
excellent starting point for development of theory of electronic band
structures of solids (and also of molecules) using methods of number theory.
Clearly, in doing so one has to use group theory in the number fields of
characteristic other than zero as explained in Appendix C.

Some uses of the WZW-type models in solid state physics are mentioned in the
book by Tsvelik [94]. Thus far they are restricted only to 1+1 dimensional
models whose exact solutions can be obtained by other methods anyway. We
would like to argue that \textsl{all} many -body problems of solid state
physics in which lattice periodicity cannot be disregarded should employ
mathematical formalism discussed in this paper.

That this is the case can be seen from the simplest example. Indeed, the
existing band theory of solids [5] is essentially the theory \textsl{one
electron} in the periodic 3d lattices, e.g. all known lattices \ $\mathcal{L}
$ are made as direct sum \ $\mathcal{L}$=$\left( \boldsymbol{Z}/m_{1}\mathbf{%
Z}\right) \oplus \left( \boldsymbol{Z}/m_{2}\mathbf{Z}\right) \otimes \left( 
\boldsymbol{Z}/m_{3}\mathbf{Z}\right) ,$ \ where $m_{1},m_{2},m_{3}$ are
respective lattice periods. All such lattices have 2 kinds of symmetry:
point-like and spatial. The affine Weyl-Coxeter reflection groups are
capable of describing these symmetries. The word "affine" accounts for
effects of translational symmetry of the lattice (as explained in Appendix A
to Part II). This symmetry is so important that the presence of potential in
the Schr\"{o}dinger equation can be often ignored [95]. The band theory
developed for the "\textsl{empty}" lattice sometimes is sufficient for good
reproduction of experimental data. For the empty lattice the energy $%
\mathcal{E}(\mathbf{k})$ of a single electron is known to be [96], Chr1.,%
\begin{equation}
\mathcal{E}(\mathbf{k})=\frac{\hbar }{2m}\left\vert \mathbf{k}+\mathbf{K}%
\right\vert ^{2},  \tag{7.27}
\end{equation}%
where $\mathbf{K}$ is \textit{any} vector of the reciprocal lattice. \ In
view of Eq.(7.25) the vector $\mathbf{k}$ can have only \textit{finite}
number of values [96]. Unlike the traditional band theory of solids dealing
with one electron in periodic lattice $\mathcal{L,}$ we would like to place
several electrons into such a lattice so that the total energy of
noninteracting (except, due to the Pauli principle) is given by 
\begin{equation}
\frac{2m}{\hbar }E^{T}=\tsum\limits_{i}\left\vert \mathbf{k}_{i}+\mathbf{K}%
\right\vert ^{2}.  \tag{7.28}
\end{equation}%
Since the vector $\mathbf{K}$ is arbitrary we choose it to be the same for
all electrons. If this is the case, the above expression can be rewritten as 
\begin{equation}
\frac{2m}{\hbar }E^{T}=\tsum\limits_{i}(\mathbf{k}_{i}^{2}+2\mathbf{k}_{i}%
\mathbf{\cdot K+K}^{2}).  \tag{7.29}
\end{equation}%
Since the total momentum of such system of electrons should be conserved
this requires us to write 
\begin{equation}
\tsum\limits_{i}\mathbf{k}_{i}=0.  \tag{7.30}
\end{equation}%
Finally, summing over all energy levels (whose number is finite), that is
taking a trace of the corresponding matrix operator, brings us back to the
equation analogous to Eq.(7.12a). The Pauli exclusion principle requires the
total wave function to be antisymmetric. This requirement is satisfied by
the factorized wave function made of product of spin and
coordinate-dependent part. In the absence of lattice periodicity the
procedure of constructing such antisymmetric total wave function for several
electrons can be found in the book by Messiah, Ref.[97]. This procedure uses
essentially the representation theory of symmetric group $S_{n}$. Presence
of lattice periodicity replaces $S_{n}$ by its affine analog as discussed in
Appendix C. The above picture can be complicated by accounting for effects
of the constant magnetic field acting on electrons in periodic lattices.
Rigorous mathematical treatment of this problem has been initiated in works
by Novikov and his collaborators [98]. The small quantum cohomology ring
discussed in earlier subsections becomes the Novikov ring under present
circumstances [82].

\bigskip

\textbf{Acknowledgement }The author would like to thank Allen Knutson (U of
California, Berkeley) for his kind permission to reproduce some figures from
his papers (with T.Tao and C.Woodward) in this paper.

\textbf{\bigskip }

{\huge \ }

\textbf{Appendix A. Details of Heisenberg's derivation of the commutator
identity }$[\hat{x},\hat{p}]=i\hbar $

\bigskip

In this Appendix we would like to provide some details of Heisenberg's
reasoning leading to the discovery of $[\hat{x},\hat{p}]=i\hbar $. \ We do
this for several reasons. First, this would be unnecessary should his
original Nobel Prize winning paper, Ref. [18], contain all details and would
be free of typographical errors. Second, although in Section 5 we mentioned
Dirac's acknowledgement (on page 177 of his book, Ref.[22]) of the fact that
light scattering experiments associated with measurement of the refractive
index (or dielectric constant) and their theoretical interpretation had lead
Heisenberg to his discovery of quantum mechanics, nowhere in the existing
literature \ on quantum mechanics were we able to find exposition using this
historic fact as the starting point for \ the development of quantum
mechanical formalism.

At the classical level consider a gas of noninteracting atoms, better just
one atom containing $N$ electrons which are assumed to scatter light
independently. The interaction between the incoming light and such an
electron is described with help of the combination \textbf{d}$=\beta $%
\textbf{E }where\textbf{\ d }is the dipole moment of the electron in the
atom, \textbf{E} is the strength of the external electric field which is
assumed to be time-dependent, and $\beta $ is the polarization tensor (in
the simplest case it is assumed to be a scalar). In the medium the strength
of the electric field changes as compared to the vacuum. By denoting it as 
\textbf{D} it is known that \textbf{D}=\textbf{E} +4$\pi \mathbf{P}$ where 
\textbf{P}=N\textbf{d. }Since, at the same time, by definition, \textbf{d}=e%
\textbf{r }we have to have an equation for\textbf{\ r. }It is given by%
\begin{equation}
\mathbf{\ddot{r}}+\omega _{0}^{2}\mathbf{r+\gamma \dot{r}=}\frac{e}{m}%
\mathbf{E(}t\mathbf{),}  \tag{A.1}
\end{equation}%
where $e$ is electron's charge and $m$ is its mass. In writing this equation
it is assumed that our electron is bound harmonically (with the basic
frequency $\omega _{0}^{2})$ and that the friction is of \ known
(electromagnetic) nature and is assumed to be small. Using Fourier
decomposition of \textbf{r}(t) we obtain,%
\begin{equation}
\mathbf{r}(\omega )=\frac{e}{m}\frac{\mathbf{E}}{\omega _{0}^{2}-\omega
^{2}+i\omega \gamma }.  \tag{A.2}
\end{equation}%
This equation allows\ us to obtain \textbf{P} and, hence, \textbf{D }as
follows\textbf{\ }:%
\begin{equation}
\mathbf{D}=\mathbf{E}+4\pi \mathbf{P}=(1+4\pi N\text{ }\frac{e^{2}}{m}\frac{1%
}{\omega _{0}^{2}-\omega ^{2}+i\omega \gamma })\mathbf{E}\equiv \varepsilon
(\omega )\mathbf{E.}  \tag{A.3}
\end{equation}%
This equation defines a complex frequency-dependent dielectric constant $%
\varepsilon (\omega ).$ From electrodynamics it can be equivalently
rewritten as $\varepsilon (\omega )=(n(\omega )-i\varkappa (\omega ))^{2}$
where $n(\omega )$ is the refractive index while $\varkappa (\omega )$ is
the coefficient of absorption. Using these facts we can write approximately%
\begin{equation}
n(\omega )=1+2\pi N\text{ }\frac{e^{2}}{m}\frac{1}{\omega _{0}^{2}-\omega
^{2}+i\omega \gamma }.  \tag{A.4}
\end{equation}%
By ignoring friction in the high frequency limit we obtain,%
\begin{equation}
n(\omega )=1-2\pi N\text{ }\frac{e^{2}}{m\omega ^{2}}.  \tag{A.5.}
\end{equation}%
To account for quantum mechanical effects, Thomas, Reich and Kuhn in 1925
(just \textbf{before} the quantum mechanics was born !) have suggested to
replace Eq.(A.4) by 
\begin{equation}
n(\omega )=1+2\pi N\text{ }\frac{e^{2}}{m}\sum\limits_{i}\frac{f_{i}}{\omega
_{i0}^{2}-\omega ^{2}},  \tag{A.6}
\end{equation}%
where, following these authors, we ignored friction and introduced the 
\textit{oscillator strength} $f_{i}.$To reconcile Eq.(A.6) with (A.5) we
have to require $\sum\limits_{i}f_{i}=1.$ This requirement is known as the 
\textit{sum rule}. \ These facts were known to Kramers and Heisenberg,
Ref.[99], where our readers can find additional details. To make our point
and to save space, we would like to reobtain the result, Eq.(A.6), quantum
mechanically \ using modern formalism. We refer our reader to Ref.[100],
pages 316-319], for additional details. Basically, we need to calculate
quantum mechanically the dipole moment \textbf{d, }that is 
\begin{equation}
\mathbf{d}_{m}=\int \psi _{m}^{\ast }e\mathbf{r}\psi _{m}d^{3}\mathbf{r.} 
\tag{A.7.}
\end{equation}%
In this expression the wave function $\psi _{m}$ is calculated with help of
the stationary perturbation theory with accuracy up to the first order in
perturbation (which is e\textbf{r}$\cdot $\textbf{E}). A short calculation
produces the following result for the oscillator strength:%
\begin{equation}
f_{km}=\frac{2m\omega _{km}}{\hbar }\left\vert \left\langle k\mid \hat{x}%
\mid m\right\rangle \right\vert ^{2}.  \tag{A.8}
\end{equation}%
This result can be equivalently rewritten as 
\begin{equation}
f_{km}=\frac{m\omega _{km}}{\hbar }\{\left\langle k\mid \hat{x}\mid
m\right\rangle ^{\ast }\left\langle k\mid \hat{x}\mid m\right\rangle
+\left\langle k\mid \hat{x}\mid m\right\rangle ^{\ast }\left\langle k\mid 
\hat{x}\mid m\right\rangle \}.  \tag{A.9}
\end{equation}%
Since, however,%
\begin{equation}
im\omega _{km}\left\langle k\mid \hat{x}\mid m\right\rangle =\left\langle
k\mid \hat{p}_{x}\mid m\right\rangle   \tag{A.10}
\end{equation}%
we can rewrite Eq.(A.9) as 
\begin{equation}
f_{km}=\frac{1}{i\hbar }\{\left\langle m\mid \hat{x}\mid k\right\rangle
\left\langle k\mid \hat{p}_{x}\mid m\right\rangle -\left\langle m\mid \hat{p}%
_{x}\mid k\right\rangle \left\langle k\mid \hat{x}\mid m\right\rangle \} 
\tag{A.11}
\end{equation}%
since $\omega _{km}=-\omega _{mk}$. Finally, we have to require $%
\sum\limits_{k}f_{km}=1.$ This is possible only if 
\begin{equation}
\frac{1}{i\hbar }\left\langle m\mid \hat{x}\hat{p}_{x}-p_{x}\hat{x}\mid
m\right\rangle =1,  \tag{A.12}
\end{equation}

QED.

\bigskip

\textbf{Appendix B}. \textbf{An} \textbf{example of detailed computation of }%
$C_{\lambda \mu }^{\nu ,d}(m,k).$

\bigskip

In this appendix we \ would like to work out an example of computation of
the Gromov-Witten invariant $C_{\lambda \mu }^{\nu ,d}(m,k)$ based on
Example 1 taken from Ref.[86]. Our calculations differ however from those in
Ref.[86] since we use the KT scheme for computation of the classical L-R
coefficients.

In Example 1 the basic Young tableaux rectangle is taken as $m\times
k=5\times 5$. From here we obtain: $m+k=N=10$ \ Hence, the length of the
rim-hooks to be used is 10. Next, we are given partitions $\lambda
=(5,4,4,2,2),\mu =(3,2,1)$ and $\nu =(2,1).$ Based on these data, we can
calculate the weights. These are respectively $\left\vert \lambda
\right\vert =17,\left\vert \mu \right\vert =6$ and $\left\vert \nu
\right\vert =3.$ Since we know already that $N=10,$ the fundamental
relation, Eq.(7.11b), that is $\left\vert \lambda \right\vert +\left\vert
\mu \right\vert =\left\vert \nu \right\vert +dN$ now leaves us no choice for 
$d$. We obtain, $d=2$. But this number gives us the number of the 10-rim
hooks to be used in our computations, that is 2. If we choose $\nu $ as the
core partition then, indeed, only 2 rim hooks will fill in the rectangle as
depicted in Fig.16 below


\begin{figure}[tbp]
\begin{center}
\includegraphics[width=3.35 in]{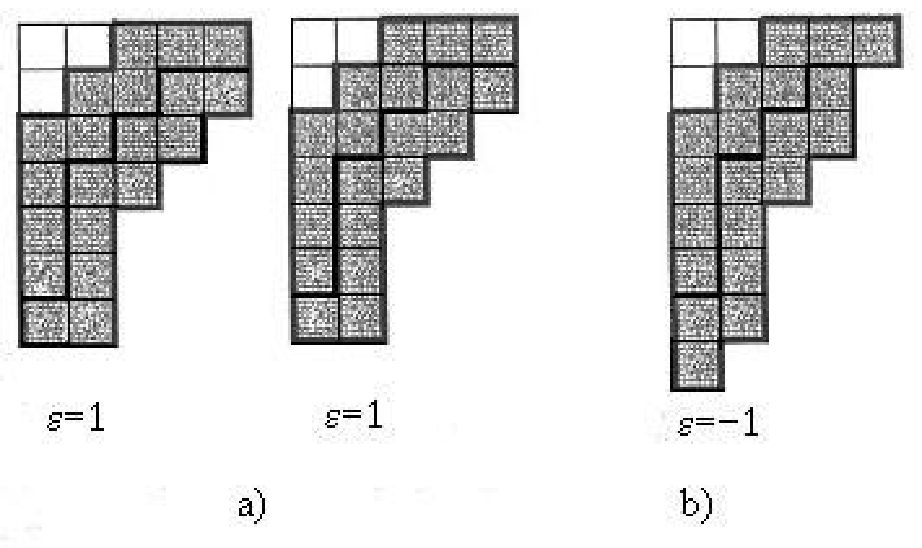}
\end{center}
\caption{$\protect\rho _{1}$ with its rim-hooks and phase factors a), the
same for $\protect\rho _{2}$ b)}
\end{figure}

The filling of the rectangle (in the present case- the square) is made in
such a way that the added rim hooks must not be added beyond the pre
assigned width of the rectangle but, at the same time, they are allowed to
occupy more height space than the rectangle can provide. This results in two
partitions $\rho _{1}=(5,5,4,3,2,2,2)$ and \ $\rho _{2}=(5,4,4,3,2,2,2,1)$ \
which can be read off directly from Fig.16. The summation over $\rho $ in
Eq.(7.20) in the present case takes place over $\rho _{1}$ and $\rho _{2}.$
The phase factor $\varepsilon (\rho /\nu )$ under this summation can be
easily computed based on the information given in Fig.16. For this purpose
we need to calculate the widths of the 10-rim hooks in both tableaux
depicted in Fig.16. For the first tableaux the widths are $w_{1}=5$ and $%
w_{2}=5$ respectively. Hence, the overall phase factor $\varepsilon (\rho
_{1}/\nu )=(-1)^{5-5}\cdot (-1)^{5-5}=1$. Analogously, for the second
tableaux the phase factor $\varepsilon (\rho _{2}/\nu )=(-1)^{5-5}\cdot
(-1)^{5-4}=-1.$To complete our calculation, we need the actual values for
the classical L-R coefficients $C_{\mu \nu }^{\rho _{1}}$ and $C_{\mu \nu
}^{\rho _{2}}.$ We calculate them graphically with help of the KT method
using Fig.6 as an example. This example indicates that we have to make some
adjustments in the initial data for partitions in order to be able to use
the KT scheme.

Hence, let us begin with calculations of $C_{\mu \nu }^{\rho _{1}}$. The
enlarged partitions having the same number of entries are $\lambda
=(5,4,4,2,2,0,0),\mu =(3,2,1,0,0,0,0)$ and $\rho _{1}=(5,5,4,3,2,2,2)$. It
should be noted that the locations of the added zeros are quite arbitrary
and that their redundancy should be discarded in actual calculations (that
is this redundancy should not affect the numerical value of $C_{\mu \nu
}^{\rho _{1}}).$ With these remarks we need to make a set of Y-shaped
tripods of the type depicted in Fig.1. By doing so we need to match together
the numbers from $\lambda $ and $\mu $ partitions in such a way that the
partition $\rho _{1}$ is obtained. Also, we must pay attention to the order
\ in which the labeling of the honeycomb is made, e.g. see Fig.3. \ After
that, the number of possibilities for such type of matching is $C_{\mu \nu
}^{\rho _{1}}.$ If we disregard redundancy of added zeros, then we
straightforwardly obtain $C_{\mu \nu }^{\rho _{1}}=2$ . To repeat this
procedure for $\rho _{2}$ we need to use the following partitions $\lambda
=(5,4,4,2,2,0,0,0),\mu =(3,2,1,0,0,0,0,0)$ and $\rho _{2}=(5,4,4,3,2,2,2,1).$
The result of matching now produces $C_{\mu \nu }^{\rho _{2}}=1$.Hence,
altogether we obtain: $C_{\lambda \mu }^{\nu }(d;m,k)=2-1=1,$ in accord with
Ref.[86].

\ 

\bigskip \textbf{Appendix C}. \textbf{Hecke algebra and Kashiwara crystals}

\ 

The purpose of this Appendix is only to provide a list of key references
needed in support of results of the main text. Obviously, the choice of
references is subjective. Nevertheless, it \ is hoped that it provides a
sufficient background level needed for reading any other literature on these
subjects.

We begin with Eq.(7.22). This result can be looked upon using theory of
k-restricted (bounded) partitions recently developed by Lapointe and Morse
[101\textbf{]}. These authors were looking at the following problem: how
fusion, Eq.(3.8), is going to change if instead of standard Schur functions $%
s_{\lambda }$ one would use $k-$bounded Schur functions $s_{\lambda }^{(k)}$
indexed by partitions $\lambda $ whose first part (i.e. $\lambda _{1})$ is
no larger than a fixed integer $k\geq 1?$ For such functions the L-R
coefficient $C_{\lambda \mu }^{\nu }$ should be replaced by $C_{\lambda \mu
}^{\nu ,k}.$ These authors managed to prove (see their Theorem 18) that, in
accord with our Eq.(7.22), for the appropriately chosen $\ k$ \ (actually
for $k=N-1$) the coefficient $C_{\lambda \mu }^{\check{\nu},N-1}$ $%
=C_{\lambda \mu }^{\nu ,d}(m,k)$ (with $N=k+m$). These results can be looked
upon from broader perspective [102].To this purpose we introduce the \textsl{%
affine} symmetric group $\hat{S}_{k}$ generated by $k$ elements $\hat{s}_{0},%
\hat{s}_{1},,..,\hat{s}_{k-1}$ and satisfying the affine Coxeter relations$%
_{{}}$%
\begin{equation}
\begin{array}{c}
\hat{s}_{i}^{2}=1, \\ 
\hat{s}_{i}\hat{s}_{j}=\hat{s}_{j}\hat{s}_{i}\text{ , }i-j=\pm \text{1}\func{%
mod}k, \\ 
\hat{s}_{i}\hat{s}_{i+1}\hat{s}_{i}=\hat{s}_{i+1}\hat{s}_{i}\hat{s}_{i+1}.%
\end{array}
\tag{C.1}
\end{equation}%
The generators $\hat{s}_{i}$ should be understood as $\hat{s}_{i\func{mod}k}$
if $i>k.$The usual symmetric group $S_{k}$ embeds into $\hat{S}_{k}$ as a
subgroup generated by $s_{0},s_{1},,..,s_{k-1}$ but this embedding is not
unique: there are many embeddings among which one has to choose the most
convenient \textbf{[}102]. The question arises: how $s_{\lambda }^{(k)}$ is
connected with representations for $\ \hat{S}_{k}?$ We would like to give an
answer in several steps. First, we recall that the Hecke algebra $H_{n}$ is
the deformation of $S_{n}$ with the \ deformation parameter $Q.$ Characters
of symmetric group $s_{\lambda }$ and those for Hecke algebra have the same
fusion algebra controlled by the L-R coefficients (as discussed in the main
text), except for the case when Q is a nontrivial root of 1. In the last
case the fusion algebra for the Hecke characters coincides with the Verlinde
algebra as demonstrated by Goodman and Wenzl [91]. Lapointe and Morse [101]
argue that the fusion algebra for $k-$bounded Schur functions $s_{\lambda
}^{(k)}$ coincides with the Verlinde algebra for the WZW models. \ Since
generators $D_{i}$ in Eq.(6.9) for the Hecke algebra $H_{n}$ are
deformations of the generators $s_{i}$ for symmetric group $S_{n},$ one can
think about deformation of generators $\hat{s}_{i}$ for the affine symmetric
group $\hat{S}_{n}.$ In Ref.[103] one can find that such a deformation is
described by the affine Hecke algebra $\hat{H}_{n}(Q)$ which is the Weyl
reflection group for the affine Lie algebra $%
\begin{array}{c}
\widehat{\mathbf{sl}}_{n}.%
\end{array}%
$ Finally, according to Ref.s[104],\ [105] representations of $\widehat{%
\mathbf{sl}}_{n}$ and those for the Hecke algebra at root of unity are
interrelated albeit \ in a very nontrivial way (as explained below). Hence, $%
s_{\lambda }^{(k)}$ (up to a constant) may coincide with representation for
the Hecke algebra at $\sqrt[n]{1}.$ $s_{\lambda }^{(k)}$ form a Schubert
basis for the cohomology ring of the affine Grassmannian as it is explained
by Lam [106]. Affine Grassmannian was recently discussed in works by
Kapustin and Witten [107] and also, independently, by Frenkel [108]. In both
cases it was discussed in connection with applications of \ methods of
number theory to string and CFT.

Because of this, we would like to explain how the number theory enters into
previous discussion. For this purpose, we would like to discuss the notion
of a crystal and a crystal base following Ref.[109]. A crystal is made of a
set $B$ endowed with the maps $\tilde{e}_{i},\tilde{f}_{i}:B\sqcup
\{0\}\rightarrow B\sqcup \{0\}$ ($i\in I),$ where $I$ is an index set. These
maps satisfy the following conditions: a) $\tilde{e}_{i}0=\tilde{f}_{i}0=0;b)
$ $\tilde{e}_{i}^{n}b=\tilde{f}_{i}^{n}b=0$ $\forall b\in B$ and $i\in I$ ,
c) $\exists b,b^{\prime }\in B$ and$\ i\in I$ such that $b^{\prime }=\tilde{f%
}_{i}b$ if and only if $b=\tilde{e}_{i}b^{\prime }.$

The above definition is too abstract to see the connection with crystals.
This deficiency is easily correctable. Using results and notations of
Appendix A (Part II) we introduce the weight lattice $P_{l}^{+}$ (typical
for the affine reflection groups) by%
\begin{equation}
P_{l}^{+}=\{\tsum\limits_{i=0}^{n}a_{i}\omega _{i}\mid a_{i}\geq 0,\text{ }%
a_{i}\in \mathbf{Z}\text{ }\forall i=1\div n;\tsum\limits_{i=0}^{n}a_{i}=l\}
\tag{C.2}
\end{equation}%
so that $B=P^{+}=\tbigcup\nolimits_{l>0}P_{l}^{+}.$ At the same time,
Eq.(C.2) has the number-theoretic meaning (e.g.see Ref.[110], Appendix) in
the case if $a_{i}\in \mathbf{F}_{q}\equiv \mathbf{Z}/q\mathbf{Z}$ with $q$
being some prime number. In this case the expansion $\beta =$ $%
\tsum\limits_{i=0}^{n}a_{i}\omega _{i}$ can be interpreted as an expansion
of a number $\beta $ which belongs to the field extension of the number
field $\mathbf{F}_{q}.$ Since such an extension corresponds to a particular
solution of the familiar (by now)\ equation $\tsum\limits_{i=0}^{n}a_{i}=l,$
different solutions of this equation represent different numbers $\beta $.
This observation is sufficient for explanation of a role of the operators $%
\tilde{e}_{i}$ and $\tilde{f}_{i}$ in the emerging picture. \ Indeed, by
analogy with Section 3 we can associate with each solution of $%
\tsum\limits_{i=0}^{n}a_{i}=l$ the Young diagram. The new element in doing
so lies in the fact that boxes in such Young tableaux should be filled with
numbers which belong to the field $\mathbf{F}_{q}$. This is rather easy to
do, e.g. see Ref.[105]. With this information, a crystal graph can be
constructed as follows. We take a set of $\beta ^{\prime }s$ as vertices of
the graph\ (so that $\beta ^{\prime }s\rightleftharpoons b^{\prime }s)$
then, the operators $\tilde{f}_{i}$ connect vertices related by $b^{\prime }=%
\tilde{f}_{i}b.$ The index $i$ represents a color so that the crystal graph
is directed and colored by colors from the set $I$. The connections with
representation theory and fusion can be figured out now based on the
observation that different $\beta ^{\prime }s$ are in one-to one
correspondence with different Young diagrams. Each such diagram encodes a
character for the respective group, e.g. in characteristic zero characters
for $S_{n}$ are the Schur functions $s_{\lambda },$etc$.$Hence, as we just
explained, the crystal graphs can be used instead of KT honeycombs (or
puzzles) for calculation of fusion coefficients [111].

Finally, we would like to provide a few additional details regarding the
actual role being played by the operators $\tilde{f}_{i}$ and $\tilde{e}_{i}$
in order to connect the results of this appendix to that of Appendix B and
Section 7.4.2. In these sections we've noticed that different Young diagrams
can be obtained from the set of core Young diagrams by adding(removing) the
appropriate rim-hooks. The core Young diagrams can be looked upon as
representing a vacuum state while the addition(removal) of the particular
rim-hook is done by the raising (lowering) operator. In the simplest case we
are dealing with addition(removal) of just one box of the Young diagram.
Hence, we can associate the addition (removal) of just one box with the
operator $\tilde{e}_{i}$ (with the operator $\tilde{f}_{i}).$ Clearly, if we
need to remove(add) a rim-hook the appropriate combination of such raising
(lowering) operators should be used. Interestingly enough, a systematic
development of the formalism we just sketched leads to the Heisenberg
commutation rule, Eq.(5.8). Details can be found in Ref.[112]. The same
formalism for $q=1$ is used for description of representations of the Hecke
algebra H$_{m}(\sqrt[n]{1})$, Ref.[105], page 279.

\textbf{Remark}. We would like to mention that the concept of a crystal was
developed by Kashiwara, e.g. see Ref.[113], in connection with obtaining the
exact solution of XXZ model in the \textsl{thermodynamic limit (that is in
the limit of infinite} \textsl{spin chain}). The same authors notice that
the whole formalism of quantum groups, vertex operators, Virasoro algebra,
etc. was developed and makes sense either for infinite chains or for finite
chains with appropriately chosen boundary conditions. In the case of finite
chains without specially chosen boundary conditions the whole apparatus of
this Appendix fails. This observation is very important for development of
our new string- theoretic formalism. It is totally consistent with results
presented in Parts I-III and will be further discussed in Part IV,
especially in connection to and comparison with recent alternative
string-theoretic developments[114].

\bigskip

{\huge \ }

{\Large References}

\bigskip

[1] A.Kholodenko, New strings for old Veneziano amplitudes I.

\ \ \ \ Analytical treatment, \ J.Geom.Phys.55 (2005) 50-74.

[2] A.Kholodenko, New strings for old Veneziano amplitudes II.

\ \ \ \ Group-theoretic treatment, J.Geom.Phys.56 (2006) 1387-1432.

[3] A.Kholodenko, New strings for old Veneziano amplitudes III.

\ \ \ \ Symplectic treatment, J.Geom.Phys.56 (2006) 1433-1472.

[4] R.Stanley, Combinatorics and Commutative Algebra,

\ \ \ \ Birkh\"{a}user, Boston, 1996.

[5] J.Ziman, Principles of the Theory of Solids,

\ \ \ \ Cambridge University Press, Cambridge, UK, 1972.

[6] S.Ghorpade,G.Lachaud, Hyperplane sections of Grassmannians

\ \ \ \ and the number of MDS linear codes,

\ \ \ \ Finite fields \ Their Appl. 7 (2001) 468-506.

[7] V. Kac, Infinite Dimensional Lie Algebrs,

\ \ \ \ Cambridge University Press, Cambridge, UK, 1990.

[8] N.Bourbaki, Groupes at Algebres de Lie

\ \ \ \ (Chapitre 4-6), Hermann, Paris,1968.

[9] R.Stanley, Enumerative Combinatorics, Vol.I,

\ \ \ \ Cambridge University Press, Cambridge, UK, 1999.

[10] H.Weyl, Das asymptotische Verteiilungsgesetz der

\ \ \ \ \ Eigenwerte lineare partieller Differentialgleichungen,

\ \ \ \ \ Math.Ann. 71 (1912) 441-479.

[11] S.Daftuar, P. Hayden, Quantum state transformations and the

\ \ \ \ \ \ Schubert calculus, Ann. Physics 315 (2005) 80--122.

[12] A.Klyachko, Quantum marginal problem and representation

\ \ \ \ \ \ of the symmetric group, arxiv.org : quant-ph/0409113.

[13] A.Horn, Eigenvalues of sums of Hermitian matrices,

\ \ \ \ \ \ Pacific J.of Math. 12 (1962) 225-241.

[14] A. Klyachko, Stable vector bundles and Hermitian operators,

\ \ \ \ \ \ Selecta Math.(N.S.) 4 (1998) 419-445.

[15] A.Knutson,T.Tao, The honeycomb model of GL$_{n}$(\textbf{C}) tensor

\ \ \ \ \ \ producsts I: Proof of the saturation conjecture,

\ \ \ \ \ \ J.AMS 12 (1999) 1055-1090.

[16] A.Knutson,T.Tao, C.Woodward, The honeycomb model of GL$_{n}$(\textbf{C})

\ \ \ \ \ \ tensor producsts II: Puzzles determine facets of the

\ \ \ \ \ \ Littlewood-Richardson cone, J.AMS 17 (2003) 19-48.

[17] S.Friedland, Finite and infinite dimensional generalizations of
Klyachko's

\ \ \ \ \ \ theorem, Linear Algebra Appl. 319 (2000) 3--22.

[18] W.Heisenberg, Collected Works, pages 382-396,

\ \ \ \ \ \ Springer-Verlag, Berlin, 1985.

[19] A.Kholodenko, New strings for old Veneziano amplitudes IV.

\ \ \ \ \ \ Connections with spin chains and 

\ \ \ \ \ \ other stochastic systems, arXiv:0805.0113.

[20] A.Knutson, T.Tao, Honeycombs and sums of Hermitian matrices,

\ \ \ \ \ \ AMS Notices 48 (2001) 175-186.

[21] P.Dirac, Fundamental equations of quantum mechanics,

\ \ \ \ \ \ Proc.Roy.Soc.(Lond) A109 (1925) 642-653.

[22] P.Dirac, Principles of Quantum Mechanics,

\ \ \ \ \ \ Clarendon Press, Oxford, 1930.

[23] B.Lidskii, Spectral polytope for the sum of two Hermitian matrices,

\ \ \ \ \ \ Funct.Analysis Applic. (in Russian) 16 (1982) 76-77.

[24] A.Zelevinsky, Littlewood-Richardson semigroups. Math. Sci. Res. Inst.

\ \ \ \ \ \ Publ., 38, Cambridge Univ. Press, Cambridge, 1999.

[25] S.Mandelstam, Dual resonance models, Phys.Reports13 (1974) 259-353.

[26] R.Bott, L.Tu, Differential Forms in Algebraic Topology,

\ \ \ \ \ \ Springer-Verlag, Berlin, 1982.

[27] A.Postnikov, Affine approach to quantum Schubert calculus,

\ \ \ \ \ \ Duke Math.J. 128\ (2005) 473-509.

[28] H.Hiller, Geometry of Coxeter Groups,

\ \ \ \ \ \ Pitman Advanced Publ.Program, Boston, 1982.

[29] W.Fulton, Young Tableaux, Cambridge U.Press,

\ \ \ \ \ \ Cambridge, 1997.

[30] L.Manivel, Symmetric Functions, Schubert Polynomials

\ \ \ \ \ \ and Degeneracy Loci, Soc. Math.de France, Paris, 1998.

[31] H.Tamvakis, The connection between representation theory

\ \ \ \ \ \ and Schubert calculus, Enseign. Math. 50 (2004) 267--286.

[32] I.Macdonald, Symmetric Functions and Orthogonal Polynomials,

\ \ \ \ \ \ AMS University Lecture Notes Vol.12, Providence, RI, 1998.

[33] \ L.Lapointe, L.Vinet, Exact operator solution of the

\ \ \ \ \ \ \ Calogero-Sutherland model, Comm. Math. Phys. 178 (1996)
425--452.

[34] \ J.Minahan, A. Polychronakos, Equivalence of two-dimensional QCD

\ \ \ \ \ \ \ and the c=1 matrix model, Phys. Lett. B 312 (1993) 155--165.

[35] \ A. Gorsky, N. Nekrasov, Relativistic Calogero-Moser model

\ \ \ \ \ \ \ as gauged WZW theory. Nuclear Phys. B 436 (1995) 582--608.

[36] \ I.Macdonald, Symmetric Functions and Hall Polynomials,

\ \ \ \ \ \ \ Clarendon Press, Oxford, 1995.

[37] \ R.Stanley, Enumerative Combinatorics, Vol.II,

\ \ \ \ \ \ \ Cambridge University Press, Cambridge, UK, 1999.

[38] \ W.Fulton, Eigenvalues, invariant factors, highest weights,

\ \ \ \ \ \ \ and Schubert calculus, Bull. Amer. Math. Soc.

\ \ \ \ \ \ \ (N.S.) 37 (2000) 209--249.

[39] \ W.Dittrich, M.Reuter, Classical and Quantum Dynamics,

\ \ \ \ \ \ \ Springer-Verlag, Berlin, 1992.

[40] \ V.Arnol'd, Mathematical Methods of Classical Mechanics,(in Russian)

\ \ \ \ \ \ \ \ Moscow, Nauka, 1974.

[41] \ M.van Leeuwen, The Littlewood-Richardson rule and related

\ \ \ \ \ \ \ \ combinatorics, MSJ Memoirs 12 (2001) 95-145.

[42] \ A.Berenstein, A. Zelevinsky, Tensor product multiplicities and

\ \ \ \ \ \ \ \ convex polytopes \ in partition space. J. Geom. Phys. 5
(1988) 453--472.

[43] \ R.King, C. Tollu, F. Toumazet, Stretched Littlewood-Richardson

\ \ \ \ \ \ \ and Kostka coefficients, CRM Proc. Lecture Notes, 34 (2004)
99--112,

\ \ \ \ \ \ \ AMS, Providence, RI, 2004.

[44] \ A.Buch, The saturation conjecture (after A.Knutson and T.Tao,

\ \ \ \ \ \ \ \ with appendix by Fulton ), Enseign.Math.46(2000) 43-60.

[45] \ I.Pak, E.Vallejo, Combinatorics and geometry of

\ \ \ \ \ \ \ Littlewood-Richardson cones,

\ \ \ \ \ \ \ European J. of Combinatorics, 26 (2005) 995-1008.

[46] \ H.Derksen, J.Weyman, On the Littlewood-Richardson polynomials,

\ \ \ \ \ \ \ J.of Algebra 255 (2002) 247-297.

[47] \ A.Kholodenko, Kontsevich-Witten model from 2+1 gravity:

\ \ \ \ \ \ \ new exact combinatorial solution, J. Geom.Physics 43 (2002)
45-91.

[48] \ M.Born, Mechanics of Atom, (In German),

\ \ \ \ \ \ \ Springer-Verlag, Berlin, 1924.

[49] \ W.Pauli, M.Born, \"{U}ber die Quantelung gest \"{o}rther mechanischer

\ \ \ \ \ \ \ \ Systeme, Z.Phys.10 (1922) 137-158.

[50] \ E.Baly, Spectroscopy, Longmans,Green and Co, London, 1905.

[51] \ M.Born, W.Heisenberg, P. Jordan, Z\"{u}r Quantenmechanick II,

\ \ \ \ \ \ \ \ Z.Phys. 35 (1926) 557-615.

[52] \ S.Adler, Quantum Theory as Emergent Phenomenon,

\ \ \ \ \ \ \ Cambridge University Press, Cambridge, 2004.

[53] \ P.Orlik, H. Terao, Arrangements and Hypergeometric Integrals,

\ \ \ \ \ \ \ MSJ Memoirs, Vol.9, Tokyo, 2001.

[54] \ K.Aomoto, Configurations and invariant Gauss-Manin connections

\ \ \ \ \ \ \ of integrals. I. Tokyo J. Math. 5 (1982) 249--287.

[55] \ V.Vassiliev, Applied Picard-Lefschetz Theory, AMS,

\ \ \ \ \ \ \ Providence, RI, 2002.

[56] \ M.Saito, B.Sturmfels, N.Takayama, Gr\"{o}bner Deformations of

\ \ \ \ \ \ \ Hypergeometric Differential Equations, Springer-Verlag,
Berlin, 2000.

[57] \ E. Opdam, Multivariable hypergeometric functions,

\ \ \ \ \ \ \ European Congress of Mathematics, Vol. I (Barcelona, 2000),
491--508,

\ \ \ \ \ \ \ Progr. Math., 201, Birkh\"{a}user, Basel, 2001.

[58] \ A.Kirillov Jr., Lectures on affine Hecke algebras and Macdonald's

\ \ \ \ \ \ \ conjectures, Bull. Amer. Math. Soc. (N.S.) 34 (1997) 251--292.

[59] \ M.R\"{o}sler, M.Voit, Markov processes related with Dunkl operators,

\ \ \ \ \ \ \ Adv.Appl.Math. 21 (1998) 575-643.

[60] \ \ J-M.Soriau, Structure of Dynamical Systems,

\ \ \ \ \ \ \ Birkh\"{a}user, Boston, 1997.

[61] \ \ S.Bates, A.Weinstein, Lectures on Geometry of Quantization,

\ \ \ \ \ \ \ AMS, Providence, RI,1997.

[62] \ \ P.Diaconis, Group Representations in Probability and Statistics,

\ \ \ \ \ \ \ \ Inst.of Math.Statistics Lecture Notes, Vol.11, Hayward,
Ca,1988.

[63] \ \ A. Lascoux, M-P Sch\"{u}tzenberger, Symmetry and flag manifolds,

\ \ \ \ \ \ \ \ LNM 996 (1980) 118-144.

[64] \ \ T.Kohno, Linear representations of braid groups and classical

\ \ \ \ \ \ \ \ Yang-Baxter equations, Cont.Math. 78 (1988) 339-363.

[65] \ \ M.Atiyah, The Geometry and Phyics of Knots,

\ \ \ \ \ \ \ \ Cambridge University Press, Cambridge, 1990.

[66] \ \ C.K.Fan, Structure of a Hecke algebra quotient,

\ \ \ \ \ \ \ \ \ J.AMS 10 (1997) 139-167.

[67] \ \ K.Kassel, Quantum Groups, Springer-Verlag, Berlin, 1995.

[68] \ \ A.Belavin, V.Drinfel'd, Solutions of classical Yang-Baxter

\ \ \ \ \ \ \ \ \ equation and simple Lie algebras, \ Funct.Anal.Appl.

\ \ \ \ \ \ \ \ 16 (1982) 159-180.

[69] \ \ P.Etingof, I.Frenkel, A.Kirillov Jr., Lectures on Representation

\ \ \ \ \ \ \ \ Theory and Knizhnik-Zamolodchikov Equations,

\ \ \ \ \ \ \ \ AMS,\ Providence, RI, 1998.

[70] \ \ O.Gleizer, A. Postnikov, Littlewood-Richardson coefficients 

\ \ \ \ \ \ \ \ via Yang-Baxter equation, 

\ \ \ \ \ \ \ \ Internat. Math. Res. Notices \ 14 (2000) 741--774.

[71] \ \ \ H.Wenzl, Hecke algebras of type $A_{n}$ and subfactors,

\ \ \ \ \ \ \ \ \ Inv.Math.92 (1988)349-383.

[72] \ \ \ \ M.Jimbo, A q-analogue of $U(\mathfrak{g}l(N+1))$ Hecke algebra,
and the

\ \ \ \ \ \ \ \ \ Yang-Baxter equation, Lett. Math. Phys. 11 (1986) 247--252.

[73] \ \ \ \ \ A.Ram, A Frobenius formula for the characters of the Hecke
algebras,

\ \ \ \ \ \ \ \ \ Invent. Math. 106 (1991) 461--488.

[74] \ \ \ \ A.Lascoux, B. Leclerc, J-Y.Thibon, Flag varieties and the 

\ \ \ \ \ \ \ \ \ Yang-Baxter equation, Lett. Math. Phys. 40 (1997) 75--90.

[75] \ \ \ \ R.King, \ B. Wybourne, Representations and traces of the Hecke

\ \ \ \ \ \ \ \ \ algebras $H_{n}(q)$ of type $A_{n-1},$ J. Math. Phys. 33
(1992) 4--14.

[76] \ \ \ \ V. Jones, Hecke algebra representations of braid groups and link

\ \ \ \ \ \ \ \ \ polynomials, Ann.Math.126 (1987) 335-388.

[77] \ \ \ J.Birman, Braids, Links , and Mapping Class Groups,

\ \ \ \ \ \ \ \ \ Princeton University Press, Princeton, 1975.

[78] \ \ \ \ A.Kholodenko, Use of quadratic differentials for description of

\ \ \ \ \ \ \ \ \ defects\ and textures in liquid crystals and 2+1 gravity,

\ \ \ \ \ \ \ \ \ J. Geom. Phys. 33 (2000) 59--102.

[79] \ \ \ \ B.Bakalov, A.Kirillov Jr., Lectures on Tensor Categories and

\ \ \ \ \ \ \ \ \ Modular Functors, AMS, Providence, RI, 2001.

[80] \ \ \ \ D.Anosov, S.Aranson, V.Arnol'd, I.Bronshtein,V.Grines,

\ \ \ \ \ \ \ \ \ \ Yu.Ilyashenko, Ordinary Differential equations and Smooth

\ \ \ \ \ \ \ \ \ \ Dynamical Systems, Springer-Verlag, Berlin, 1997.

[81] \ \ \ \ M.Ablowitz, P.Clarkson, Solitons, Nonlinear Evolution 

\ \ \ \ \ \ \ \ \ Equations and Inverse Scattering, 

\ \ \ \ \ \ \ \ \ Cambridge University Press, Cambridge, 1991.

[82] \ \ \ \ D.McDuff, D.Salamon, J-Holomorphic Curves and 

\ \ \ \ \ \ \ \ \ \ Symplectic Topology, AMS, Providence, RI, 2004.

[83] \ \ \ \ M.Walton, Tensor products and fusion rules,

\ \ \ \ \ \ \ \ \ \ \ Canadian J.of Physics 72 (1994) 727-736.

[84] \ \ \ \ J.Humphreys, Introduction to Lie Algebras and Representation

\ \ \ \ \ \ \ \ \ \ \ Theory, Springer-Verlag, Berlin, 1972.

[85] \ \ \ \ \ V.Guillemin, E.Rassart, Signature quantization and
representations

\ \ \ \ \ \ \ \ \ \ \ of compact Lie groups, Proc. Natl. Acad. Sci.

\ \ \ \ \ \ \ \ \ \ \ USA 101 (2004) 10884--10889.

[86] \ \ \ \ \ \ A.Bertram, I. Ciocan-Fontanine, W. Fulton, Quantum
multiplication

\ \ \ \ \ \ \ \ \ \ \ of Schur polynomials. J. Algebra 219 (1999) 728--746.

[87] \ \ \ \ \ D.Cox, S. Katz, Mirror Symmetry and Algebraic Geometry,

\ \ \ \ \ \ \ \ \ \ Mathematical Surveys and Monographs, Vol. 68,

\ \ \ \ \ \ \ \ \ \ AMS, Providence, RI, 1999.

[88] \ \ \ \ W.Fulton, R. Pandharipande, Notes on stable maps and quantum

\ \ \ \ \ \ \ \ \ \ \ cohomology, Proc. Sympos. Pure Math. Vol 62 (1995)
45-96.

[89] \ \ \ \ \ \ \ P.McNamara, Cylindric skew Schur functions, arxiv:
math.CO/0410301.

[90] \ \ \ \ \ E.Witten, The Verlinde algebra and the cohomology 

\ \ \ \ \ \ \ \ \ \ of the Grassmannian, in Geometry, Topology and Physics

\ \ \ \ \ \ \ \ \ \ for Raoul Bott, pp357-422., International Press Inc.,
Boston, 1995.

\ [91] \ \ \ \ \ F.Goodman, H. Wenzl, Littlewood-Richardson coefficients for
Hecke

\ \ \ \ \ \ \ \ \ \ \ algebras at roots of unity, Adv. Math. 82 (1990)
244--265.

\ [92] \ \ \ \ \ V.Knizhnik, A. Zamolodchikov, Current algebra and
Wess-Zumino

\ \ \ \ \ \ \ \ \ \ \ model in two dimensions, Nuclear Phys. B 247 (1984)
83--103.

\ [93] \ \ \ \ A.Terras, Fourier Analysis on Finite Groups and Applications,

\ \ \ \ \ \ \ \ \ \ \ Cambridge University Press, Cambridge, 1999.

\ \ [94] \ \ \ \ A.Tsvelik, Quantum Field Theory in Condensed Matter Physics,

\ \ \ \ \ \ \ \ \ \ \ Second edition, Cambridge University Press, Cambridge,
2003.

\ \ [95] \ \ \ \ F.Herman, Theoretical investigation of the electronic
energy band

\ \ \ \ \ \ \ \ \ \ \ structure of solids, rev.Mod.Phys. 30 (1958) 102-121.

\ \ [96] \  \ \ J.Callaway, Energy Band Theory, Academic Press, New York,
1964.

\ \ [97] \ \ \ A.Messiah, Quantum Mechanics, Vol.2, North Holland,

\ \ \ \ \ \ \ \ \ \ \ Amsterdam, 1962.

\ \ [98] \ \ \ \ S.Novikov, A.Maltsev, Topological phenomena in normal
metals,

\ \ \ \ \ \ \ \ \ \ \ \ arxiv: cond-mat/9709007.

\ \ [99] \ \ \ \ \ H.Kramers, W.Heisenberg, On the dispersion of radiation
by atoms

\ \ \ \ \ \ \ \ \ \ \ \ (In German), Z.fur.Phys.31 (1925) 681-708.

[100] \ \ \ \ \ A.Davydov, Quantum Mechanics, Pergamon, Oxford, 1965.

[101] \ \ \ \ \ L.Lapointe, J.Morse, Quantum cohomology and the k-Schur
basis,

\ \ \ \ \ \ \ \ \ \ \ \ \ arxiv: math.CO/0501529.

[102] \ \ \ \ \ \ \ T.Lam, Affine Stanley symmetric functions,
arxiv:math.CO/0501335.

[103] \ \ \ \ \ M. Kashiwara, T. Miwa, E.Stern, Decomposition of q-deformed

\ \ \ \ \ \ \ \ \ \ \ \ Fock spaces, arxiv: math.q-alg/9508006.

[104] \ \ \ \ \ A.Lascoux, B.Leclerc, J-Y.Thibon, Hecke algebras at rooths of

\ \ \ \ \ \ \ \ \ \ \ \ unity and crystal bases of quantum affine algebras,

\ \ \ \ \ \ \ \ \ \ \ Comm.Math.Phys. 181 (1996) 205-263.

[105] \ \ \ \ O.Foda, B.Leclerc, M.Okado, J-Y.Thibon, T.Welsch,

\ \ \ \ \ \ \ \ \ \ \ \ Combinatorics of solvable lattice models and modular

\ \ \ \ \ \ \ \ \ \ \  representation of Hecke algebras, in Geometric
Analysis

\ \ \ \ \ \ \ \ \ \  \ and Lie Theory in Mathematical Physics, pp. 243-290,

\ \ \ \ \ \ \ \ \ \ \ \ Cambridge University Press, Cambridge, 1998.

[106] \ \ \ \ T.Lam, Schubert polynomials and affine Grassmannian,

\ \ \ \ \ \ \ \ \ \ \ \ arxiv: math.CO/0603125.

[107] \ \ \ \ \ A.Kapustin, E.Witten, Electric-magnetic duality and Langlands

\ \ \ \ \ \ \ \ \ \ \ \ \ program, arxiv: hep-th/0604151.

[108] \ \ \ \ \ E.Frenkel, Lectures on the Langlands program and conformal

\ \ \ \ \ \ \ \ \ \ \ \ \ field theory, arxiv: hep-th/0512172.

[109] \ \ \ \ \ S.Kang, M.Kashiwara, K.Misra,T.Miwa, T.Nakashima,

\ \ \ \ \ \ \ \ \ \ \ \ \ A.Nakayashiki, Perfect crystals of quantum affine
Lie algebras,

\ \ \ \ \ \ \ \ \ \ \ \ Duke Math. Journal 68 (1992) 499-608.

[110] \ \ \ \ \ A.Kholodenko, "New" Veneziano amplitudes from "Old" Fermat

\ \ \ \ \ \ \ \ \ \ \ \ \ (hyper) surfaces, in

\ \ \ \ \ \ \ \ \ \  \ \ Trends in Mathematical Physics Research, pp1-94,

\ \ \ \ \ \ \ \ \ \ \ \ \ Nova Science Publishers, New York, 2004.

[111] \ \ \ \ \ G.Fourier, P.Littlemann, Weyl modules, Demazure modules,

\ \ \ \ \ \ \ \ \ \ \ \ \ K-R modules, crystals, fusion products and limit
constructions,

\ \ \ \ \ \ \ \ \ \ \ \ \ arxiv: math.RT/0509276.

[112] \ \ \ \ \ T.Lam, Ribbon tableaux and the Heisenberg algebra,

\ \ \ \ \ \ \ \ \ \ \ \ \ Math. Z. 250 (2005) 685--710.

[113] \ \ \ \ \ B.Davies, O.Foda, M.Jimbo, T.Miwa, A.Nakayashiki,

\ \ \ \ \ \ \ \ \ \ \ \ \ Diagonalization of the XXZ Hamiltonian by vertex
operators,

\ \ \ \ \ \ \ \ \ \ \ \ \ arxiv: hep-th/9204064.

[114] \ \ \ \ \ A.Gorsky, Gauge theories as string theories: the first
results,

\ \ \ \ \ \ \ \ \ \ \ \ Physics Uspekhi 175 (2005) 1145-1162.

\bigskip

\end{document}